\documentclass[preprint]{imsart}

\usepackage{amsthm,amsmath,natbib}

\usepackage{epsfig, amssymb}
\usepackage{multirow}
\usepackage{dsfont}
\startlocaldefs
\usepackage{graphics}
\usepackage{subcaption}

\newtheorem{thm}{Theorem}
\newtheorem{defi}{Definition}
\newtheorem{Prop}{Proposition}

\newcommand{\beq}{\begin{equation}}
\newcommand{\eeq}{\end{equation}}
\newcommand{\ba}{\begin{array}}
\newcommand{\ea}{\end{array}}

\def\a{\alpha}

\def\e{\varepsilon}

\endlocaldefs

\begin{document}

\begin{frontmatter}

\title{Copula Correlation: An Equitable Dependence Measure and Extension of Pearson's Correlation.}
\runtitle{Copula Correlation}


\begin{aug}
\author{\fnms{A. Adam} \snm{Ding} \thanksref{t1}\corref{}\ead[label=e1]{a.ding@neu.edu}}
\and
\author{\fnms{Yi} \snm{Li}\ead[label=e2]{li.yi3@husky.neu.edu}}
\thankstext{t1}{This research project is supported by NSF grant CCF-1442728}
 \runauthor{Ding and Li}

  \affiliation{Department of Mathematics, Northeastern University}

  \address{567 Lake Hall, \\
  360 Huntington Ave., \\
  Boston, MA 02115\\
          \printead{e1,e2}}

\end{aug}






\begin{abstract}
In {\it Science}, \citet{Reshef2011MIC} proposed the concept of equitability for measures of dependence between two random variables. To this end, they proposed a novel measure, the maximal information coefficient (MIC). Recently a PNAS paper \citep{Kinney2014} gave a mathematical definition for equitability. They proved that MIC in fact is not equitable, while a fundamental information theoretic measure, the mutual information (MI), is self-equitable. In this paper, we show that MI also does not correctly reflect the proportion of deterministic signals hidden in noisy data. We propose a new equitability definition based on this scenario. The copula correlation (Ccor), based on the $L_1$-distance of copula density, is shown to be equitable under both definitions. We also prove theoretically that Ccor is much easier to estimate than MI. Numerical studies illustrate the properties of the measures.
\end{abstract}



\end{frontmatter}

\section{INTRODUCTION}

With the advance of modern technology, the size of available data keeps exploding. Data mining is increasingly used to keep up with the trend, and to explore complex relationships among a vast number of variables. The nonlinear relationships are as important as the linear relationship in data exploration. Hence the traditional measure such as Pearson's linear correlation coefficient is no longer adequate for today's big data analysis. \citet{Reshef2011MIC} proposed the concept of equitability. That is, a dependence measure should give equal importance to linear and nonlinear relationships. For this purpose, they proposed a novel maximal information coefficient (MIC) measure.

The MIC measure stimulated great interest and further studies in the statistical community. \citet{Speed2011comment} praised it as ``a correlation for the 21st century". It has been quickly adopted by many researchers in data analysis. However, its mathematical and statistical properties are still not studied very well. There are also criticisms on the measure based on those properties.

MIC has been criticised for its low power in detecting dependence \citep{simon2011comment, SantosBrfBioinfo2013CompDepMeas, Heller2013IndTest}, in comparison to existing measures and tests. Particularly, \citet{simon2011comment} recommended the distance correlation (dcor) by~\citet{szekely2007dcor} over MIC. However, dcor does not have the equitable property. The equitable dependence measure is needed to properly rank the strength of relationships in data exploration. As we will discuss in detail later, the equitability is a different feature from the power of dependence testing.

\citet{Kinney2014} gives a strict mathematical definition of R$^2$-equitability described in \citet{Reshef2011MIC}. They discovered that no non-trivial statistic can be R$^2$-equitable, thus MIC is in fact not R$^2$-equitable. They further proposed a replacement definition of self-equitability. Interestingly, the MIC is also not self-equitable. \citet{Kinney2014}  recommended a fundamental measure from information theory, the mutual information (MI), which is self-equitable.

While the estimation of MI has been studied extensively in the literature, practitioners are often frustrated by the unreliability of these estimation~\citep{Fernandes2010MIhardEst, Reshef2011MIC}. We show that this is in fact due to a problem in the MI measure's definition: it does not correctly reflect the strength of deterministic relationships hidden in noise. We propose a new equitability definition to clarify the issue.

We relate the study of equitability to another popular line of research on the copula -- a joint probability distribution with uniform marginals. Sklar's Theorem decomposes any joint probability distribution into two components: the marginal distributions and the copula. The copula captures all the dependence information among the variables. Hence an equitable dependence measure should be copula-based. The copula-based dependence measures have been studied for a long time.  An earlier classic work by \citet{Schweizer1981MeasDep} proved many mathematical properties for several copula-based dependence measures. With the advance of modern computing power, there are renewed high interest in copula-based dependence measures~\citep{Schmid2010CopMeasures, Poczos12, NIPS2013Lopez-PazRandDepCoef}.

Using copula, we mathematically define the robust-equitability condition: a dependence measure should equal the proportion of deterministic relationship (linear or nonlinear) hidden in uniform background noise. Hence such measures equal Pearson's correlation for linear relationship hidden in uniform background noise, and  extend Pearson's correlation to all deterministic relationships hidden in uniform background noise.
We propose a new robust-equitable measure, the copula correlation (Ccor), which is defined as half the $L_1$-distance of the copula density function from independence. This measure was used as a test statistic for independence testing before~\citep{ChanJTSA1992NonparaTestDep,Tjostheim1996MeasDep,Bagnato2013TestInd}. For discrete random variables, it is also called as the Kolmogorov dependence measure in the pattern recognition literature~\citep{Vilmansen1972Dep,Vilmansen1973Feature,Ekdahl2006BoundLoss} and as the Mortara
dependence index~\citep{Bagnato2013TestInd}. We consider the measure for continuous variables, and refer to it as the copula correlation. The name emphasizes the facts that it is a copula-based dependence measure, and that it is an extension of Pearson's correlation. The $L_1$-distance based statistics are robust in many statistical application. The $L_1$-distance based dependence measure here is robust to mixture of some deterministic data with continuous data, properly reflect the dependence strength in the mixture.

We shall show that Ccor is both self-equitable and robust-equitable. On the other hand, MI is not robust-equitable. This also provides insights on the difficulty to estimate MI.
Some authors \citep{pal2010estimation, liu2012exponential} studied the convergence of MI estimators by imposing the H\"{o}lder condition on the copula density. This H\"{o}lder condition, while being a standard condition for density estimations, does not hold for any commonly used copula \citep{omelka2009copula,segers2012copula}.
Under a more realistic H\"{o}lder condition on the bounded region of copula density, we provide a theoretical proof that the mutual information (MI)'s minimax risk is infinite. This provides a theoretical explanation on the statistical difficulty of estimating MI observed by practitioners. In contrast, Ccor is consistently estimable under the same condition.

Section~\ref{sec:meas_backgr} prepares the notations by defining several dependence measures and relating equitability to the copula. A weak-equitability definition is introduced which relates to copula-based measures. We define our new measure Ccor and review some existing dependence measures in literature, including MIC, MI, dcor, etc. We review the copula-based measures by~\citet{Schweizer1981MeasDep}, and their modified version of R\'enyi's Axioms~\citep{Renyi1959MeasDep}. We clarify the relationship between these Axioms and the equitability.
Section~\ref{sec:equit} reviews the equitability definitions of \cite{Kinney2014}, and studies the self-equitability of these dependence measures. The self-equitable measures such as MI may not reflect the proportion of deterministic signal in data correctly. This motivates our definition of equitable extension of the Pearson's linear correlation coefficient. Section~\ref{sec:cop-equit} mathematically formulate this into our robust-equitability definition. Ccor is the only measure proven to be both self-equitable and robust-equitable. Multivariate extension is also discussed. Section~\ref{sec:EstErr} further studies the convergence of estimators for the two self-equitable measures MI and Ccor. Ccor is shown to be easier to estimate theoretically than MI. This and its equitability provide the desirable theoretical properties for the applications of Ccor in big data  exploration. The estimation of MI have been studied extensively in literature. MI can be estimated using methods including kernel density estimation (KDE) method~\citep{Moon1995MIkde}, the $k$-nearest-neighbor (KNN) method~\citep{kraskov2004estimating},  maximum likelihood estimation of density ratio method~\citep{suzuki2009mutual}, etc. We advocate that more attention should be paid to estimating Ccor instead. In this paper, we propose a KDE-based estimator for Ccor. Section~\ref{sec:Numerical} compares the numerical performance of this estimator $\widetilde{Ccor}$ with other dependence measures through simulation studies and a real data analysis. The Ccor is shown to rank the strength of dependence relationship better than other measures. It also provides good performance in the real data. We end the paper with proofs and summary discussions.

\section{COPULA AND DEPENDENCE MEASURES}\label{sec:meas_backgr}

We review several classes of dependence measures $D(X;Y)$ between two random variables $X$ and $Y$  in the literature, and introduce our proposed new measure.
For simplicity, we will focus on the dependence measures for two continuous univariate random variables $X$ and $Y$ in most of the paper. The multivariate extension will be discussed in Section~\ref{sec:multivariate}.

\subsection{Weak-equitability and Copula-based Dependence Measures}\label{sec:weak-equit}

The most commonly used dependence measure is Pearson's linear correlation coefficient $\rho(X;Y)= Cov(X,Y)/\sqrt{Var(X)Var(Y)}$ where $Cov(X,Y)$ denotes the covariance between $X$ and $Y$, and $Var(X)$ denotes the variance of $X$. The linear correlation coefficient $\rho$ is good at characterizing linear relationships between $X$ and $Y$:  $|\rho|=1$ for perfectly deterministic linear relationship, and $\rho=0$ when $X$ and $Y$ are independent. However, it does not measure the nonlinear relationships between $X$ and $Y$ well.

To motivate the equitability concept, we can look at three examples in the left half of Table~\ref{tab:CopulaTrans}, where the two continuous random variable $X$ and $Y$ are related by deterministic relationships: linear in (A); nonlinear in (B) and (C). These examples illustrate two deficiencies for Pearson's linear correlation coefficient $\rho(X;Y)$:
\begin{itemize}
  \item [(D1)] It is not invariant to monotone transformations of the variables. The value would change, say, using a logarithm/exponential scale. The $\rho$ value is lower in example (B) than (A) of Table~\ref{tab:CopulaTrans} under a logarithm transformation of $X$.
  \item [(D2)] $\rho$ does not treat all deterministic relationship equally, and can not capture some non-monotone nonlinear relationships. In example (C), $\rho=0$ for $X$ and $Y$ related by the nonlinear relationship $Y=\cos(4 \pi X)$, in contrast to $\rho=1$ in the linear relationship of example (A).
\end{itemize}

\begin{table} [htbp]
\begin{center}
\small{
\begin{tabular}{llllll}
    \hline
 \multicolumn{3}{c}{raw data scale} & \multicolumn{3}{c}{copula transformation} \\
  \parbox[c]{18pt}{\includegraphics[scale=0.1]{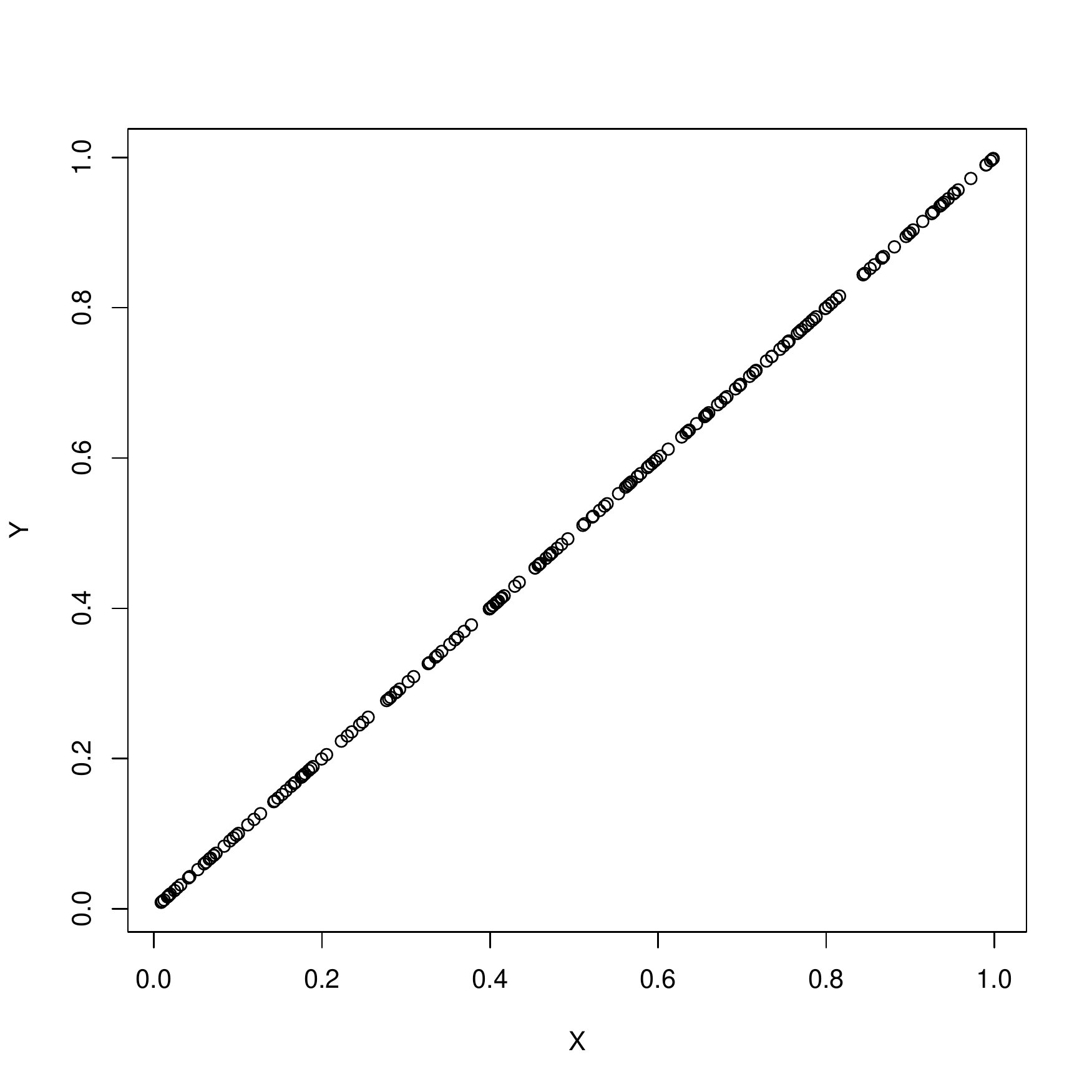}} \ & \parbox[c]{18pt}{\includegraphics[scale=0.1]{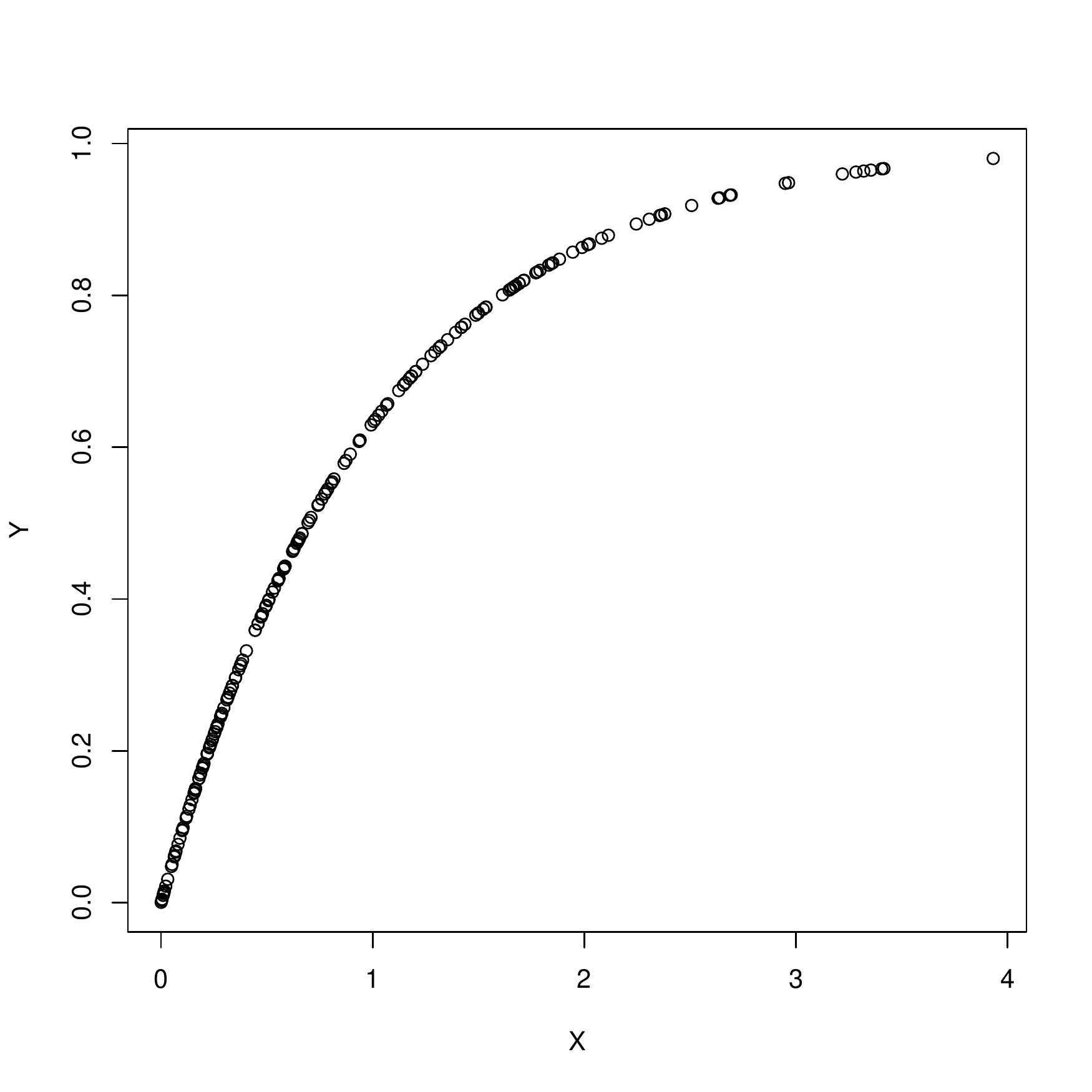}} \ & \parbox[c]{18pt}{\includegraphics[scale=0.1]{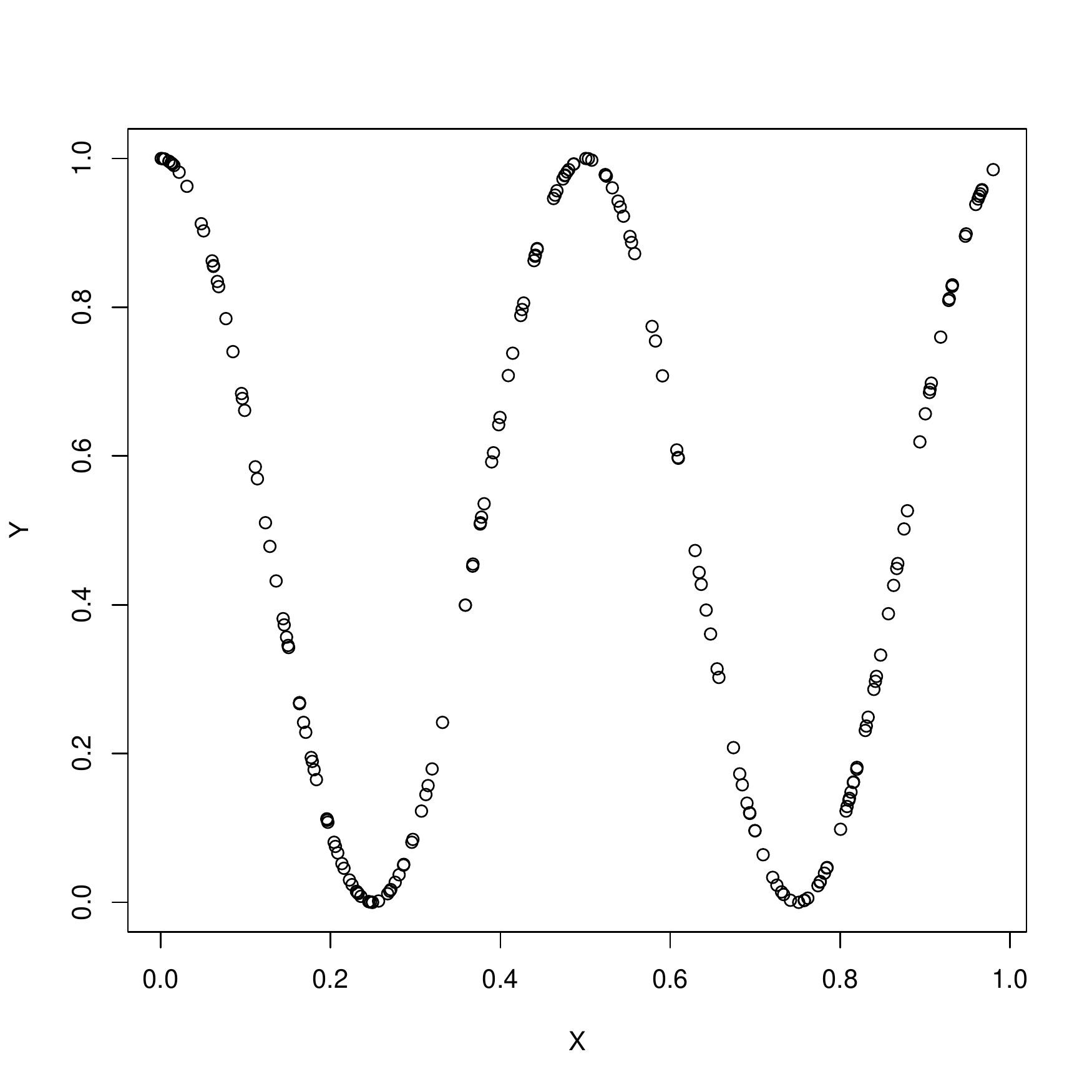}} \ &
 \parbox[c]{18pt}{\includegraphics[scale=0.1]{lin1}} \ &
 \parbox[c]{18pt}{\includegraphics[scale=0.1]{lin1}} \ &
 \parbox[c]{18pt}{\includegraphics[scale=0.1]{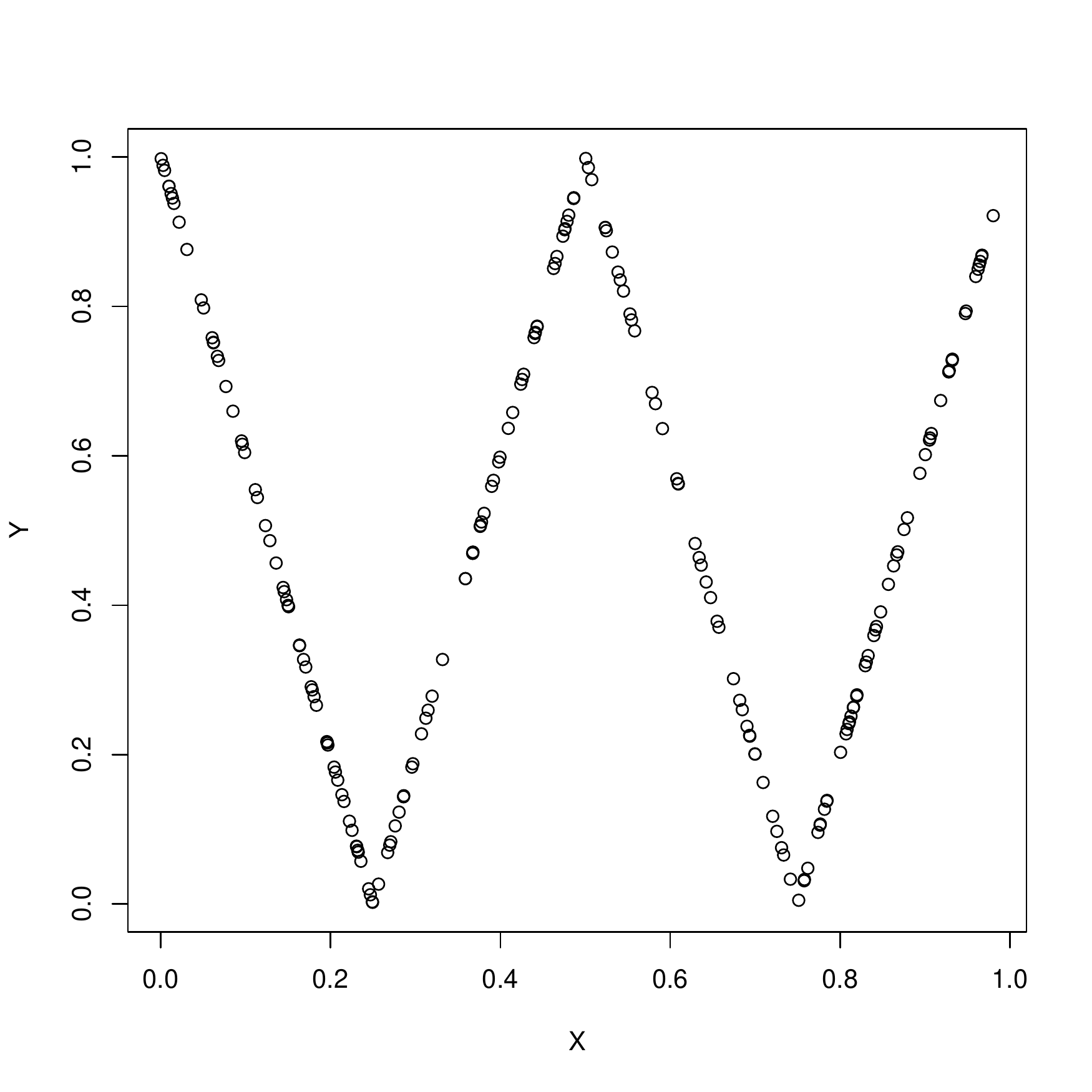}} \\
 A. $\rho=1$ \ \ & B. $\rho=0.866$ & C. $\rho=0$ \ \ & A. $\rho=1$ \ \ & B. $\rho=1$ \ \ & C. $\rho=0$ \ \ \\
    \hline
\end{tabular}
}
\end{center}
\caption{Pearson's linear correlation on three functional relationships.}
\label{tab:CopulaTrans}
\end{table}

\cite{Kinney2014} mathematically defines equitability of a dependence measure $D[X; Y]$ through its invariance under certain transformations of the random variables $X$ and $Y$. The deficiency (D1) above provides the original motivation for invariance consideration. For example, if we change the unit of $X$ (or $Y$), the values of $X$ (or $Y$) changes by a constant multiple, but should not affect the dependence measure $D[X; Y]$ at all. Similarly, if we apply a monotone transformation on $X$ (e.g. the commonly used logarithmic or exponential transformation), then the dependence with $Y$ should not be affected and the measure $D[X; Y]$ should remain the same.
For dependence scanning in data mining/variable selection, invariance to monotone transformations of the variables is very important, since we do not know beforehand the appropriate scale of each variable.
This leads to our following definition of weak-equitability.
\begin{defi}\label{weak-equit}
A dependence measure $D[X; Y]$ is weakly-equitable if and only if
$D[X; Y] = D[f(X); Y]$ whenever $f$ is a strictly monotone continuous deterministic function.
\end{defi}

The weak-equitability property relates to the popular copula concept. The Sklar's theorem ensures that, for any joint distribution function $F_{X,Y}(x,y)=Pr(X \le x, Y \le y)$, there exists a copula $C$ -- a probability distribution on the unit square $\mathcal{I}^2 = [0,1] \times [0,1]$ -- such that
\beq \label{sklar}
F_{X,Y}(x,y) = C[F_X(x), F_Y(y)] \qquad \mbox{for all } x, y.
\eeq
Here $F_X(x)=Pr(X \le x)$ and $F_Y(y)=Pr(Y \le y)$ are the marginal cumulative distribution functions (CDFs) of $X$ and $Y$ respectively. The copula $C$ captures all the dependence between $X$ and $Y$.

\begin{figure}[h]
	\centering
        \begin{subfigure}[b]{0.3\textwidth}
               \includegraphics[width=\textwidth]{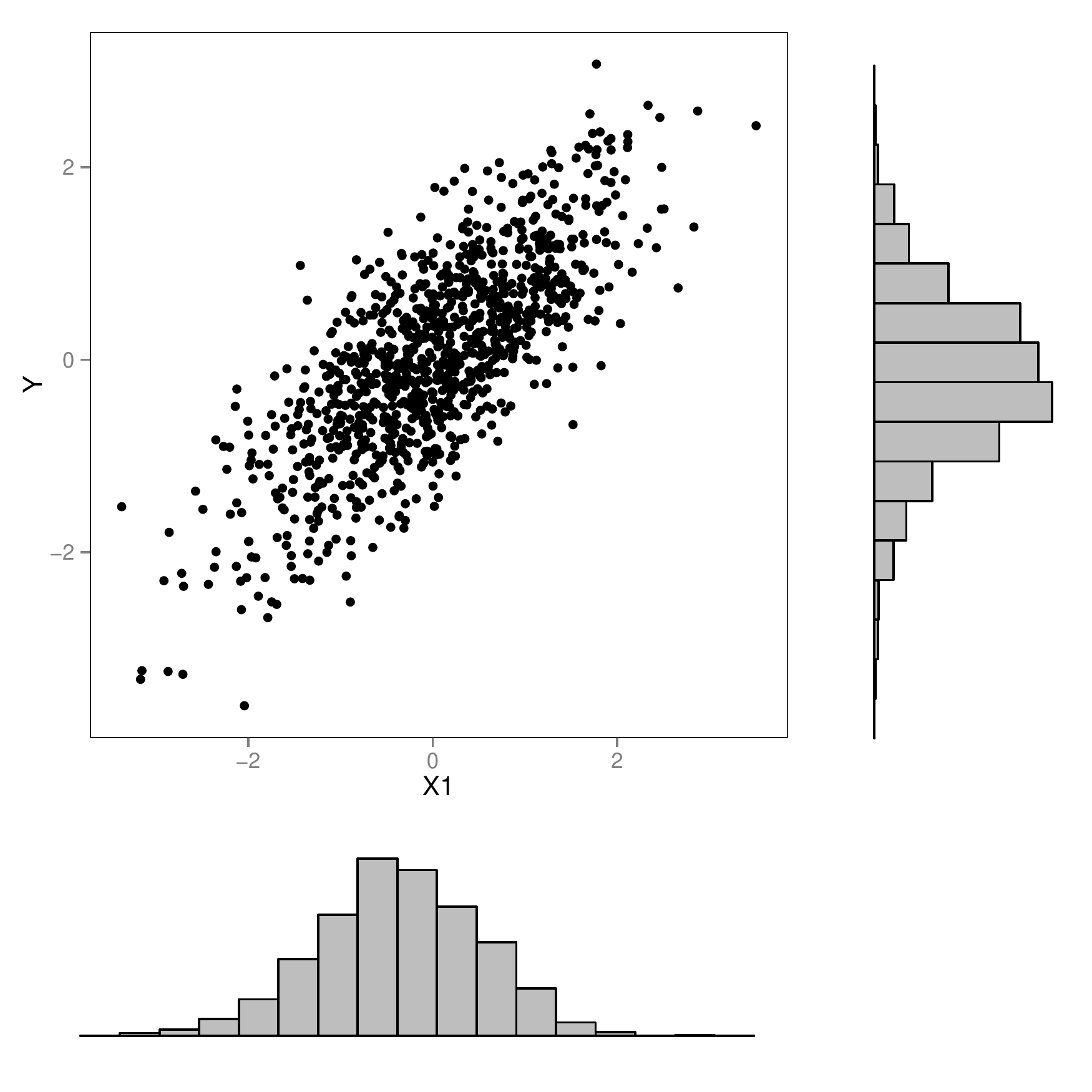}
                \caption{Bivariate Gaussian}
                \label{fig:CopulaTrans.a}
        \end{subfigure}
 	\quad
         \begin{subfigure}[b]{0.3\textwidth}
               \includegraphics[width=\textwidth]{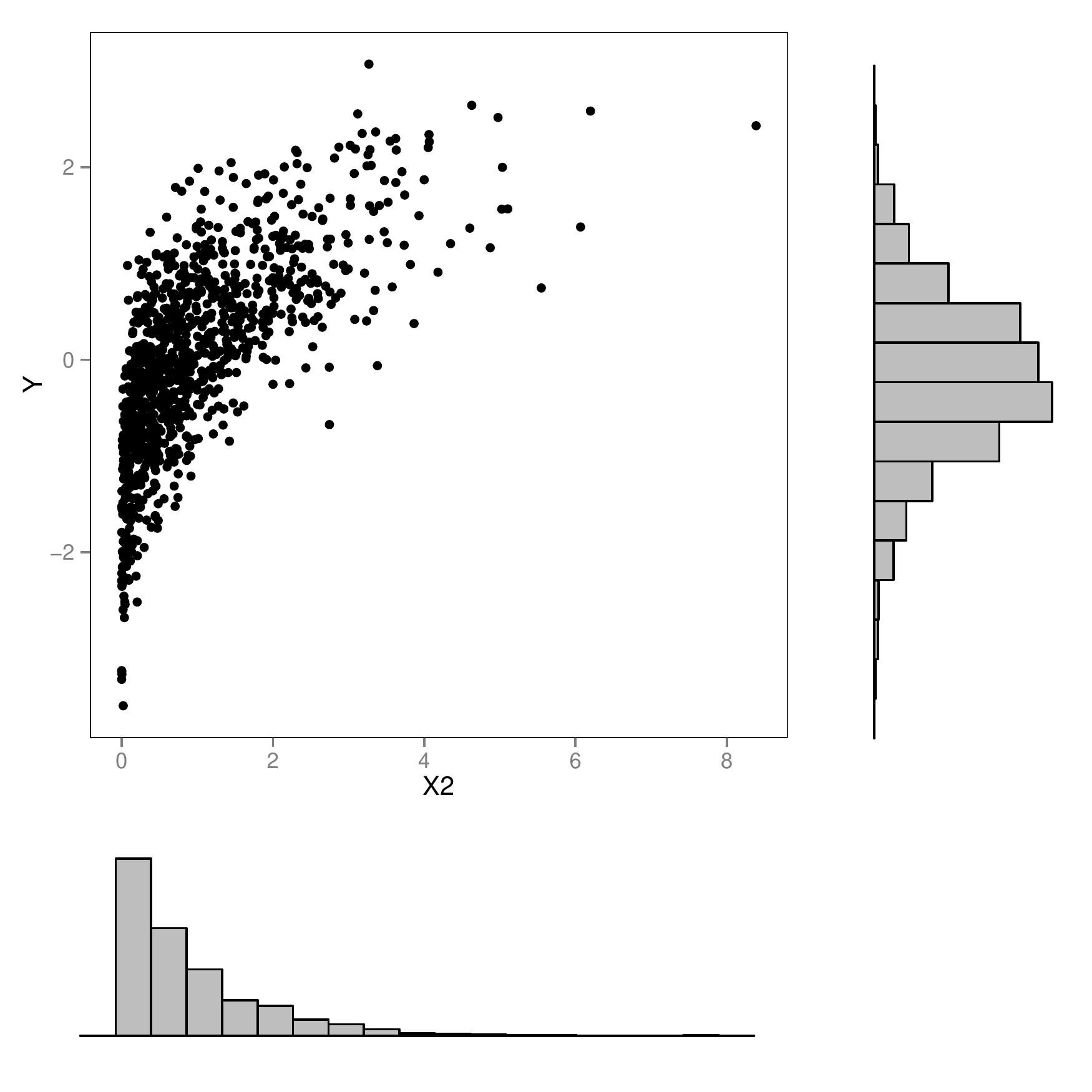}
                \caption{Different marginals}
                \label{fig:CopulaTrans.b}
        \end{subfigure}
	\quad
         \begin{subfigure}[b]{0.3\textwidth}
               \includegraphics[width=\textwidth]{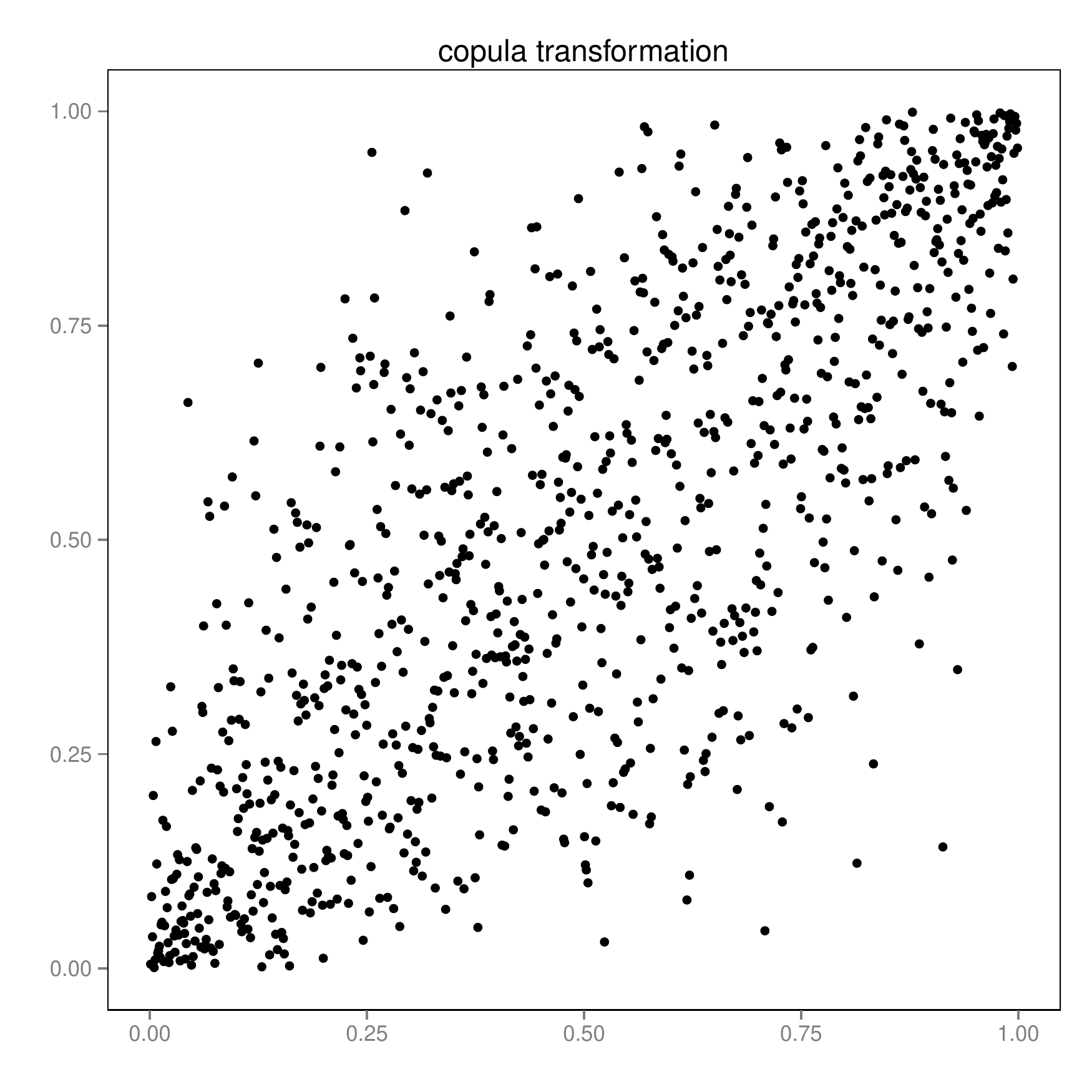}
                \caption{The Guassian copula}
                \label{fig:CopulaTrans.c}
        \end{subfigure}
 	\caption{(a) Bivariate Gaussian data with $\rho=0.75$. (b) The data with exponential marginal for $X$. (c) The Gaussian copula. The first two distributions both have the same copula as in (c).}
	\label{fig:CopulaTrans}
\end{figure}

The copula decomposition separates the dependence (copula) from any marginal effects. Figure~\ref{fig:CopulaTrans} shows the data from two distributions with different marginals but the same dependence structure.

We call a dependence measure $D[X; Y]$ {\it symmetric} if $D[X; Y] = D[Y; X]$ for all random variables $X$ and $Y$. Then a symmetric weakly-equitable measure satisfies the monotone-invariance property: $D[X; Y]$ is invariant to strictly monotone continuous transformations both for $X$ and for $Y$. A symmetric dependence measure $D[X; Y]$ is weakly-equitable if and only if $D[X; Y]$ depends on the copula $C(u,v)$ only and is not affected by the marginals $F_X(x)$ and $F_Y(y)$. In other words, the symmetric weakly-equitable dependence measures are defined on the copula-transformed, uniformly distributed, variables $U=F_X(X)$ and $V=F_Y(Y)$.
The right half of Table~\ref{tab:CopulaTrans} shows the copula-transformed variables for Examples (A), (B) and (C) in contrast to the original variables on the left. Calculating the linear correlation coefficient on the copula-transformed variables leads to the Spearman's Rho, which is weakly-equitable. This remedies the first deficiency  (D1) above, as shown in Examples (A) and (B) in Table~\ref{tab:CopulaTrans} after copula-transformation. The deficiency (D2) is still not solved by copula-transformation in example (C). We will address this in section~\ref{sec:equit},  as this relates to the equitability concept of treating all deterministic relationships equally.

\subsection{R\'enyi's Axioms for Nonlinear Dependence Measures}\label{sec:renyi}
\citet{Schweizer1981MeasDep} showed that several copula-based dependence measures $D[X; Y]$ satisfy a modified version of R\'enyi's Axioms on two continuously distributed random variables $X$ and $Y$.
\begin{enumerate}
  \item[A1.] $D[X; Y]$ is defined for any $X$ and $Y$.
  \item[A2.] $D[X; Y] = D[Y; X]$.
  \item[A3.] $0 \le D[X; Y] \le 1$.
  \item[A4.] $D(X;Y)=0$ if and only if $X$ and $Y$ are statistically independent.
  \item[A5.] $D(X;Y)=1$ if and only if each of $X$, $Y$ is a.s. a strictly monotone function of the other.
  \item[A6.] If $f$ and $g$ are strictly monotone a.s. on $Range(X)$ and $Range(Y)$, respectively, then $D[f(X); g(Y)] = D[X; Y]$.
  \item[A7.] If the joint distribution of $X$ and $Y$ is bivariate Gaussian, with linear correlation coefficient $\rho$, then $D[X; Y]$ is a strictly increasing function of $|\rho|$.
\end{enumerate}

\citet{Renyi1959MeasDep}'s original axioms differ from the \citet{Schweizer1981MeasDep}'s version in that: (i) They were not restricted to continuously distributed random variables; (ii) Axiom A5, A6 and A7 are replaced by:
\begin{enumerate}
 \item[A5a.] $D(X;Y)=1$ if either $X=f(Y)$ or $Y=g(X)$ for some Borel-measurable functions $f$ and $g$.
  \item[A6a.] If $f$ and $g$ are Borel-measurable, one-one mappings of the real line into itself then $D[f(X); g(Y)] = D[X; Y]$.
  \item[A7a.] If the joint distribution of $X$ and $Y$ is bivariate Gaussian, with linear correlation coefficient $\rho$, then $D[X; Y]=|\rho|$.
\end{enumerate}

We will mostly stick with continuous random variables as in~\citet{Schweizer1981MeasDep} so that we can relate to the copula representation. But we will also discuss the original A5a, A6a and A7a as they relate to the discussions on the equitability concept. The original R\'enyi's Axioms are too strong for nonparametric measures~\citep{Schweizer1981MeasDep}. The only known measure shown to satisfy all seven original R\'enyi's Axioms is the R\'enyi's maximum correlation coefficient (Rcor). The Rcor has a number of major drawbacks, e.g., it equals 1 too often and is generally not effectively computable~\citep{Schweizer1981MeasDep,szekely2009dcor}. We will discuss this more in section~\ref{sec:equit}. In section~\ref{sec:Numerical}, we will numerically study a recently proposed estimator for Rcor by \cite{NIPS2013Lopez-PazRandDepCoef}.

The Axiom A4 partially addresses the deficiency (D2) in the example (C) above. The Axiom A2 states that the measure is symmetric. Hence under Axiom A2, the weak-equitability Definition~\ref{weak-equit} is equivalent to the Axiom A6.
The self-equitability definition~\citep{Kinney2014} is stronger than Axiom A6 (weak-equitability), and is weaker than the original Axiom A6a.

\subsection{Some Dependence Measures and Independence Characterization}\label{sec:measures}

One common class of copula-based measures are the concordance measures~\citep[chapter 5]{Nelsen2006Copula}. In the bivariate case, let $c(u,v)=(\partial^2/\partial u \partial v)C(u,v)$ denote the density function of the copula $C(u,v)$, for $(u,v) \in \mathcal{I}^2$. Then Spearman's Rho is $\rho = - 3 + 12 \iint_{\mathcal{I}^2} C(u,v)dudv$; Kendall's Tau is $\tau = -1 + 4 \iint_{\mathcal{I}^2} c(u,v)C(u,v)dudv $; Gini's Gamma is $\gamma=2 \iint_{\mathcal{I}^2} (|u+v-1| - |u-v|) c(u,v)du dv$; Blomqvist's Beta is $\gamma= -1 + 4 C(0.5, 0.5)$.

However, those concordance measures all suffer from the deficiency (D2) above: they all equal zero for the deterministic relationship in example (C) of Table~\ref{tab:CopulaTrans}. Naturally we want dependence measures satisfies R\'enyi's Axiom A4. Several classes of dependence measures satisfies Axiom A4 using different but equivalent mathematical characterizations of the statistical independence between $X$ and $Y$ with a similar form:
\beq\label{ind}
f_{X,Y}(x,y)=f_X(x)f_Y(y) \qquad \mbox{for all } x,y.
\eeq Here the $f_{X,Y}$ can be either joint CDF $F_{X,Y}(x,y)$, or joint characteristic function $\phi_{X,Y}(s,t)=E[e^{i(X s + Y t)}]$ with $E[\cdot]$ denoting the expectation, or joint probability density function $p_{X,Y}$. Then $f_X$ and $f_Y$ are the corresponding marginal functions: CDFs $F_X$ and $F_Y$, or characteristic functions $\phi_X(s)=E[e^{i X s}]$ and $\phi_Y(t)=E[e^{i Y t}]$, or probability density functions $p_X$ and $p_Y$.

Due to the characterization (\ref{ind}), it is natural to define $D(X;Y)$ through a discrepancy measure between the joint function $f_{X,Y}$ and the product of marginal functions $f_X f_Y$. Such types of $D(X;Y)$ would equal to zero if and only if $f_{X,Y}=f_X f_Y$ always, i.e., $X$ and $Y$ are independent.

The first class of dependence measures use CDFs in the characterization (\ref{ind}). Denote the independence copula $\Pi = C(u,v) = uv$ on $\mathcal{I}^2$. Then using $L_{\infty}$ and $L_2$ distance between $C$ and $\Pi$, we get the commonly used Kolmogorov-Smirnov criterion $KS(X;Y) = \max_{\mathcal{I}^2} |C(u,v) - \Pi(u,v)|$ and Cram{\'e}r-von Mises criterion $CVM(X;Y)=\iint_{\mathcal{I}^2} [C(u,v) - \Pi(u,v)]^2 du dv$. These criteria are often used for independence testing~\citep{Genest2004testInd, Genest2007, Kojadinovic2009}.

We notice that, to satisfy the Axiom A3: $0 \le D(X;Y) \le 1$, $KS$ and $CVM$ need to be scaled with appropriate constants. The scaling does not affect the results for independence testing, but only affects the numerical values of the dependence measures. \citet{Schweizer1981MeasDep} studied dependence measures in this class using $L_p$ distance. The $L_1$, $L_2$ and $L_\infty$ distance result in, respectively, the Wolf's $\sigma$, Hoeffding's $\Phi^2$  and Wolf's $\kappa$ measures:
\beq \label{sigma}
\sigma(X;Y) = 12 \iint_{\mathcal{I}^2} |C(u,v) - \Pi(u,v)| du dv,
\eeq
\beq \label{Phi^2}
\Phi^2(X;Y) = 90 \iint_{\mathcal{I}^2} [C(u,v) - \Pi(u,v)]^2 du dv = 90 CVM(X;Y),
\eeq
\beq \label{KS}
\kappa(X;Y) = 4 \max_{\mathcal{I}^2} |C(u,v) - \Pi(u,v)| = 4 KS(X;Y).
\eeq
This class of dependence measures satisfies the modified R\'enyi's Axioms 1-7~\citep{Schweizer1981MeasDep}.

For the second class of dependence measures, using the characteristic functions in the characterization (\ref{ind}) can lead to the {\it distance covariance}~\citep{szekely2007dcor,szekely2009dcor}.
\beq \label{dCov}
\operatorname{dCov}^2(X;Y)= \iint\limits_{s,t} \frac{| \phi_{X,Y}(s, t) - \phi_X(s)\phi_Y(t)|^2}{|s|^{2} |t|^{2}} dt ds.
\eeq
To satisfy the Axiom A3, the {\it distance correlation} is defined as
\beq \label{dcor}
\operatorname{dcor}(X;Y)=  \frac{dCov(X;Y)}{\sqrt{dCov(X;X)dCov(Y;Y)}}.
\eeq
The $dcor$ does not satisfy the Axiom A6. This can be remedied by defining the distance correlation on the copula-transformed variables $U$ and $V$. That is, we use the rank-based version of $dcor$ that replaces $\phi_{X,Y}$, $\phi_X$ and $\phi_Y$ with $\phi_{U,V}$, $\phi_U$ and $\phi_V$ in (\ref{dCov}). This will be assumed in the rest of the paper.

The third class of dependence measures use the probability density functions $p_{X,Y}$, $p_{X}$ and $p_{Y}$ in the characterization (\ref{ind}). Then the copula-based version involves only the copula density $c(u,v)$. This class includes many information-theoretical measures such as the R\'enyi's mutual information
\beq \label{RenyiMI}
MI_{\a}(X;Y) = \frac{1}{\a-1} \log [ \iint_{\mathcal{I}^2} c^\a(u,v) du dv], \qquad \a > 0 \mbox{ and } \a \ne 1.
\eeq
In the limit of $\a \to 1$, $MI_1$ becomes the popular Shannon's mutual information (MI) criterion
\beq \label{MI}
MI(X;Y) = \iint_{\mathcal{I}^2} \log [c(u,v)] c(u,v) du dv.
\eeq
MI is the recommended measure in \citet{Kinney2014}. For Axiom A3, we can define mutual information correlation~\citep{Joe1989Entropy}
\beq \label{MIcor}
MIcor= \sqrt{1-e^{-2MI}}.
\eeq
We use the name $MIcor$ to indicate it as the scaled version of MI. It is also known as the Linfoot correlation in literature~\citep{Speed2011comment}.

Other information measures include Tsallis entropy~\citep{Tsallis1988entropy}:
\beq \label{TsallisDelta}
\Delta_{\a}(X;Y) = \frac{1}{1-\a} [1 - \iint_{\mathcal{I}^2} c^\a(u,v) du dv], \qquad \a \ne 0,1.
\eeq
In the limit of $\a \to 1$, $\Delta_1$ becomes MI. When $\a = 1/2$, $\Delta_{1/2}= \iint_{\mathcal{I}^2} 2 [1 -c^{1/2}(u,v)] du dv$ becomes the Hellinger distance. The scaled version is the Hellinger dependence measure~\citep{Tjostheim1996MeasDep, granger2004dependence} $H(X;Y) = \Delta_{1/2}/2$.

Also in this class are measures using $L_p$ distance between the copula density $c(u,v)$ and the independence copula density $\pi(u,v) \equiv 1$. Hence we call them the {\it Copula-Distance}
\beq \label{CD}
 CD_\a = \iint_{\mathcal{I}^2} |c(u,v)-1|^\a du dv, \qquad \a >0.
\eeq
Again, we can scale $CD_\a$ to satisfy Axiom A3. $CD_2$ is the Pearson's $\phi^2$ with its scaled version being ${\phi}cor = \sqrt{CD_2/(1+CD_2)}$ \citep{Joe1989Entropy}.

Particularly, we call the scaled version of $CD_1$ as {\it copula correlation}
\beq \label{Ccor}
Ccor = \frac{1}{2} CD_1 =  \frac{1}{2} \iint_{\mathcal{I}^2} |c(u,v)-1| du dv.
\eeq

We defined the third class of dependence measures through the copula density $c(u,v)$. For some important cases such as when $Y$ is a deterministic function of $X$, the copula density $c(u,v)$ does not exist with respect to the two-dimensional Lebesgue measure. That is, the copula $C(u,v)$ contains a singular component~\citep[page 27]{Nelsen2006Copula}.  For the copula with a singular component, we define the dependence measures on it as the limits of dependence measures on continuous copulas approaching it. Let $\{C_1, C_2, ...\}$ be a sequence of continuous copulas that converges to the copula $C$. The convergence can be defined in any distance for probability distributions, and we take the $L_1$-distance here. That is, $\lim\limits_{m \to \infty} \|C_m -C\|_1 := \lim\limits_{m \to \infty} \sup\limits_{A} |C_m (A) - C(A)| = 0$, where the supreme is taken over all Borel sets $A$. Then the dependence measure $D[X;Y]$'s value under copula $C(u,v)$ is defined as $D[X;Y|C] := \lim\limits_{m \to \infty} D[X;Y|C_m]$. Using such a definition, if $Y$ is a deterministic function of $X$, then clearly $MI = \infty$, $MIcor=1$, ${\phi}cor=1$ and $Ccor=1$.

\subsection{Parameters, Estimators and MIC}\label{sec:MIC}
The dependence measures in Section~\ref{sec:measures} are all parameters. Sometimes the same names also refer to the corresponding sample statistics. Let $(X_1,Y_1)$, ..., $(X_n,Y_n)$ be a random sample of size $n$ from the joint distribution of $(X,Y)$. Then the sample statistic $\rho_n = \sum_{i=1}^n (X_i - \bar X)(Y_i - \bar Y)/\sqrt{\sum_{i=1}^n (X_i - \bar X)^2 \sum_{i=1}^n(Y_i - \bar Y)^2}$ is also called Pearson's correlation coefficient. In fact, $\rho_n$ is an estimator for $\rho$, and converges at the parametric rate of $n^{-1/2}$. The first two classes of measures have natural empirical estimators, replacing CDFs and characteristic functions by their empirical versions. Particularly, \citet{szekely2007dcor} showed that the resulting $dcor_n$ statistic is the sample correlation of centered distances between pairs of $(X_i,Y_i)$ and $(X_j,Y_j)$. The last class of dependence measures use the probability density functions instead, and are harder to estimate. For continuous $X$ and $Y$, simply plugging in empirical density functions may not result in good estimators for the dependence measures. However, we will see in section~\ref{sec:equit} that the first two class of measures do not have the equitability property. Hence we need to study the harder-to-estimate measures such as MIcor and Ccor.

The MIC introduced in~\citet{Reshef2011MIC} is in fact a definition of a sample statistic, not a parameter. On the data set $(X_1,Y_1)$, ..., $(X_n,Y_n)$, they first consider putting these $n$ data points into a grid $G$ of $b_X \times b_Y$ bins. Then the mutual information $MI_G$ for the grid is computed from the empirical frequencies of the data on the grid. The MIC statistic is defined as the maximum value of $MI_G/\log[min(b_X, b_Y)]$ over all possible grids $G$ with the total number of bins $b_X b_Y$ bounded by $B=n^{0.6}$. That is,
\beq \label{MIC}
MIC_n = \max_{b_X b_Y <B} \frac{MI_G}{\log[min(b_X, b_Y)]}
\eeq The $MIC_n$ is always bounded between $0$ and $1$ since $0 \le MI_G \le \log[min(b_X, b_Y)]$.

The corresponding parameter MIC for the joint distribution of $X$ and $Y$ can be defined as the limit of the sample statistic for big sample size $MIC =\lim_{n \to \infty} MIC_n$. We notice that this definition depends on the tuning parameter $B$ and the implicit assumption that the limit exists. Hence the MIC parameter may change with different selection of $B(n)$. This is in contrast to the usual statistical literature, where the parameter definition is fixed but its estimator may contain some tuning parameter $B(n)$. Because the MIC parameter is only defined as a limit, the theoretical study on its mathematical properties is very hard.

As we introduce the strict mathematical definition for the equitability in next subsection~\ref{sec:equit}, we can see that equitability should be a property for the parameter but not for the statistic.

\section{Equitable measures}\label{sec:equit_main}
\subsection{$R^2$-Equitability and Self-equitability}\label{sec:equit}

We first describe the theoretical results on equitability by \citet{Kinney2014}.
\citet{Reshef2011MIC} proposed that an equitable measure should treat all deterministic relationships similarly under noisy situations. Particularly, they focused on the nonlinear regression setting for motivation: $Y=f(X)+ \e $, where $\e $ denotes the random noise that is independent of $X$ conditional on $f(X)$.
The squared Pearson's coefficient $R^2$ reflects the proportion of variance in $Y$ explained by the regression on $X$.
They want the nonlinear dependence measure to be close to $R^2$ regardless of the specific form of $f(\cdot)$.
To formalize this concept, \citet{Kinney2014} used the condition ``$X \leftrightarrow f(X) \leftrightarrow Y$ forms a Markov chain" to characterize the nonlinear regression model. This condition means, in the model $Y=f(X)+\e$ with deterministic $f$, $\e$ is the random noise variable which may depend on $f(X)$ as long as $\e$ has no additional dependence on $X$. Then \citet{Kinney2014} defined the $R^2$-equitability as
\begin{defi} \label{R2-equit}
A dependence measure $D[X; Y]$ is $R^2$-equitable if and only if,
$D[X; Y] =g(R^2[f(X); Y])$.
Here, $g$ is a function that does not depend on the distribution $p_{X,Y}$,  $f$ is a deterministic function and $X \leftrightarrow f(X) \leftrightarrow Y$ forms a Markov chain.
\end{defi}
Given the joint distribution $p_{X,Y}$, the function $f$ in the regression model $Y=f(X)+ \e $ is not uniquely specified. This implies that any $R^2$-equitable measure must be a trivial constant measure. Therefore, \citet{Kinney2014} proposed a new replacement definition of equitability by extending the invariance property (of the weakly-equitability or Axiom A6) in the regression model.
\begin{defi}\label{self-equit}
A dependence measure $D[X; Y]$ is self-equitable if and only if
$D[X; Y] = D[f(X); Y]$ whenever $f$ is a deterministic function and $X \leftrightarrow f(X) \leftrightarrow Y$ forms a Markov chain.
\end{defi}
The self-equitability turned out to be characterized by a commonly used inequality in information theory.
\begin{defi} \label{DPI}
A dependence measure $D[X; Y]$ satisfies the {\bf Data Processing Inequality}
(DPI) if and only if $D[X; Y] \ge D[X; Z]$
whenever the random variables X, Y, Z form a Markov chain $X \leftrightarrow Y \leftrightarrow Z$.
\end{defi}
\citet[SI, Theorem~3]{Kinney2014} showed that every DPI-satisfying measure is self-equitable.
\citet[SI, Theorem~4]{Kinney2014} proved that measures of the following form must satisfy DPI:
$$D_g(X;Y) = \iint g(\frac{p_{X,Y}(x,y)}{p_X(x)p_Y(y)})p_X(x)p_Y(y)dx dy,$$ with $g$ a convex function on the nonnegative real numbers. In term of copula density, $D_g(X;Y) = \iint_{\mathcal{I}^2} g[c(u,v)]dudv$.

Therefore, due to the convexity of functions $|x-1|^\a$ (when $\a \ge 1$) and $1-x^\a$ (when $\a \le 1$) on $x>0$, we get the following proposition.
\begin{Prop}\label{thm:CD-equit}
The Copula-Distance $CD_\a$ with $\a \ge 1$ and the Tsallis entropy $\Delta_\a$ with $\a \le 1$ are self-equitable.
\end{Prop}

As a direct result of Proposition~\ref{thm:CD-equit}, the copula correlation $Ccor = CD_1/2$ and the Hellinger dependence measure $H=\Delta_{1/2}/2$ are both self-equitable.

The R\'enyi's Axiom A6a is a stronger condition than the self-equitability as no Markov Chain condition is required. Therefore, R\'enyi's maximum correlation coefficient Rcor is also self-equitable. However, Rcor equals one too often. We illustrate this deficiency of Rcor, and the self-equitability of the dependence measures on some examples of simple probability distributions on the unit square. These examples are modified from those in \citet{Kinney2014}, and the results are displayed in Table~\ref{tab:ExampleCor}.

\begin{table} [htbp]
\begin{center}
\small{
\begin{tabular}{c|cccccccccc}
    \hline
{\bf Examples} &   {\bf MIcor} & {\bf Ccor} & {\bf $\phi$cor} & {\bf Rcor} & {\bf MIC} & {\bf dcor} & {\bf $\kappa$ } & {\bf $\Phi^2$} & {\bf $\sigma$ } \\
    \hline \hline
A \parbox[c]{36pt}{\includegraphics[scale=0.1]{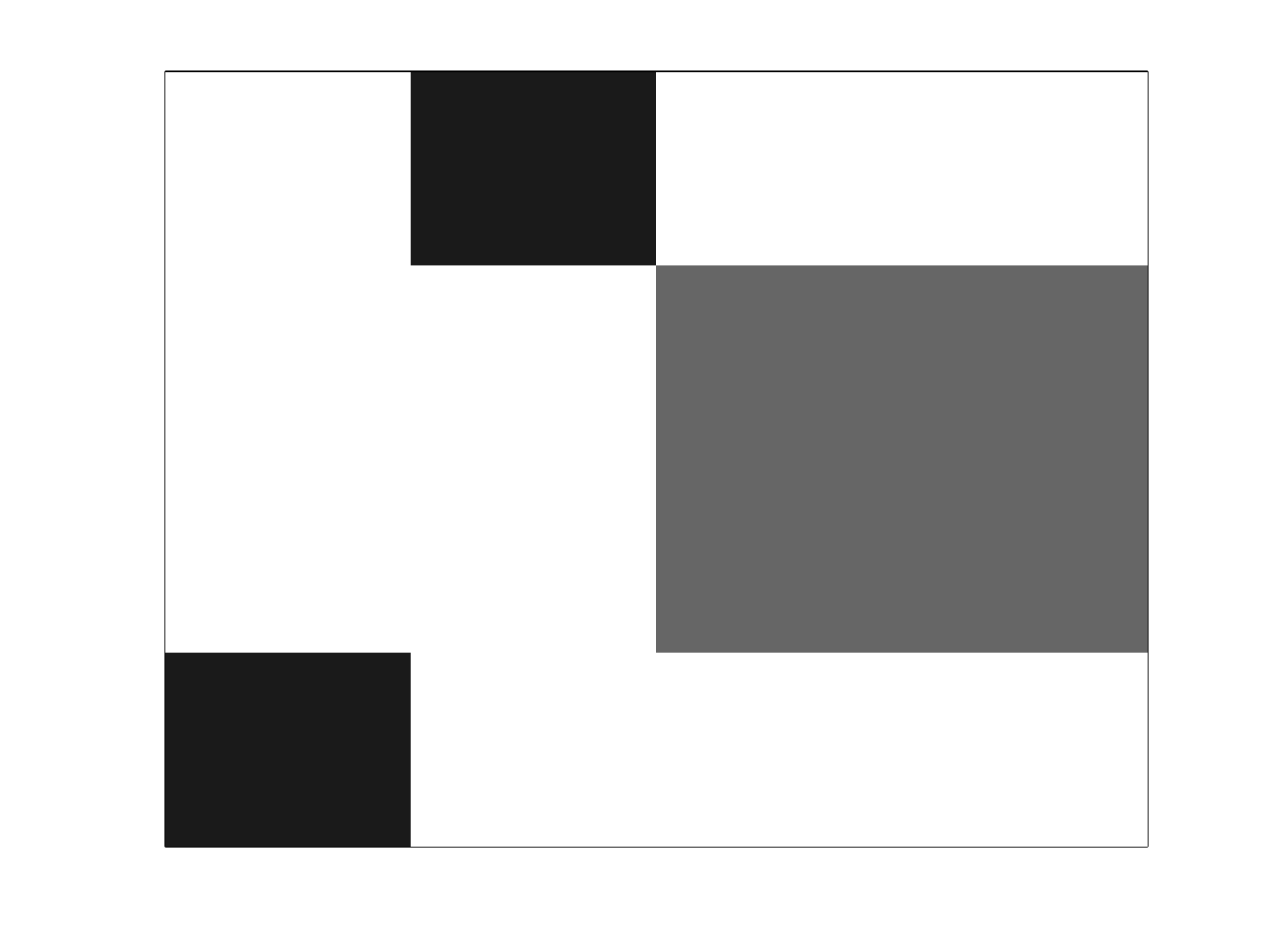}} & 0.94 & 0.63 & 0.82 & 1    & 1 & 0.56 & 0.75 & 0.31 & 0.53 \\
B \parbox[c]{36pt}{\includegraphics[scale=0.1]{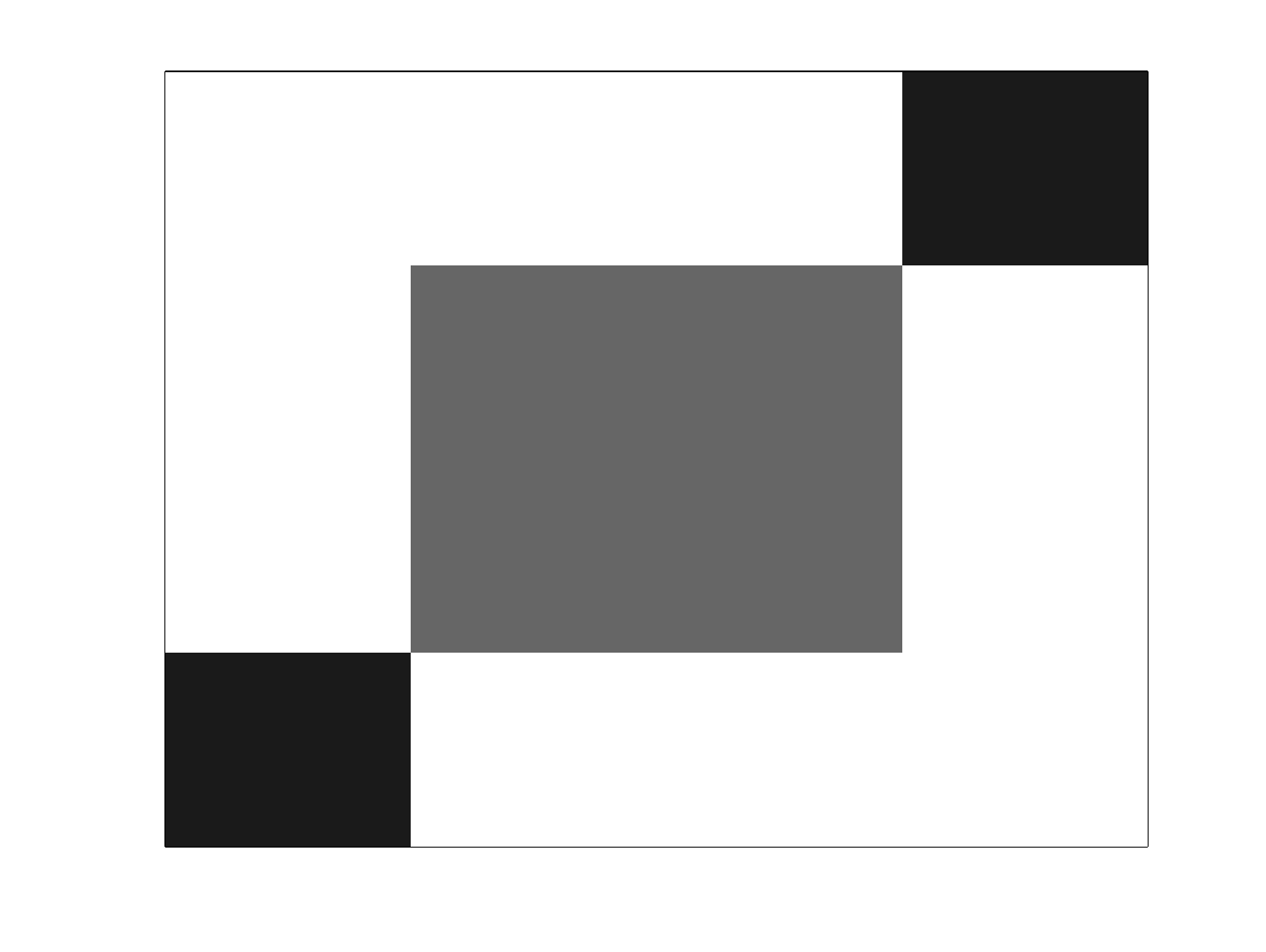}} & 0.94 & 0.63 & 0.82 & 1 & 0.95 & 0.82 & 0.75 & 0.66 & 0.84 \\
C \parbox[c]{36pt}{\includegraphics[scale=0.1]{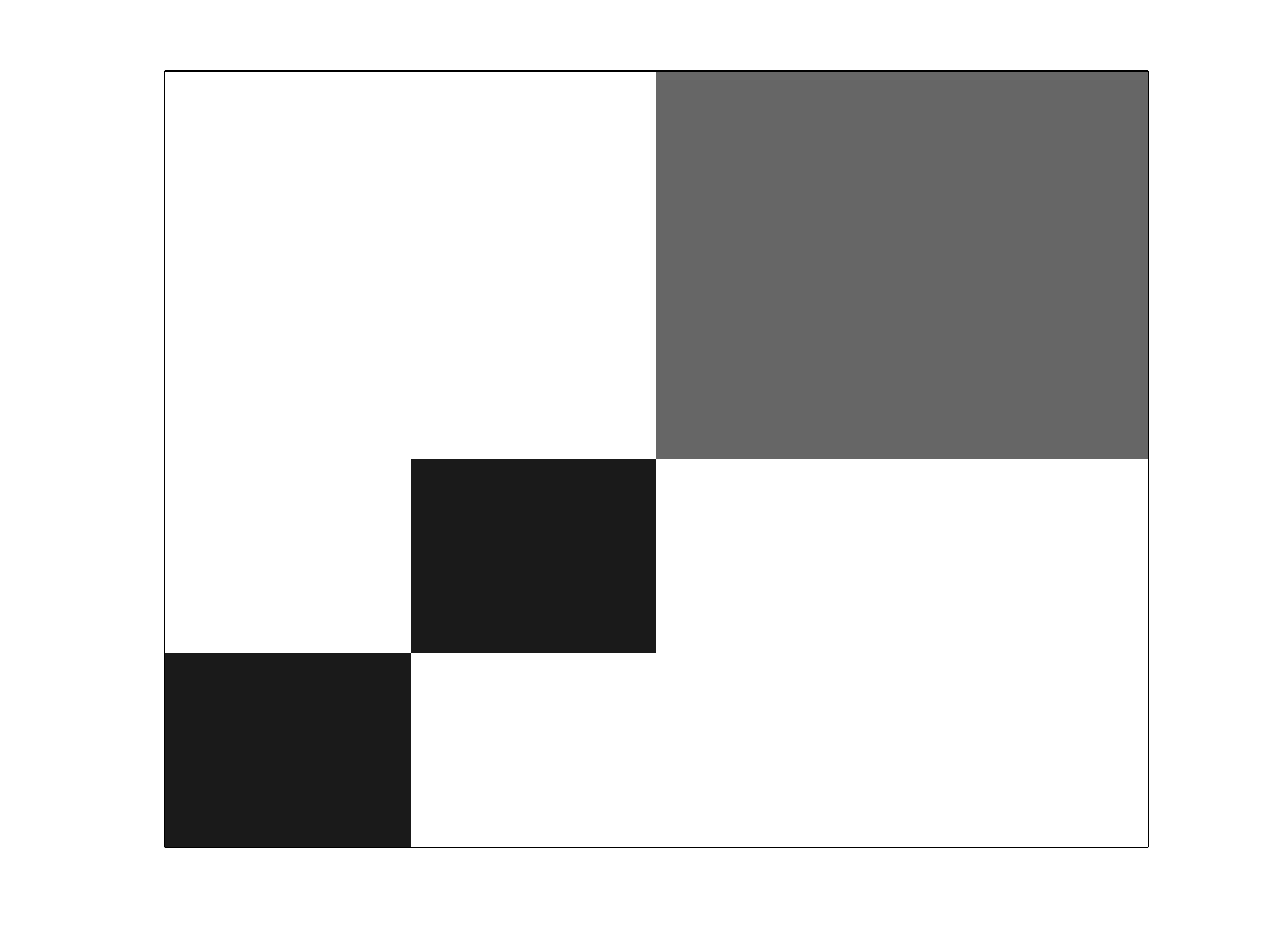}} & 0.94 & 0.63 & 0.82 & 1    & 1 & 0.87 & 1    & 0.75 & 0.84 \\
D \parbox[c]{36pt}{\includegraphics[scale=0.1]{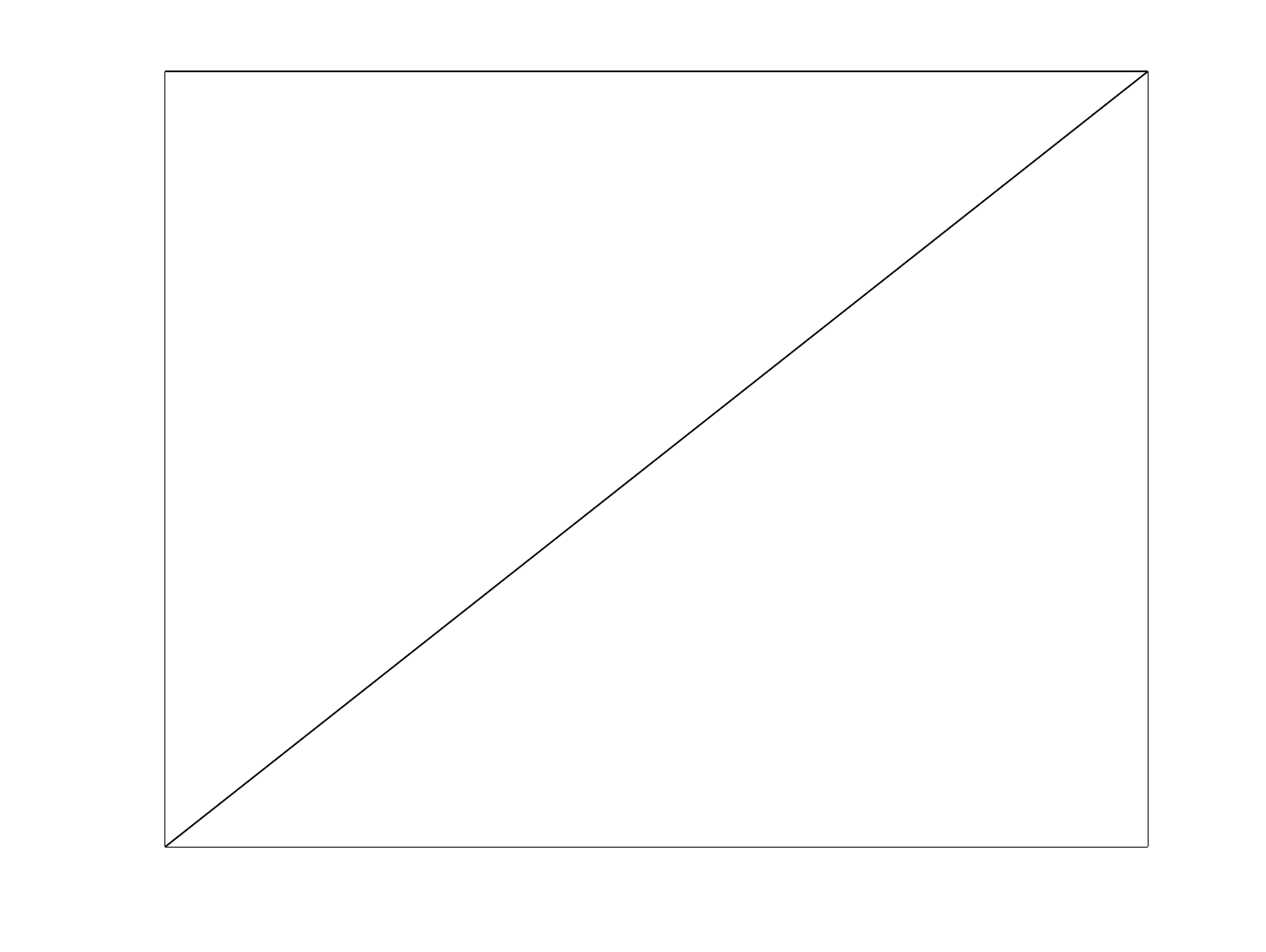}} & 1    & 1    & 1    & 1    & 1 & 1    & 1    & 1    & 1 \\
E \parbox[c]{36pt}{\includegraphics[scale=0.1]{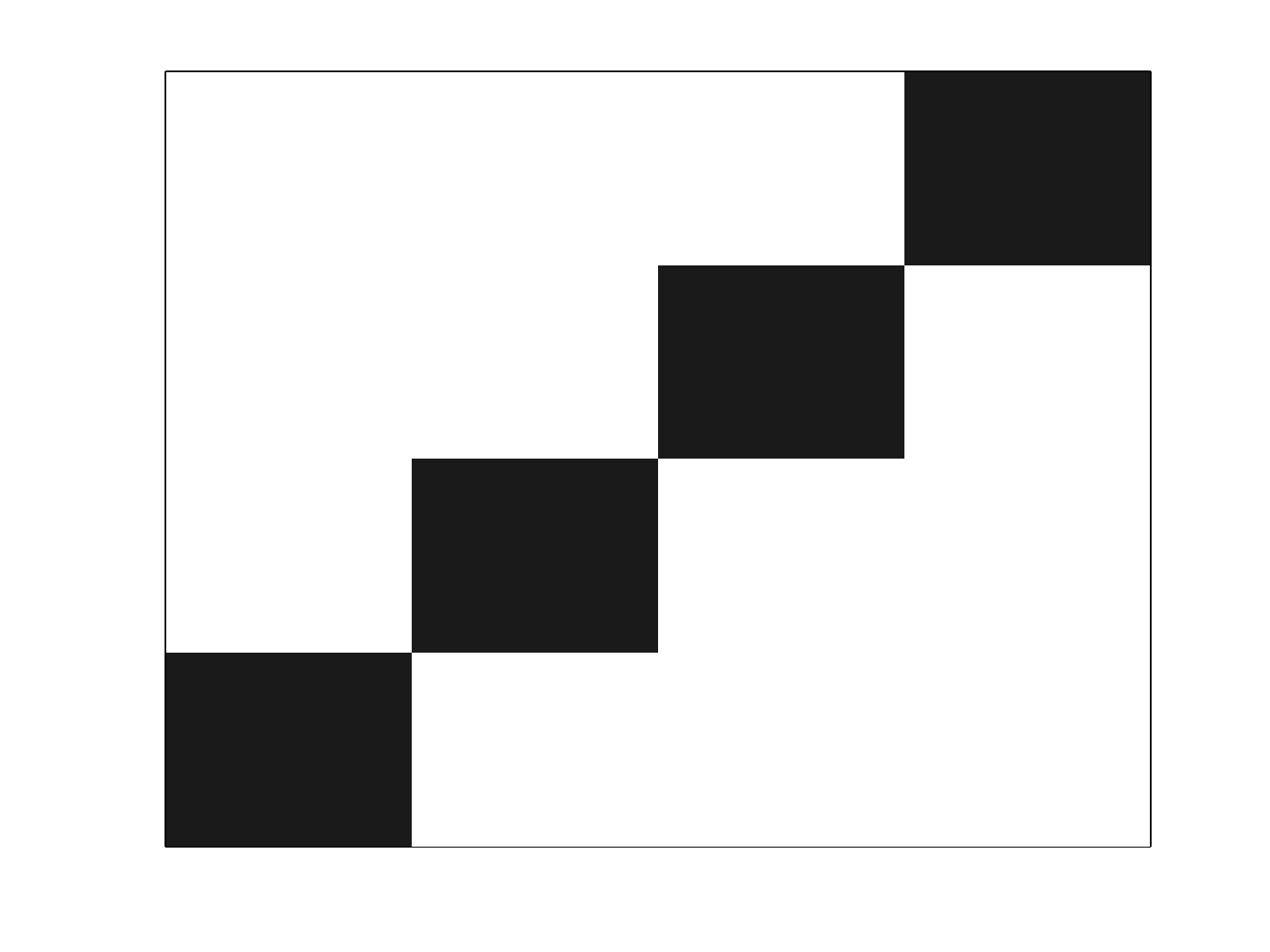}} & 0.97 & 0.75 & 0.87 & 1    & 1 & 0.94 & 1    & 0.88 & 0.94 \\
F \parbox[c]{36pt}{\includegraphics[scale=0.1]{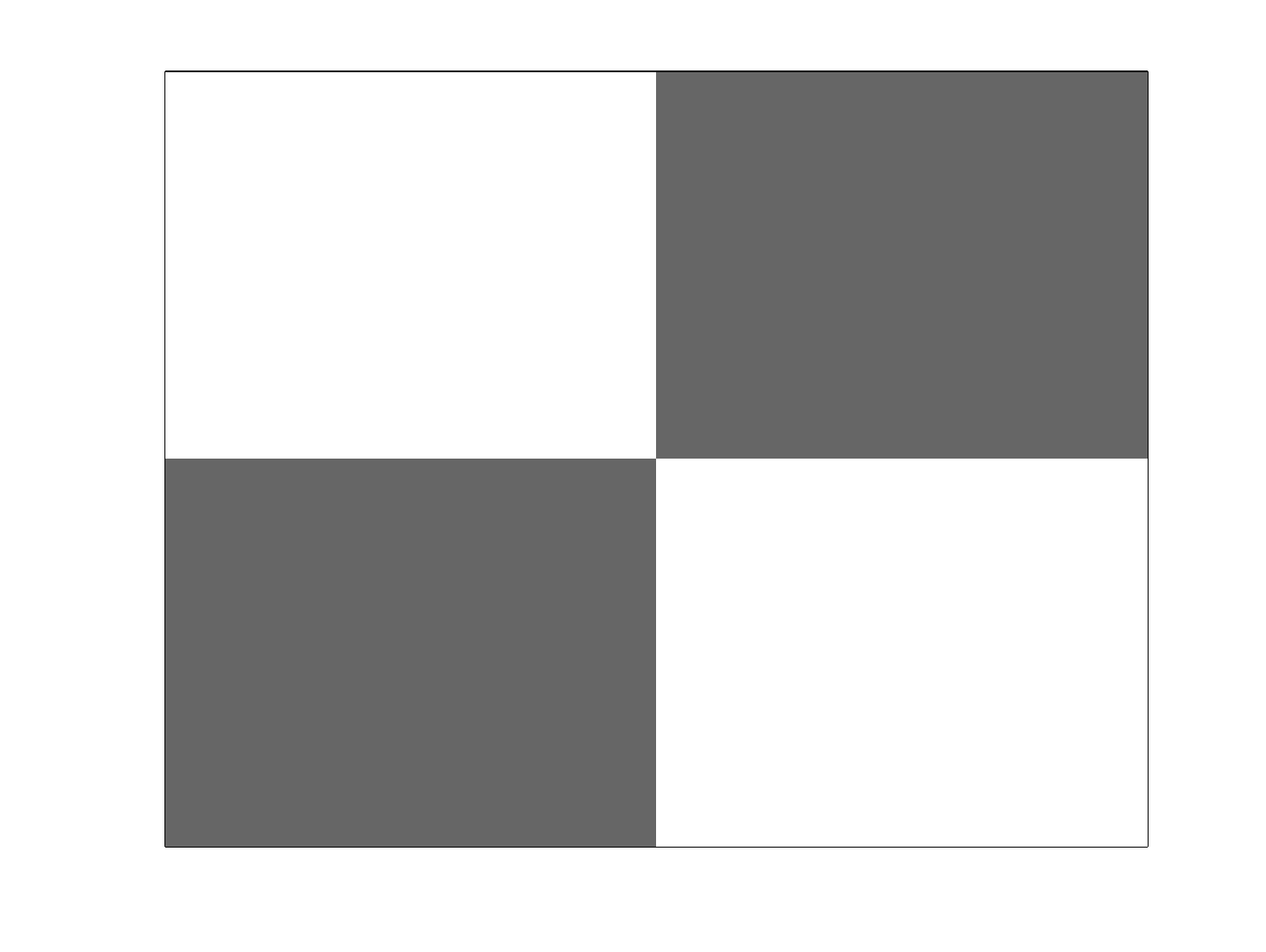}} & 0.87 & 0.50 & 0.71 & 1    & 1 & 0.79 & 1    & 0.63 & 0.75  \\
    \hline
\end{tabular}
}
\end{center}
\caption{The values of several dependence measures on some example distributions. For each example distribution, the graph shows its probability density function: the white regions have zero density, the shaded regions have constant densities. The dark regions have densities twice as big as the densities on the light grey regions. }
\label{tab:ExampleCor}
\end{table}

A self-equitable measure will equal the same value in the first three examples A, B and C in Table~\ref{tab:ExampleCor} due to the existence of an invertible transformation satisfying the Markov chain condition~\citep{Kinney2014}. We can see that MIcor (or MI), Ccor, $\phi$cor (or $CD_2$) and Rcor all remain constants for the first three examples A, B and C. In contrast, the MIC, dcor, and those measures of the first class ($\kappa$, $\Phi^2$ and $\sigma$) are not self-equitable.

The next three examples D, E and F show increasing noise levels. However, Rcor, MIC and $\kappa$ always equal one across Examples D, E and F, failing to correctly reflect the noise levels here. Particularly, Rcor equals one in all six examples here, failing to distinguish the strengths of deterministic signals among them.

\subsection{robust-equitability}\label{sec:cop-equit}
An equitable dependence measure should reflect the strength of the deterministic signal in data, regardless of the relationship form. However, what quantity is the proper measure for the signal's strength? \cite{Reshef2011MIC} proposed to use the nonlinear $R^2$ to measure the signal strength, which could not lead to a proper equitability definition  \citep{Kinney2014}. One reason for the failure is the incompatibility of the nonlinear regression model $Y=f(X) + \e$  with the joint Gaussian distribution. (The $R^2$ is the natural measure for Gaussian distribution as in  R\'{e}nyi's Axiom A7). However, $Y=f(X) + \e$ would result in the joint Gaussian distribution only for linear $f(x)$ but not for any nonlinear $f(x)$.

For a better equitability definition, we consider a different situation: a mixture distribution with $p$ proportion of deterministic relationship $Y=f(X)$ hidden in continuous background noise. This situation can be mathematically rigorously expressed through the mixture-copula. The copula can always be separated into a singular component and an absolutely continuous component~\citep[page 27]{Nelsen2006Copula}. The absolutely continuous component corresponds to the background noise. The independent background noise must corresponds to the independence copula $\Pi(u,v)=uv$ (the uniform distribution on the unit square). Therefore, the data $(X,Y)$ with $p$ proportion of hidden deterministic relationship $Y=f(X)$ have copula $C = p C_s + (1-p) \Pi$. Here $C_s$ is a singular copula representing the deterministic relationship, so that its support $\mathcal{S}$  has Lebesgue  measure zero. Clearly the signal strength in this situation should equal to $p$, regardless of the specific form of deterministic relationship. Hence we have the following equitability definition.

\begin{defi}\label{mix-cop-eqit}
A dependence measure $D[X; Y]$ is robust-equitable if and only if $D[X; Y] = p$ whenever $(X,Y)$ follows a distribution whose copula is  $C = p C_s + (1-p) \Pi$, for a singular copula $C_s$.
\end{defi}

We note that a robust-equitable measure is an extension for the Pearson's linear correlation. When the $p$ proportion of the deterministic relationship is linear, $C_s$ has the support on the diagonal of the unit square, and hence $p=|\rho|$. A robust-equitable dependence measure treat the linear hidden deterministic relationship the same as a nonlinear one. For the dependence measures mentioned above, only the copula correlation is known to be robust-equitable.

\begin{Prop}\label{thm:Ccor-mix-cop-equit}
The copula correlation $Ccor$ is robust-equitable.
\end{Prop}
The Proposition~\ref{thm:Ccor-mix-cop-equit} comes directly from calculation that $$Ccor= [p \int_{\mathcal{S}}C(du,dv) + \int_{\mathcal{I}^2 \setminus \mathcal{S}} |(1-p)-1| du dv]/2 = [p(1) + p]/2=p. $$

Most self-equitable measures discussed above are not robust-equitable.
Direct calculations show that the mutual information $MI$ and copula distance $CD_\a$ for $\a>1$ all equal to $\infty$ for the mixture copula with $p>0$. Hence they are not robust-equitable, neither are their scaled version ($MIcor$ and other scaled version such as $\phi cor$ all equal to $1$). On the mixture copula, the Tsallis entropy $\Delta_\a = [1-(1-p)^\a]/(1-\a)$ for $\a < 1$. Hence the Tsallis entropies are also not robust-equitable.

We do not have a proof on whether Rcor is robust-equitable. However Rcor has many drawbacks as mentioned earlier. As shown in the examples in Table~\ref{tab:ExampleCor}, Rcor equals one too often. Because Rcor's definition involve taking the supreme over all Borel functions, its theoretical properties are often hard to analyze. Another drawback of Rcor is that it is very difficult to estimate. There is no commonly accepted estimator for Rcor.

\begin{center}
\begin{figure}[h]
	\centering
\begin{center}
        \begin{subfigure}[b]{0.225\textwidth}
               \includegraphics[width=\textwidth]{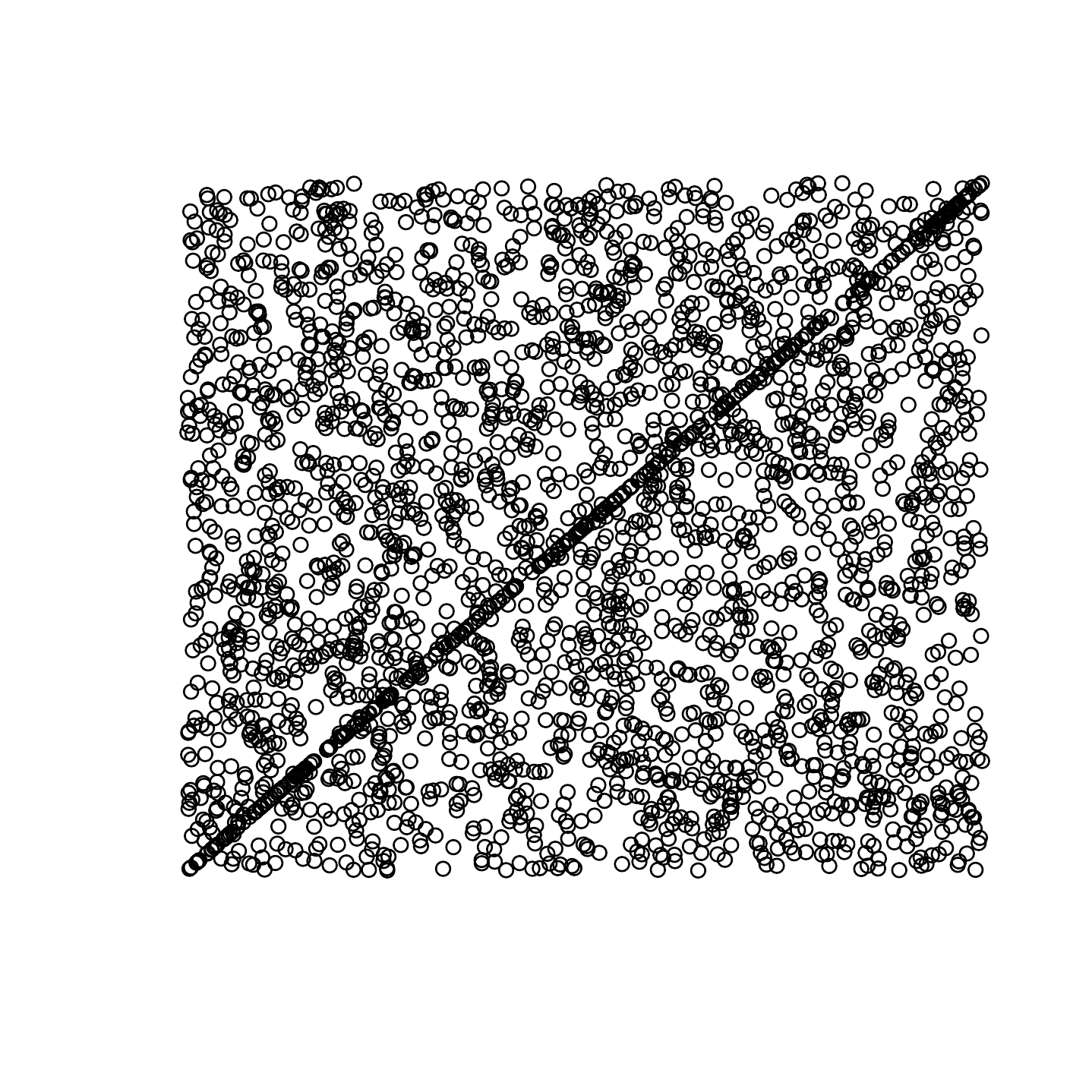}
                \caption{$\rho=0.1$, $MI=\infty$, $Ccor=0.1$ }
                \label{fig:equit.a}
        \end{subfigure}
 	\quad
         \begin{subfigure}[b]{0.225\textwidth}
               \includegraphics[width=\textwidth]{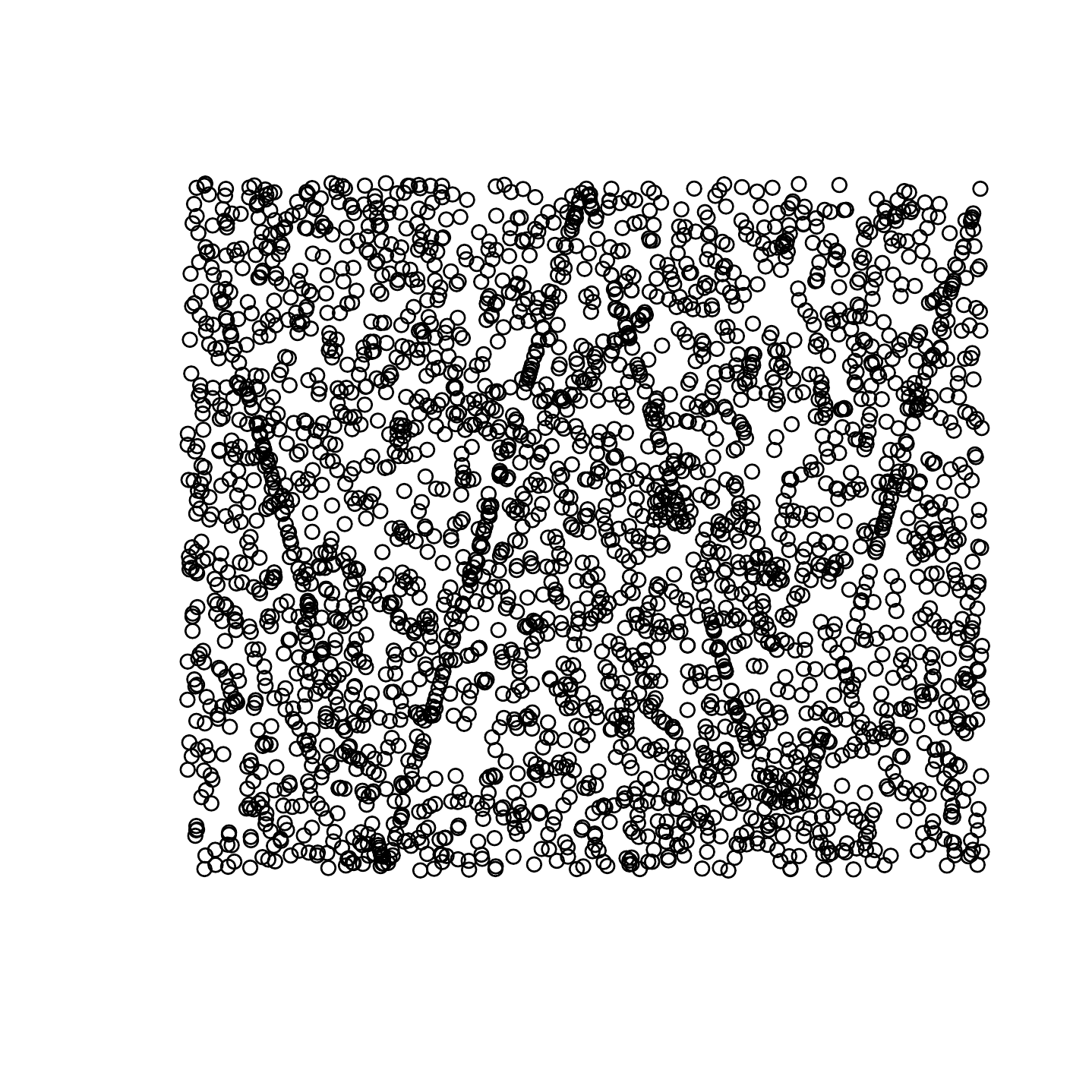}
                \caption{$\rho=0$, $MI=\infty$, $Ccor=0.1$}
                \label{fig:equit.b}
        \end{subfigure}
	\quad
         \begin{subfigure}[b]{0.225\textwidth}
               \includegraphics[width=\textwidth]{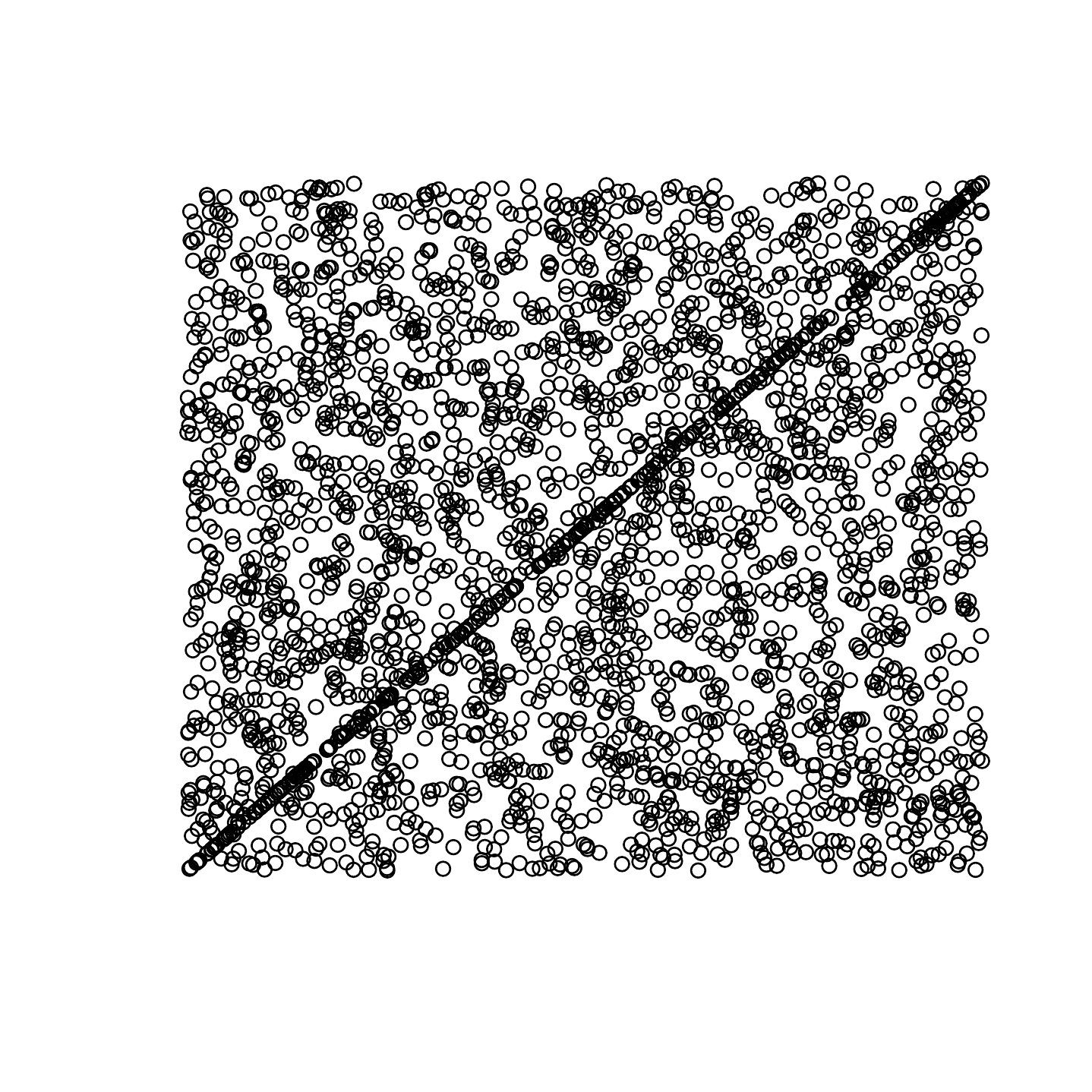}
                \caption{$\rho=0.1$, $MI=1$, $Ccor=0.1$}
                \label{fig:equit.c}
        \end{subfigure}	
        \quad
         \begin{subfigure}[b]{0.225\textwidth}
               \includegraphics[width=\textwidth]{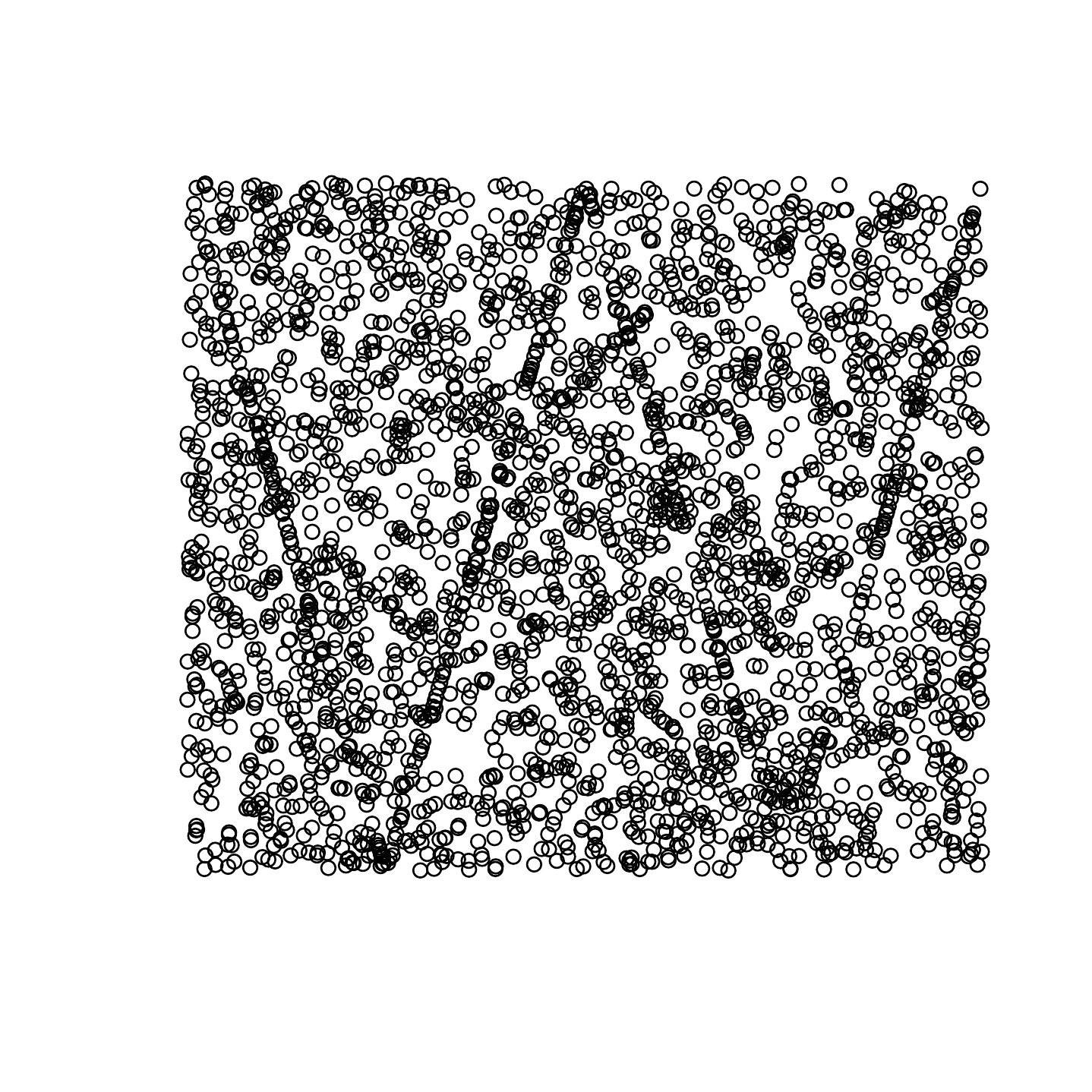}
                \caption{$\rho=0$, $MI=1$, $Ccor=0.1$}
                \label{fig:equit.d}
        \end{subfigure}
        \end{center}
 	\caption{(a) and (b): $10\%$ data on a deterministic curve hidden in background noise. (c) and (d): the $10\%$ nearly deterministic data on a narrow strip around the curve. }
	\label{fig:equit}
\end{figure}
\end{center}

The difference between self-equitable and robust-equitable measures is illustrated through examples in Figure~\ref{fig:equit}. Figures~\ref{fig:equit.a} and \ref{fig:equit.b} shows $10\%$ of data coming from two deterministic curves, and in Figures~\ref{fig:equit.c} and \ref{fig:equit.d} the $10\%$ of data is nearly deterministic around the curve in a very small strip of area $0.1/exp(10)=4.5 \times 10^{-6}$. In Figure~\ref{fig:equit}, MI and Ccor are self-equitable, (their values are the same on (a) and (b), and the same on (c) and (d)),
whereas Pearson's correlation coefficient $\rho$ is not. However, the data distributions in  (a) and (b) ($MI=\infty$) are in fact very close to the corresponding cases of (c) and (d) ($MI=1$), Ccor reflects this with $Ccor=0.1$ (differ only in $10^{-6}$ order) in all cases but MI does not.

From the examples, we see that self-equitability is not sufficient for a good dependence measure. While self-equitability ensures the measure's invariance under transformation between Figures~\ref{fig:equit.a} and Figures~\ref{fig:equit.b}, MI would equal to $\infty$, an unreasonable value for those cases. In fact, MI would equal to $\infty$ for an arbitrarily tiny amount of hidden deterministic relationship in the data. Therefore, its value is very unstable. This instability makes the consistent estimation of MI impossible as we will show in Section~\ref{sec:EstErr}.

\subsection{Multivariate Extensions}\label{sec:multivariate}

We have so far concentrated on the simple bivariate case. The dependence measure can be extended to the multivariate case.

There are two possible directions of extending dependence measures to the multivariate case. In the first direction, we are interested in any dependence among $d$ variables $X_1$, ..., $X_d$. Therefore, the divergence of their joint distribution from the independent joint distribution (the product of marginals) can be used to measure such dependence. \citet{Schmid2010CopMeasures} provided higher-dimensional extension of many copula-based dependence measures along this direction. We define a multivariate version $Ccor1$ as the half $L_1$ distance between the $d$-dimensional joint copula density from the independent copula density:
\beq \label{Ccor1}
\ba{cl}
Ccor1  & =  \frac{1}{2} \int |c(u_1, ..., u_d)-1| du_1 ... du_d  \\
& =  \frac{1}{2} \int |p_{X_1, ..., X_d} (x_1, ..., x_d) - p_{X_1}(x_1) ... p_{X_d}(x_d)| dx_1 ... dx_d  .
 \ea
\eeq
The corresponding robust-equitability definition becomes
\begin{defi}\label{mix-cop-eqit1}
A dependence measure $D[X_1,...,X_d]$ is robust-equitable if and only if $D[X_1, ..., X_d] = p$ whenever $(X_1, ..., X_d)$ follows a distribution whose copula is  $C = p C_s + (1-p) \Pi$, for a singular copula $C_s$.
\end{defi}
Here $\Pi(u_1, ..., u_d)$ is the independence copula of dimension $d$.

It is easy to check that $Ccor1$ is robust-equitable for this $d$-dimensional extension.

In the second direction, we can divide the $d$-dimensional vector into a $q$-dimensional vector $\vec X$ and $r$-dimensional vector $\vec Y$ with $q+r=d$. And we want a dependence measure between $\vec X$ and $\vec Y$, not caring about the dependence within $\vec X$ or within $\vec Y$. The dcor \citep{szekely2009dcor} is a dependence measure of this type. Along this direction, we define the multivariate version $Ccor2$ for $\vec X =(X_1, ..., X_q)$ and $\vec Y = (Y_1, ..., Y_r)$ as
\beq \label{Ccor2}
\ba{cl}
Ccor2 & = \frac{1}{2} \int |p_{\vec X, \vec Y} (\vec x, \vec y) - p_{\vec X}(\vec x) p_{\vec Y}(\vec y)| dx_1 ... dx_q dy_1 ... dy_r \\
&= \frac{1}{2} \int |c(\vec u, \vec v) - c_{\vec X}(\vec u) c_{\vec Y}(\vec v)| du_1 ... du_q dv_1 ... dv_r  .
\ea
\eeq
Here $c_{\vec X}$ and $c_{\vec Y}$ are the copula densities for $\vec X$ and $\vec Y$ respectively.
The robust-equitability definition in this direction of extension is
\begin{defi}\label{mix-cop-eqit2}
A dependence measure $D[\vec X; \vec Y]$ is robust-equitable if and only if $D[\vec X; \vec Y] = p$ whenever $(\vec X, \vec Y)$ follows a distribution whose copula is  $C = p C_s + (1-p) C_{\vec X} \times C_{\vec Y}$, for a singular copula $C_s$.
\end{defi}
Here $C_{\vec X}$ and $C_{\vec Y}$ are the $q$-dimensional and $r$-dimensional copulas of $\vec X$ and $\vec Y$ respectively. The measure $Ccor2$ is robust-equitable under this definition.
\section{STATISTICAL ERROR IN THE DEPENDENCE MEASURE ESTIMATION}\label{sec:EstErr}
We now turn our attention to the statistical errors in estimating the dependence measures. Particularly we focus on the two self-equitable measures MI and Ccor.

First, we point out that the first class of dependence measures are generally estimable at the parametric rate of $n^{-1/2}$. These measures, including Hoeffding's $\Phi^2$, Wolf's $\sigma$ and $\kappa$, are defined through the CDFs. We use the notations $\Phi^2(C)$, $\sigma(C)$ and $\kappa(C)$ to emphasize that they are functionals of the copula function $C(u,v)$. Then we can estimate them by plug-in estimators $\hat \Phi^2=\Phi^2(C_n)$, $\hat \sigma=\sigma(C_n)$ and $\hat \kappa=\kappa(C_n)$, where $C_n(u,v)$ denotes the empirical estimator for the copula function $C(u,v)$. Since $C_n(u,v)$ converges to $C(u,v)$ at the parametric rate of $n^{-1/2}$~\citep{omelka2009copula,segers2012copula}, $\Phi^2$, $\sigma$ and $\kappa$ can also be estimated at the parametric rate of $n^{-1/2}$.

However, the self-equitable measures come from the third class of dependence measures which involves the density function. Hence the parametric rate of convergence $n^{-1/2}$ can only be achieved with the plug-in density estimator for discrete distributions, e.g., for $\widehat{MI}=MI(c_n)$~\citep{Joe1989Entropy}. The convergence rate involving continuous distributions need more care. We consider the estimation of MI and Ccor respectively in the next two subsections~\ref{sec:MIbad} and \ref{sec:Ccor.consistent}.

\subsection{The Mutual Information Is Not Consistently Estimable}\label{sec:MIbad} The estimation of MI has been studied extensively in literature. Over all distributions, even discrete ones, no uniform rate of convergence is possible for MI~\citep{antos2001convergence,paninski2003estimation}. On the other hand, many estimators were shown to converge to MI for every distribution. These two results are not contradictory, but rather common phenomenon for many parameters. The first result is about the uniform convergence over all distributions while the second result is about the pointwise convergence for each distribution. The first restriction is too strong while the second restriction is too weak. The difficulty of estimating a parameter needs to be studied for uniform convergence over a properly chosen family.

As MI is defined through the copula density, it is natural to consider the families generally used in density estimation literature. Starting from~\citet{Farrell1972BestRate}, it is standard to study the minimax rate of convergence for density estimation over the class of functions whose $m$-th derivatives satisfy the H\"{o}lder condition. Since the minimax convergence rate usually is achieved by the kernel estimator, it is also the optimal convergence rate of density estimation under those H\"{o}lder classes. Generally, with the H\"{o}lder condition imposed on the $m$-th derivatives, the optimal rate of convergence for two-dimensional kernel density estimator is $n^{-(m+1)/(2m+4)}$~\citep{silverman1986density,scott1992multivariate}.

Therefore, when studying the convergence of MI estimators, it is very attempting to impose the H\"{o}lder condition on the $m$-th derivatives of the copula density. In fact, under the H\"{o}lder condition on the copula density itself (i.e., on the $0$-th derivative), \citet{liu2012exponential} showed that the kernel density estimation (KDE) based MI estimator converges at the parametric rate of $n^{-1/2}$. \citet{pal2010estimation} also considered similar H\"{o}lder condition when they studied the convergence of $k$-nearest-neighbor (KNN) based MI estimator. However, we argue that such conditions are too strong for copula density, thus these results do not reflect the true difficulty of MI estimation.

Specifically, the H\"{o}lder condition on the copula density means
  \beq\label{eq:holder}
 |c(u_1,v_1) - c(u_2,v_2)| \le M_1 \| (u_1 - u_2, v_1-v_2) \|
  \eeq   for a constant $M_1$ and all $u_1,v_1,u_2,v_2$ values between $0$ and $1$. Here and in the following $\|\cdot\|$ refers to the Euclidean norm. However, this H\"{o}lder condition (\ref{eq:holder}) would exclude all commonly used continuous copula densities since they are unbounded~\citep{omelka2009copula,segers2012copula}. Therefore, we need to consider the minimax convergence rate under a less restrictive condition.

When $c(u,v)$ is unbounded, the H\"{o}lder condition can not hold for the region where $c(u,v)$ is big. Hence we impose it only on the region where the copula density is small. Specifically, we assume that the H\"{o}lder condition (\ref{eq:holder}) holds only on the region $A_M = \{(u,v): c(u,v)<M\}$ for a constant $M>1$. That is, $|c(u_1,v_1) - c(u_2,v_2)| \le M_1 \| (u_1 - u_2, v_1-v_2) \|$ whenever $(u_1,v_1) \in A_M$ and $(u_2,v_2) \in A_M$. Then this condition is satisfied by all common continuous copulas in the book by~\citet{Nelsen2006Copula}. For example, all Gaussian copulas satisfy the H\"{o}lder condition (\ref{eq:holder}) on $A_M$ for some constants $M>1$ and $M_1>0$. But no Gaussian copulas, except the independence copula $\Pi$, satisfy the H\"{o}lder condition (\ref{eq:holder}) over the whole $\mathcal{I}^2$.

If (\ref{eq:holder}) holds on $A_M$ for any particular $M$ and $M_1$ values, then (\ref{eq:holder}) holds on $A_M$ also for all smaller $M$ values and for all bigger $M_1$ values. Without loss of generality, we assume that $M$ is close to $1$ and $M_1$ is a big constant.

Let $\mathfrak{C}$ denotes the class of continuous copulas whose density satisfies the H\"{o}lder condition  (\ref{eq:holder}) on $A_M$. We can then study the minimax risk of estimating $MI(C)$ for $C \in \mathfrak{C}$.
Without loss of generality, we consider the data set $\{(U_1,V_1), ..., (U_n,V_n)\}$ consisting of independent observations from a copula distribution $C \in \mathfrak{C}$.

\begin{thm}\label{thm:MI_rate}
Let $\widehat{MI}_n$ be any estimator of the mutual information $MI$ in equation (\ref{MI}) based on the observations $(U_1,V_1)$, ..., $(U_n,V_n)$ from a copula distribution $C \in \mathfrak{C}$. And let $\widehat{MIcor}_n$ be any estimator of the $MIcor$ in equation (\ref{MIcor}). Then
\begin{equation} \label{rate}
\ba{ccl}
 \sup\limits_{C \in  \mathfrak{C}} E[|\widehat{MI}_n(C) - MI(C)|] &=&  \infty, \mbox{ and }\\
  \sup\limits_{C \in  \mathfrak{C}} E[|\widehat{MIcor}_n(C) - MIcor(C)|] &\ge&  a_2 >0,
\ea
\end{equation}
for a positive constant $a_2$.
\end{thm}

The proof of Theorem~\ref{thm:MI_rate} uses a method of Le Cam~\citep{lecam1973convergence,lecam1986book} by finding a pair of hardest to estimate copulas. That is, we can find a pair of copulas $C_1$ and $C_2$ in the class $\mathfrak{C}$ such that $C_1$ and $C_2$ are arbitrarily close in Hellinger distance but their mutual information are very different. Then no estimator can estimate MI well at both copulas $C_1$ and $C_2$, leading to a lower bound for the minimax risk. Detailed proof is provided in Section~\ref{sec:proof.mi}.

In the literature, MI are estimated using methods including kernel density estimation (KDE) \citep{Moon1995MIkde}, the $k$-nearest-neighbor (KNN) \citep{kraskov2004estimating}, maximum likelihood estimation of density ratio~\citep{suzuki2009mutual}. There are also other density estimation based MI estimators~\citep{Blumentritt2012MImeasEst} that use the Beta kernel density estimation~\citep{Chen1999BetaKernel} and the Bernstein estimator~\citep{Bouezmarni2013BernsteinEstCopDens}.

No matter which MI estimator above is used, Theorem~\ref{thm:MI_rate} states that its minimax risk over the family $\mathfrak{C}$ is infinite. Also, the scaled version for estimating MIcor have minimax risk bounded away from zero. That is, the MI and MIcor can not be estimated consistently over the class $\mathfrak{C}$. This inconsistency is not specific to an estimation method. The estimation difficulty comes from the instability of MI due to its definition, as shown by the huge difference in MI values in Figures~\ref{fig:equit.a} and ~\ref{fig:equit.c} for two virtually same probability distributions.

Mathematically, MI is unstable because it overweighs the region with large density $c(u,v)$ values. From equation (\ref{MI}), $MI$ is the expectation of $\log[c(u,v)]$ under the true copula distribution $c(u,v)$. In contrast, the $Ccor$ in (\ref{Ccor}) takes the expectation at the independence case $\Pi$ instead. This allows consistent estimation of $Ccor$ over the family $\mathfrak{C}$, as shown in the next subsection~\ref{sec:Ccor.consistent}.

\subsection{The Consistent Estimation Of Copula Correlation}\label{sec:Ccor.consistent}
 The proposed copula correlation measure $Ccor$ can be consistently estimated since the region of large copula density values has little effect on it. To see this, we derive an alternative expression of $Ccor$ (\ref{Ccor}). Let $x_+ = \max(x,0)$ denote the non-negative part of $x$. Then
$$\int\limits_{0}^1 \int\limits_{0}^1 [c(u,v)-1]_+ dudv - \int\limits_{0}^1 \int\limits_{0}^1 [1-c(u,v)]_+ dudv = \int\limits_{0}^1 \int\limits_{0}^1 [c(u,v)-1]dudv = 1 -1 =0.$$
Hence $\int\limits_{0}^1 \int\limits_{0}^1 [c(u,v)-1]_+ dudv = \int\limits_{0}^1 \int\limits_{0}^1 [1-c(u,v)]_+ dudv$. Therefore,
 $$
 \ba{cl}
 \int\limits_{0}^1 \int\limits_{0}^1 |c(u,v)-1| dudv & = \int\limits_{0}^1 \int\limits_{0}^1 [c(u,v)-1]_+ dudv + \int\limits_{0}^1 \int\limits_{0}^1 [1-c(u,v)]_+ dudv \\
 & = 2 \int\limits_{0}^1 \int\limits_{0}^1 [1-c(u,v)]_+ dudv.
 \ea
 $$
Then we arrive at the alternative expression
\beq \label{Ccor_+}
Ccor = \frac{1}{2}\int\limits_{0}^1 \int\limits_{0}^1 |c(u,v)-1| dudv = \int\limits_{0}^1 \int\limits_{0}^1 [1-c(u,v)]_+ dudv.
 \eeq
In the new expression (\ref{Ccor_+}), $Ccor$ only depends on $[1-c(u,v)]_+$ which is nonzero only when $c(u,v)<1$. To estimate $Ccor$ well, we only need the density estimator $c_n(u,v)$ to be good for points $(u,v)$ with low copula density. Specifically, we consider the plug-in estimator
\beq \label{Ccor.est}
\widehat{Ccor} = Ccor(c_n) = \int\limits_{0}^1 \int\limits_{0}^1 [1-  c_n(u,v)]_+ dudv,
\eeq where ${c}_n(u,v)=\frac{1}{nh^2} \sum\limits_{i=1}^n K(\frac{u - U_i}{h})K(\frac{v - V_i}{h})$ is a kernel density estimator with kernel $K(\cdot)$ and bandwidth $h$.

To analyze the statistical error of $\widehat{Ccor}$, we can look at the error in the low copula density region separately from the error in the high copula density region. Specifically, let $M_2$ be a constant between $1$ and $M$, say, $M_2=(M+1)/2$. Then we can separate the unit square into the low copula density region $A_{M_2} = \{(u,v): c(u,v) \le M_2\}$ and the high copula density region $A_{M_2}^c = \{(u,v): c(u,v) >  M_2\}$. We now have $Ccor = T_1(c) + T_2(c)$ where $T_1(c)= \iint\limits_{A_{M_2}} [1-c(u,v)]_+ du dv$ and $T_2(c) = \iint\limits_{A_{M_2}^c} [1-c(u,v)]_+ du dv$. Since the H\"{o}lder condition (\ref{eq:holder}) holds on $A_M$, the classical error rate $O(h+(nh^2)^{-1/2})$ for the kernel density estimator holds for $|c_n(u,v)- c(u,v)|$ on  the low copula density region $A_{M_2}$. Hence the error $|T_1(c_n) - T_1(c)|$ is also bounded by $O(h+(nh^2)^{-1/2})$. While the density estimation error $|c_n(u,v)- c(u,v)|$ can be unbounded on the high copula density region $A_{M_2}^c$, it only propagates into error for $\widehat{Ccor}$ when $c_n(u,v)<1$. We can show that the overall propagated error $|T_2(c_n) - T_2(c)| $ is controlled at a  higher order $O((nh^2)^{-1})$. Therefore, the error rate of $\widehat{Ccor}$ can be controlled by the classical kernel density estimation error rate as summarized in the following Theorem~\ref{thm:Ccor}.

\begin{thm}\label{thm:Ccor}
Let ${c}_n(u,v)=\frac{1}{nh^2} \sum\limits_{i=1}^n K(\frac{u - U_i}{h})K(\frac{v - V_i}{h})$ be a kernel estimation of the copula density based on observations $(U_1,V_1)$, ..., $(U_n,V_n)$. We assume the following conditions
\begin{enumerate}
  \item The bandwidth $h \to 0$ and $nh^2 \to \infty$.
  \item The kernel $K$ has compact support $[-1,1]$.
  \item $\int_{-\infty}^\infty K(x) dx = 1$, $\int_{-\infty}^\infty x K(x) dx = 0$ and $\mu_2 = \int_{-\infty}^\infty x^2 K(x) dx >0$.
\end{enumerate}
Then the plug-in estimator $\widehat{Ccor}=Ccor(c_n) $ in (\ref{Ccor.est}) has a risk bound
\begin{equation} \label{Ccor.err}
 \sup_{C \in  \mathfrak{C}} E[|\widehat{Ccor} - Ccor|] \le 2\sqrt{M_1} h + \frac{2\mu_2}{\sqrt{nh^2}} + \frac{M_5}{nh^2}
\end{equation}
for some finite constant $M_5>0$.
\end{thm}

The detailed proofs for Theorem~\ref{thm:Ccor} are provided in Section~\ref{sec:proof.ccor}. From (\ref{Ccor.err}), if we choose the bandwidth $h=n^{-1/4}$, then $\widehat{Ccor}$ converges to the true value $Ccor$ at the rate of $O(n^{-1/4})$. Thus $Ccor$ can be consistently estimated, in contrast to the results on $MI$ and $MIcor$ in subsection~\ref{sec:MIbad}.

The Theorem~\ref{thm:Ccor} provides only an upper bound for the statistical error of the plug-in estimator $\widehat{Ccor}$. The actual error may be lower. In fact, the error $|T_1(c) - T_1(\hat c_n)|$ can be controlled at $O(n^{-1/2})$ using kernel density estimator $c_n$ \citep{BickelRitov2003}. Here we did not find the optimal rate of convergence. But the upper bound already shows that $Ccor$ is much easier to estimate than $MI$ and $MIcor$. Similar to classical kernel density estimation theory, assuming that the H\"{o}lder condition holds on $A_M$ for the $m$-th derivatives of the copula density, the upper bound on the convergence rate can be further improved to $O(n^{-(m+1)/(2m+4)})$.

The technical conditions $1-3$ in Theorem~\ref{thm:Ccor} are classical conditions on the bandwidth and the kernel. We have used the bivariate product kernel for technical simplicity. Other variations of the conditions in the literature may be used. For example, it is possible to relax the compact support condition~2 to allow using the Gaussian kernel.

Further adjustment is needed for a practical estimator for Ccor. In practice, the $(U_i,V_i)$'s are not observed. From the raw data of $(X_i,Y_i)$'s, $i=1,...,n$, it is conventional to estimate $(\hat U_i = R_{X,i}/(n+1), \hat V_i = R_{Y,i}/(n+1))$, and then calculate $\widehat{Ccor}$ using $(\hat U_i, \hat V_i)$'s. Here $R_{X,i}$ is the rank of $X_i$ among all Xs, and $R_{Y,i}$ is the rank of $Y_i$ among all Ys. We will use the square kernel $K(u)K(v)=\mathds{1}\{|u|<1\}\mathds{1}\{|v|<1\}$ and $h=0.25n^{-1/4}$ in practice. The bandwidth $h=0.25n^{-1/4}$ is selected through numerical study detailed in the supplemental Section~\ref{sec:supplemental}.

Also, for any fixed sample size $n$ and fixed bandwidth $h$, the estimator $\widehat{Ccor}$ can never reach the value of $1$ and $0$. Thus we make a finite-sample linear correction
\beq\label{Ccor.est.corrected} \widetilde{Ccor} =  (\widehat{Ccor} - Cmin)/(Cmax-Cmin). \eeq  Here $Cmax$ and $Cmin$ are respectively the maximum and minimum possible values of $\widehat{Ccor}$ given the $n$ and $h$ values. We use $\widetilde{Ccor}$ in the numerical study of Section~\ref{sec:Numerical}.

Extra effort is needed to prove the risk bound for $\widetilde{Ccor}$ using $(\hat U_i, \hat V_i)$'s. We did not do that here. The purpose of Theorem~\ref{thm:Ccor} is to show that Ccor is fundamentally easier to estimate than MI. The risk bound on $\widehat{Ccor}$ suffices for that purpose.

\section{NUMERICAL STUDIES}\label{sec:Numerical}
In this section, we conduct several numerical studies on the finite sample properties of the proposed Ccor, and compare it with several other measures. We first compare the equitability of different correlation measures in simulation studies in subsection~\ref{sec:equana}. Subsection~\ref{sec:power} compares the computation time and the power of the independence tests based on these dependence measures. Finally, we apply Ccor to a data set of social, economic, health, and political indicators from the World Health Organization (WHO) in subsection~\ref{sec:WHO}. This WHO data set is analyzed by~\citet{Reshef2011MIC}, and is available from their website http://www.exploredata.net. We used their MINE package from the same website to calculate MIC.

\subsection{Equitability Analysis}\label{sec:equana}
The main purpose of a dependence measure is to rank the strength of dependence within pairs of random variables. The Pearson's correlation ranks the pairs of related variables based on the strength of linear relationship within each pair. An equitable dependence measure does not prefer linear relationship nor any other particular types of relationship. The  equitable dependence measure should treat all types of relationship equally, and do the ranking purely on the strength of contained deterministic relationship.

\begin{center}
\begin{figure}[htb]
\includegraphics[scale=0.6]{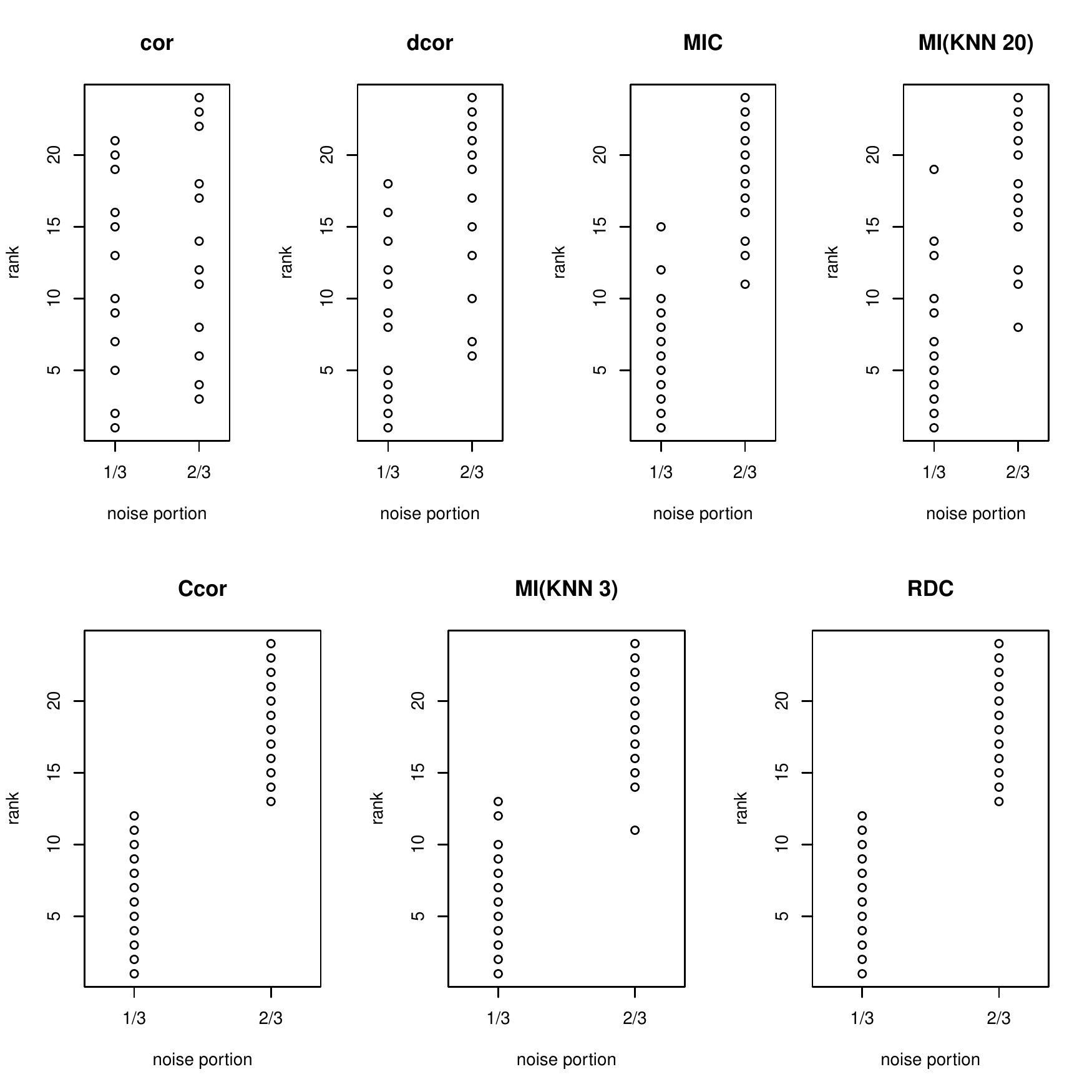}
\caption{Ranking the data sets using various dependence measures against the noise levels.}
\label{fig:PlotRS}
\end{figure}
\end{center}

We check the performance of various dependence measures in this respect with a simulation study. We generate bivariate data sets each with a deterministic relationship hidden in the uniform background noise. We generate data with two noise proportions $1-p$ at $1/3$ and $2/3$. Six different deterministic relationships, linear and nonlinear ones, are used in the simulation. These relationships are specified in the supplemental section~\ref{sec:functions}. The data sets are of two sample sizes $n=200$ and $n=2000$. We calculate the dependence measures on these data sets, and rank the data sets in order according to each dependence measure. An equitable dependence measure would separate the data sets purely based on the noise levels $1-p$. Figure~\ref{fig:PlotRS} shows the ranking by Ccor and other dependence measures reviewed.
We can see that the Pearson correlation (cor), distance correlation (dcor) and MIC all do not separate the two noise levels well. For the mutual information MI, we considered the KNN estimators with different tuning parameters $K=20$ and $K=3$ as in \cite{Kinney2014}. \cite{Kinney2014} showed that the estimator MI(KNN20) is more powerful when used to test independence, while the estimator MI(KNN3) shows better self-equitability in finite sample. In Figure~\ref{fig:PlotRS}, the MI(KNN3) separates the two noise levels much better than MI(KNN20). The Ccor and RDC do the best job at separating the two noise levels. The RDC (Randomized Dependence Coefficient) is proposed by~\citet{NIPS2013Lopez-PazRandDepCoef} as an estimator for Rcor.

In section~\ref{sec:equit}, we showed that MI is not robust-equitable. Its theoretical value is defined as infinity in those cases. So as sample size increases, the value of its estimator will increase.  Hence MI ranks higher those data sets with larger sample size, rather than ranking purely by the noise level.

\begin{center}
\begin{figure}[htb]
\includegraphics[scale=0.5]{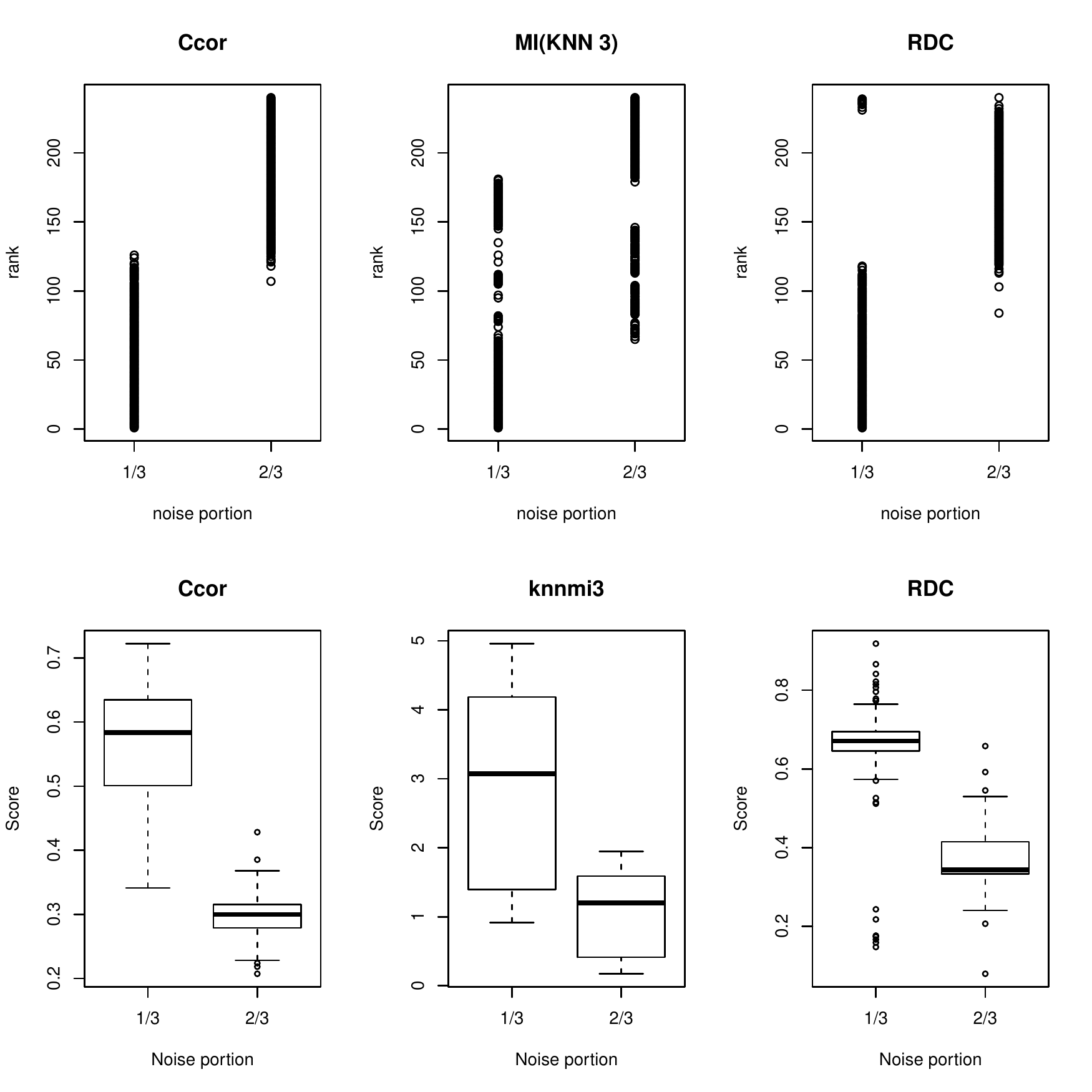}
\caption{Ranking the data sets using three dependence measures against the noise levels.}
\label{fig:PlotRS2}
\end{figure}
\end{center}

We use a bigger simulation to study further the three good measures in Figure~\ref{fig:PlotRS}, namely Ccor, MI(KNN3) and RDC. For bigger simulation, we ignore the more computationally intensive dcor and MIC which already perform badly. We repeat the simulation with sample sizes $n=200$ and $n=20000$, doing ten simulation runs at each combination of the noise levels and function types as before. The result is given by Figure~\ref{fig:PlotRS2}. We also plotted the box-plots of the dependence measures for more detailed information on the ranking.

The MI(KNN3) does not separate the noise levels anymore. This confirms that no estimator of MI reflects the strength of deterministic signal well.

\begin{center}
\begin{figure}[htbp]
	\centering
\begin{center}
         \begin{subfigure}[b]{0.3\textwidth}
               \includegraphics[width=\textwidth]{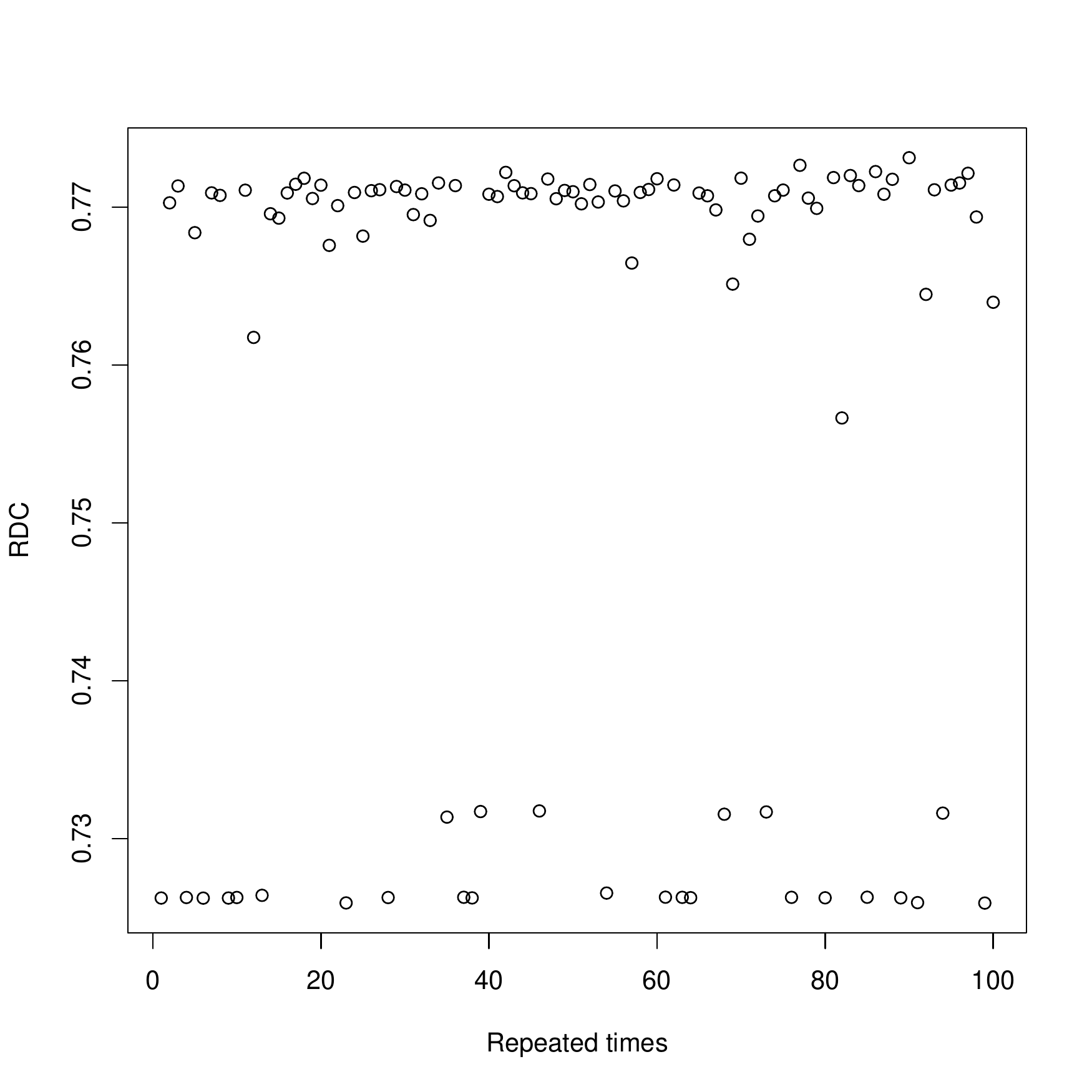}
                \caption{}
                \label{fig:RDCa}
        \end{subfigure}
	\quad
         \begin{subfigure}[b]{0.3\textwidth}
               \includegraphics[width=\textwidth]{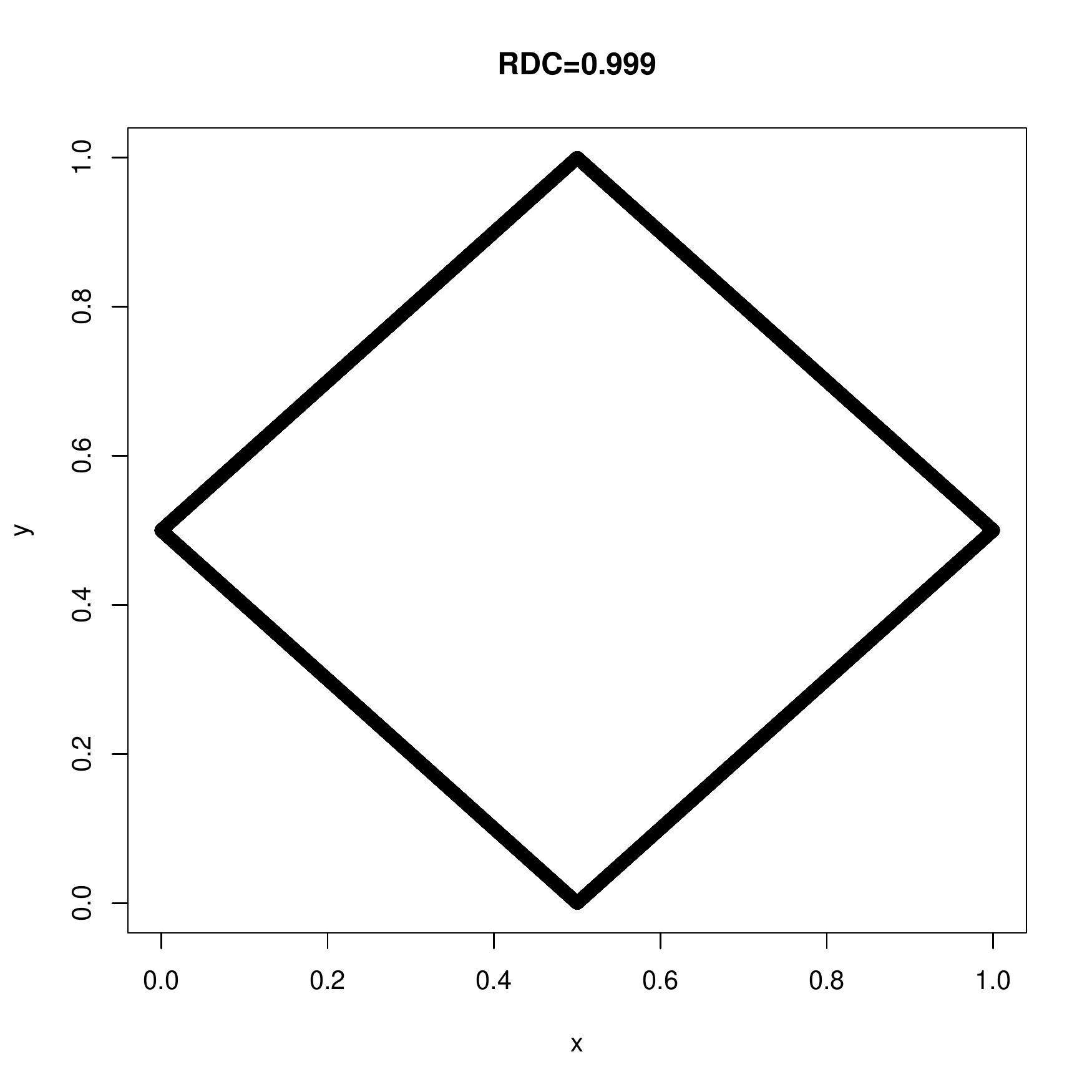}
                \caption{}
                \label{fig:RDCb}
        \end{subfigure}	
        \quad
         \begin{subfigure}[b]{0.3\textwidth}
               \includegraphics[width=\textwidth]{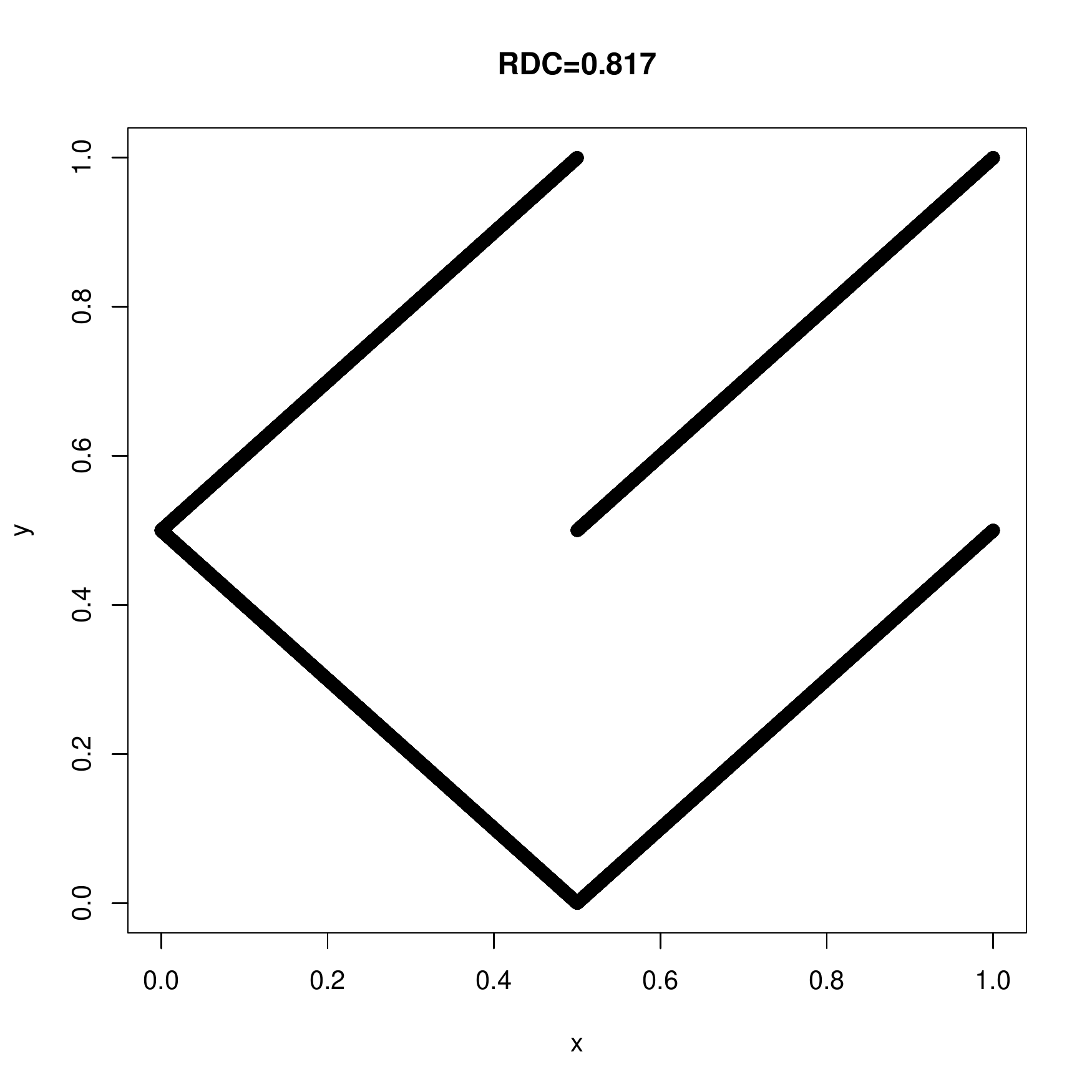}
               \caption{}
                \label{fig:RDCc}
        \end{subfigure}
        \end{center}
 	\caption{RDC drawback: (a) random RDC values of different runs on the same data set ($n=1,000$); (b) and (c)  plots two data sets ($n=100,000$ each) with different deterministic relationships and their RDC estimates.}
 \label{fig:PlotRDC}
\end{figure}
\end{center}

The Ccor still does a pretty good job at separating the noise levels, reflecting its good equitability property. RDC does much worse in separating the noise levels. RDC is an estimator for the Rcor. While we can not prove if Rcor is robust-equitable or not, one of its drawbacks mentioned earlier is that no good estimator exists. RDC also has problems as an estimator. For one, it is a randomized estimator. As Figure~\ref{fig:RDCa} shows over a fixed data set with sample size $n=1,000$, its value varies a lot over different runs. Also, sometimes it has trouble converging to the true value of Rcor. Figures~\ref{fig:RDCb} and \ref{fig:RDCc} show two data sets generated from two different deterministic relationships. In both case, Rcor=1. However, only in the first case~\ref{fig:RDCb} RDC gets close to one,  for a very large sample size $n=100,000$. For the second case in Figure~\ref{fig:RDCc}, even when $n=100,000$, RDC remains below $0.82$, far from $Rcor=1$.

\subsection{Comparison of Powers and Computation Times}\label{sec:power}
Here we conduct simulation to compare powers of independence tests corresponding to various dependence measures, similar to those in \citet{simon2011comment} and \cite{Kinney2014}. We compare the tests based the empirical estimates of linear correlation (cor), our copula correlation (Ccor), MIC, distance correlation (dcor), two versions of MI estimators and RDC. The two MI estimators (MI03 and MI20) are those KNN estimators in \cite{Kinney2014} with tuning parameters $K=3$ and $K=20$ respectively. We also included comparison to the (HHG) test of \cite{Heller2013IndTest}. The dcor, RDC and the HHG tests were calculated using the R packages contributed by those authors.

Similar to \citet{simon2011comment}, we simulated data sets of sample size $n=320$ from the regression model $Y=f(X)+\e$ with Gaussian error $\e \sim N(0,\sigma^2)$, with different bivariate functional relationships $Y=f(X)$. We used nine bivariate relationships from literature \citep{newton2009introDcor,Reshef2011MIC,Heller2013IndTest,Kinney2014}, listed in Table~\ref{tab:power-data}.
\begin{table} [htbp]
\begin{center}
\small{
\begin{tabular}{ccc|ccc}
    \hline
Type & $f(x)$ & Data & Type & $f(x)$ & Data \\
\hline \hline
Linear & $x$  & \includegraphics[scale=0.04]{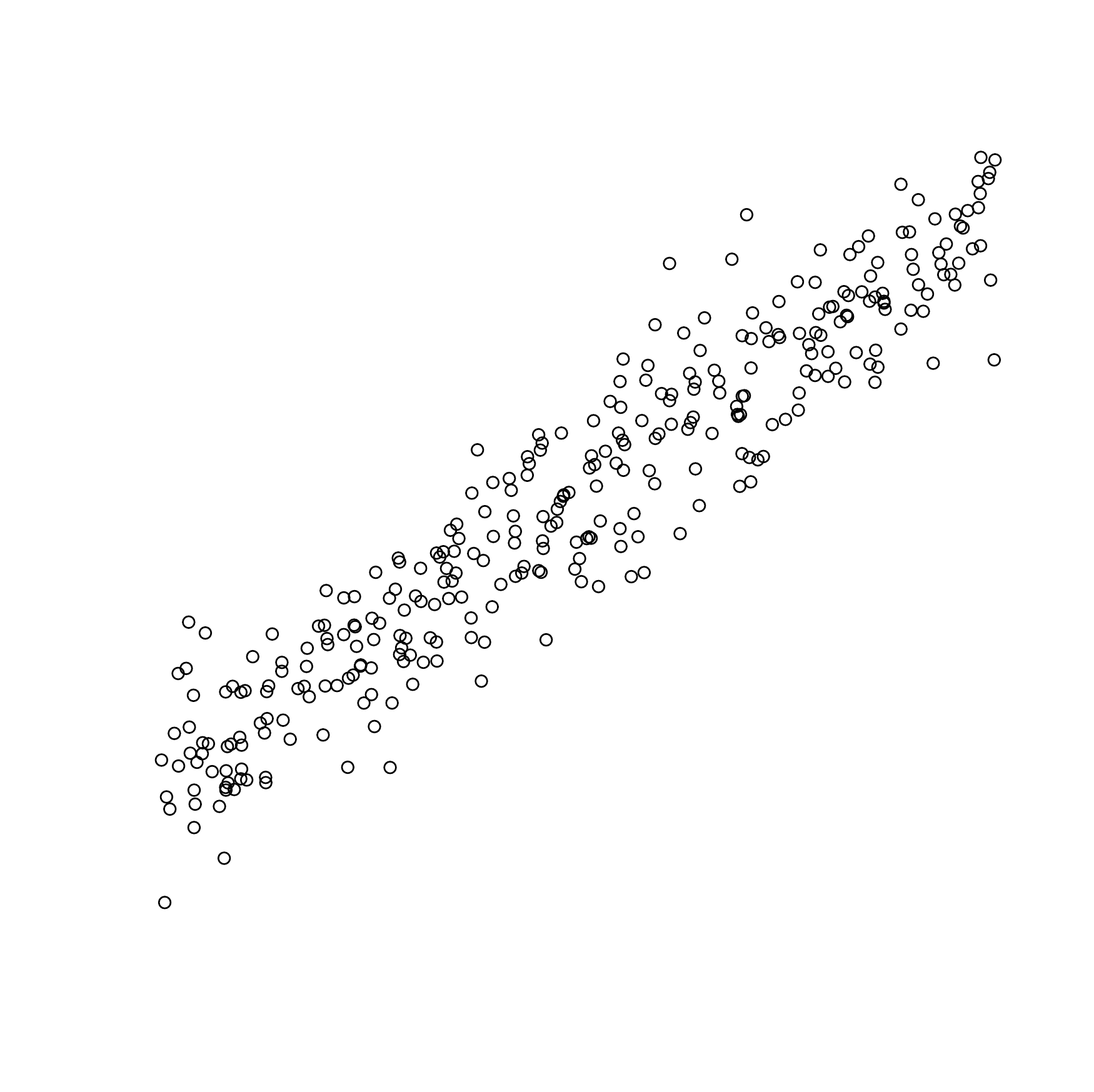} & Circle & $\pm \sqrt{\frac{1}{4}-(x-\frac{1}{2})^2}$ & \includegraphics[scale=0.04]{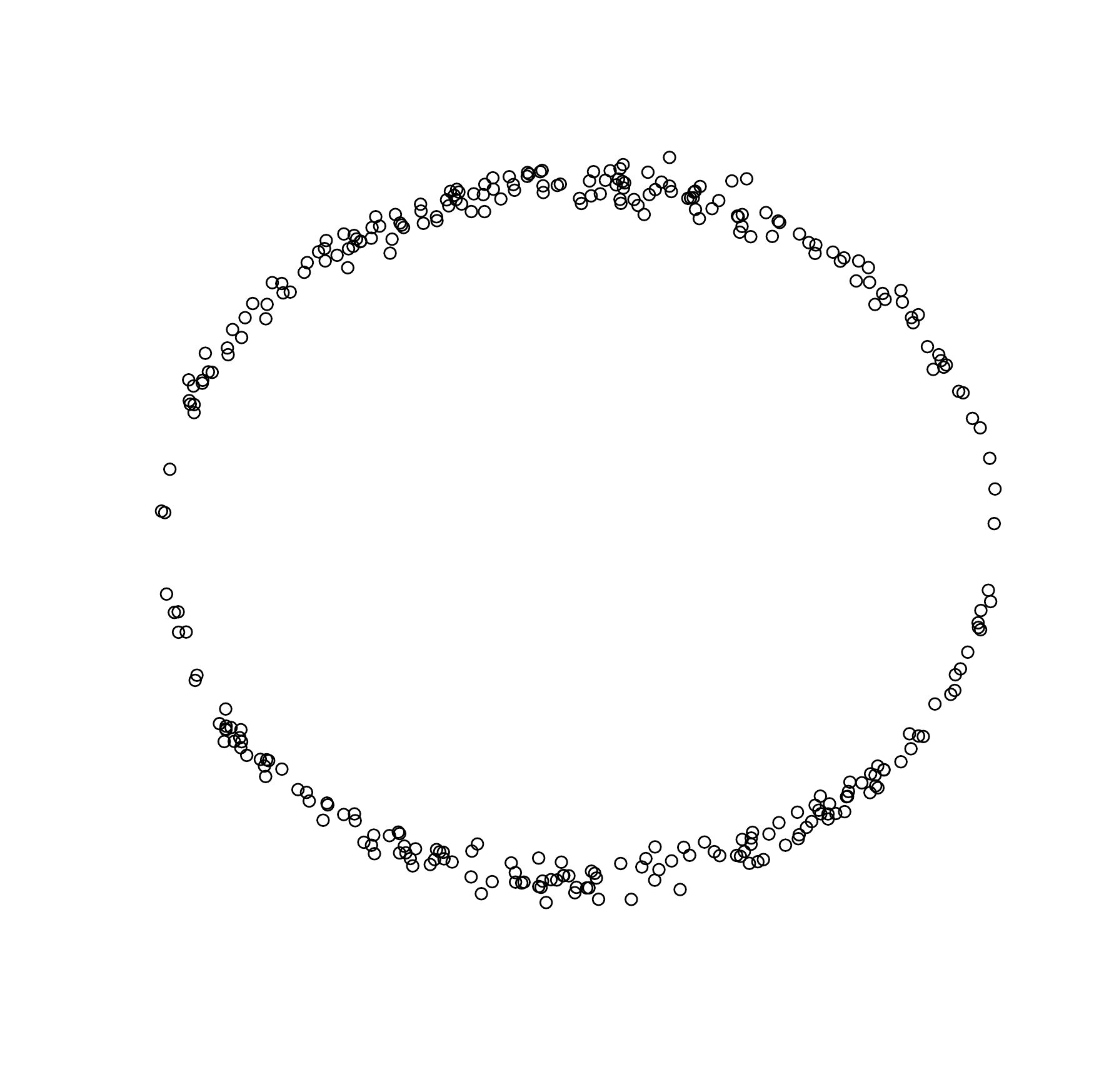}  \\
Parabolic & $4(x-\frac{1}{2})^2$ & \includegraphics[scale=0.04]{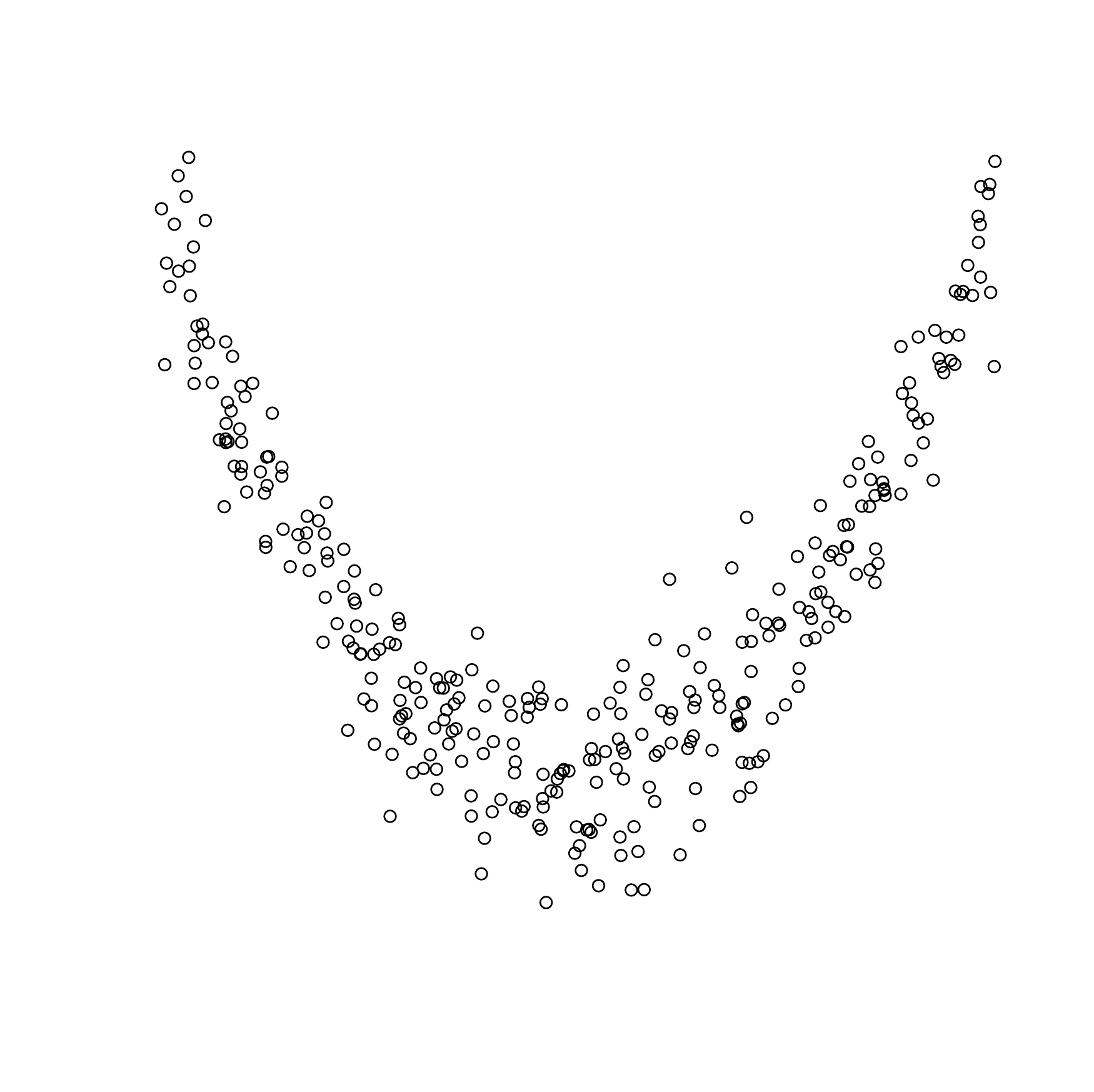}   & Cross & $\pm(x-\frac{1}{2})$ & \includegraphics[scale=0.04]{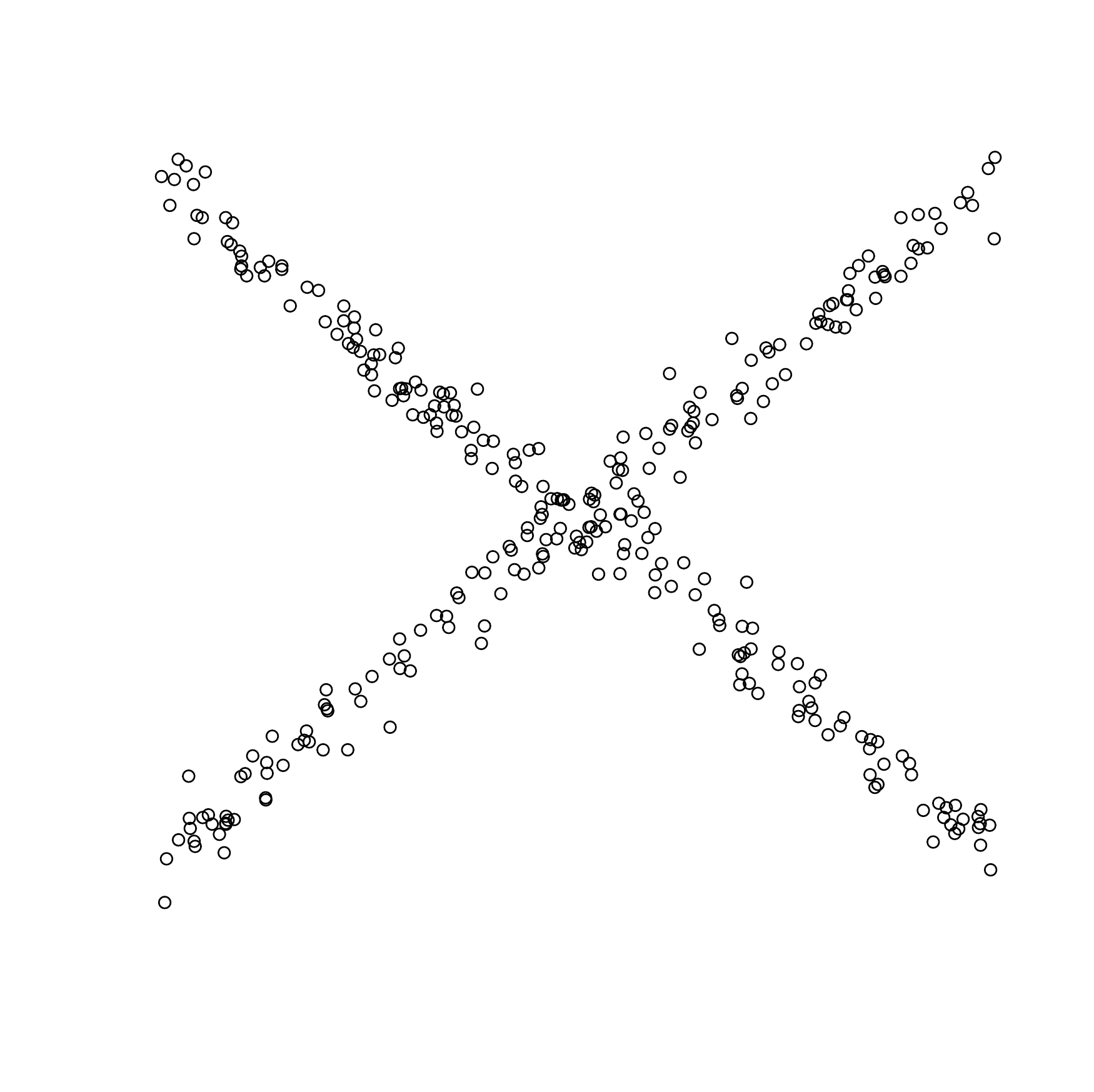} \\
Sin(4pix) &  $sin(4 \pi x)$ & \includegraphics[scale=0.04]{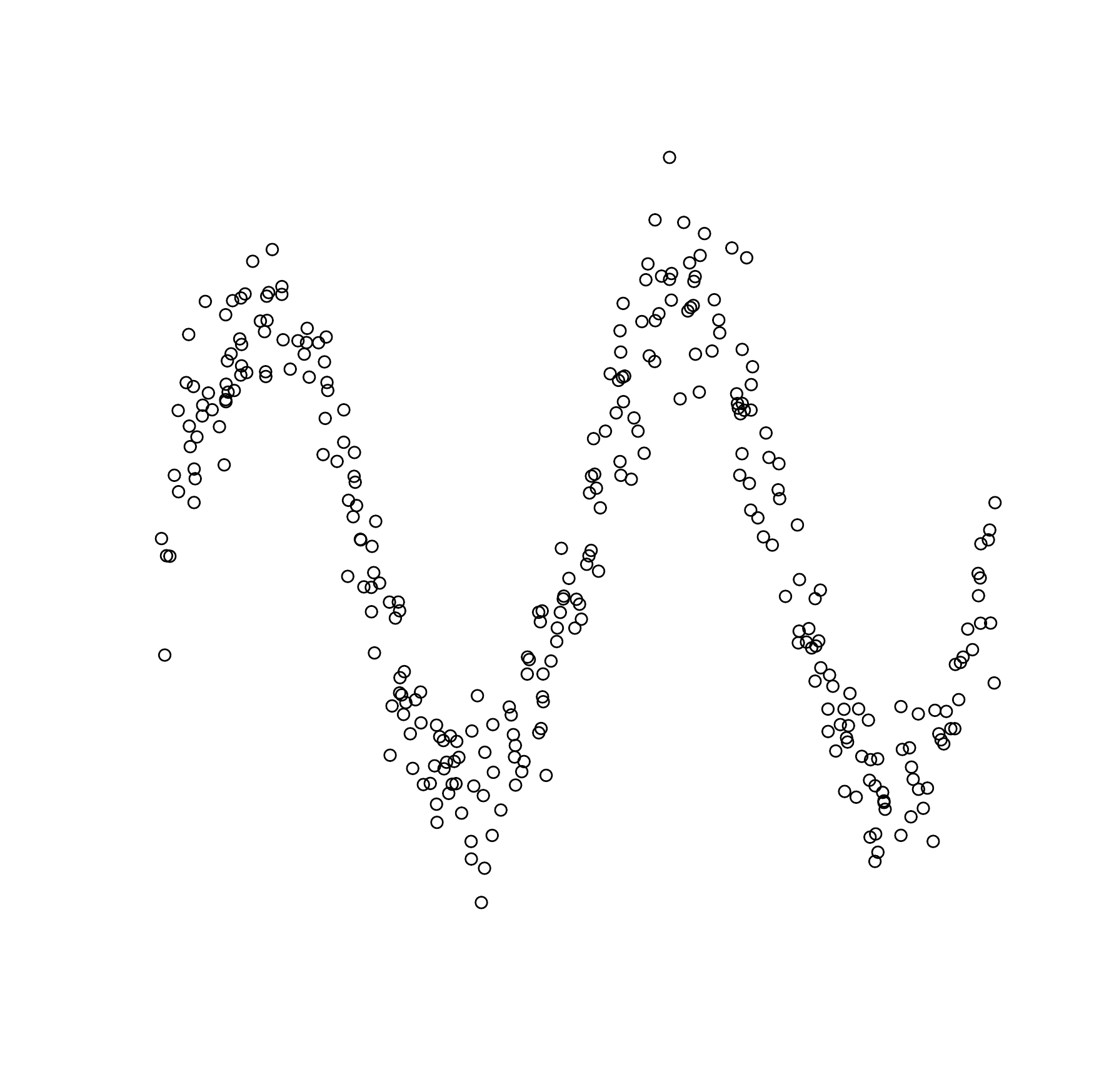}  & W & $4[(2x-1)^2-\frac{1}{2}]^2 $  & \includegraphics[scale=0.04]{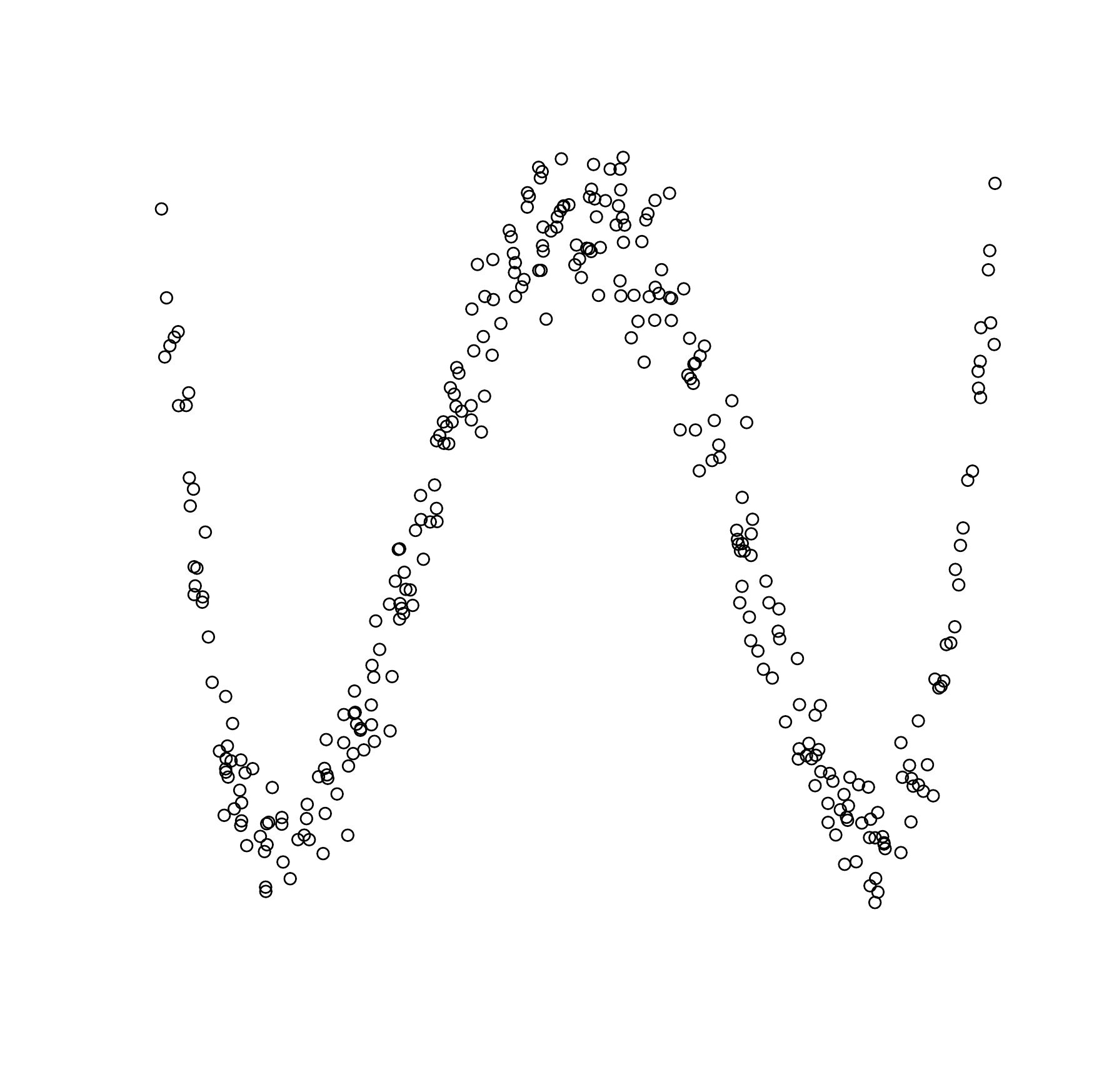} \\
Sin(16pix) & $sin(16 \pi x)$ & \includegraphics[scale=0.04]{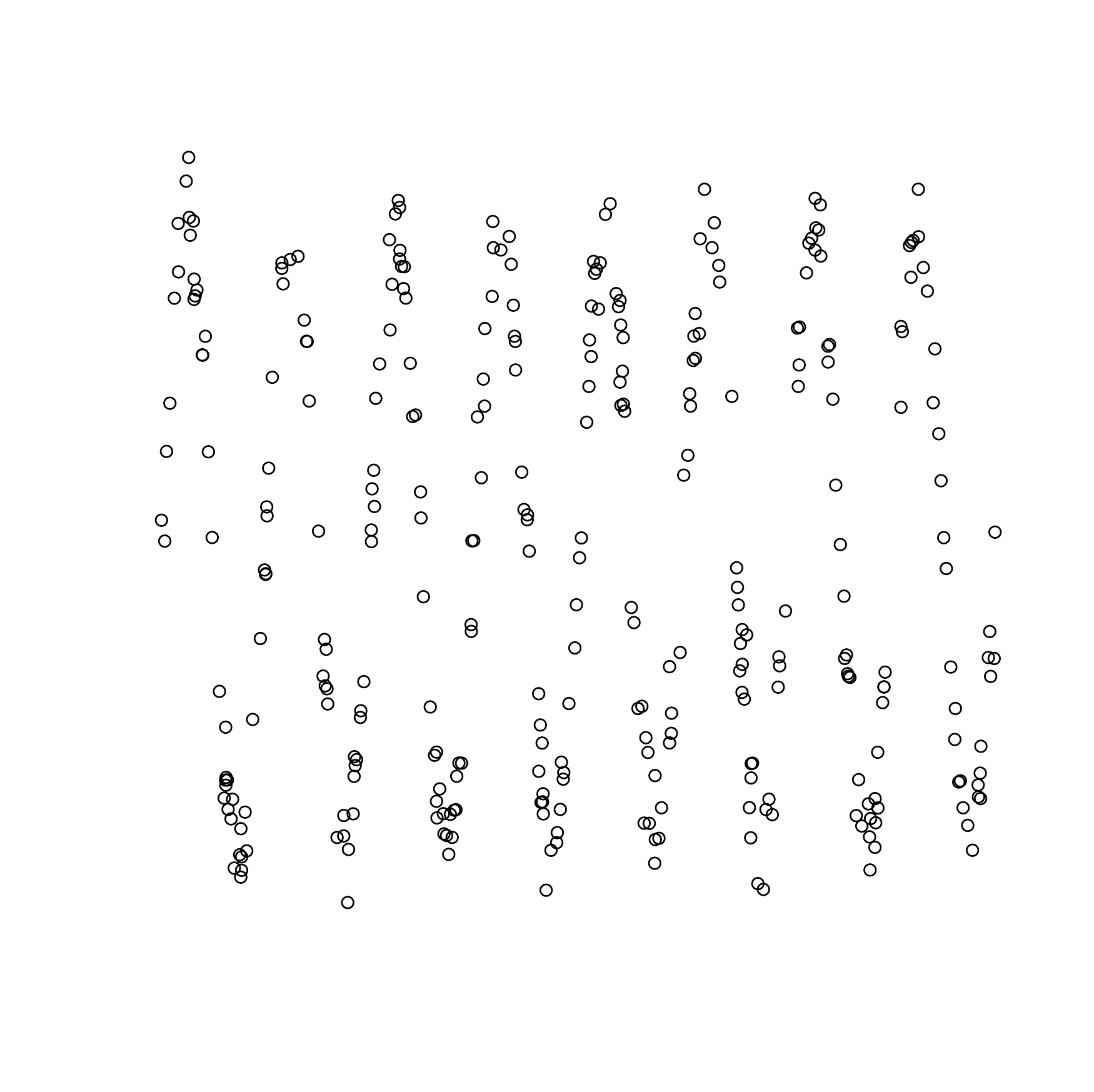}  & X para & $\pm 4(x-\frac{1}{2})^2 $ &  \includegraphics[scale=0.04]{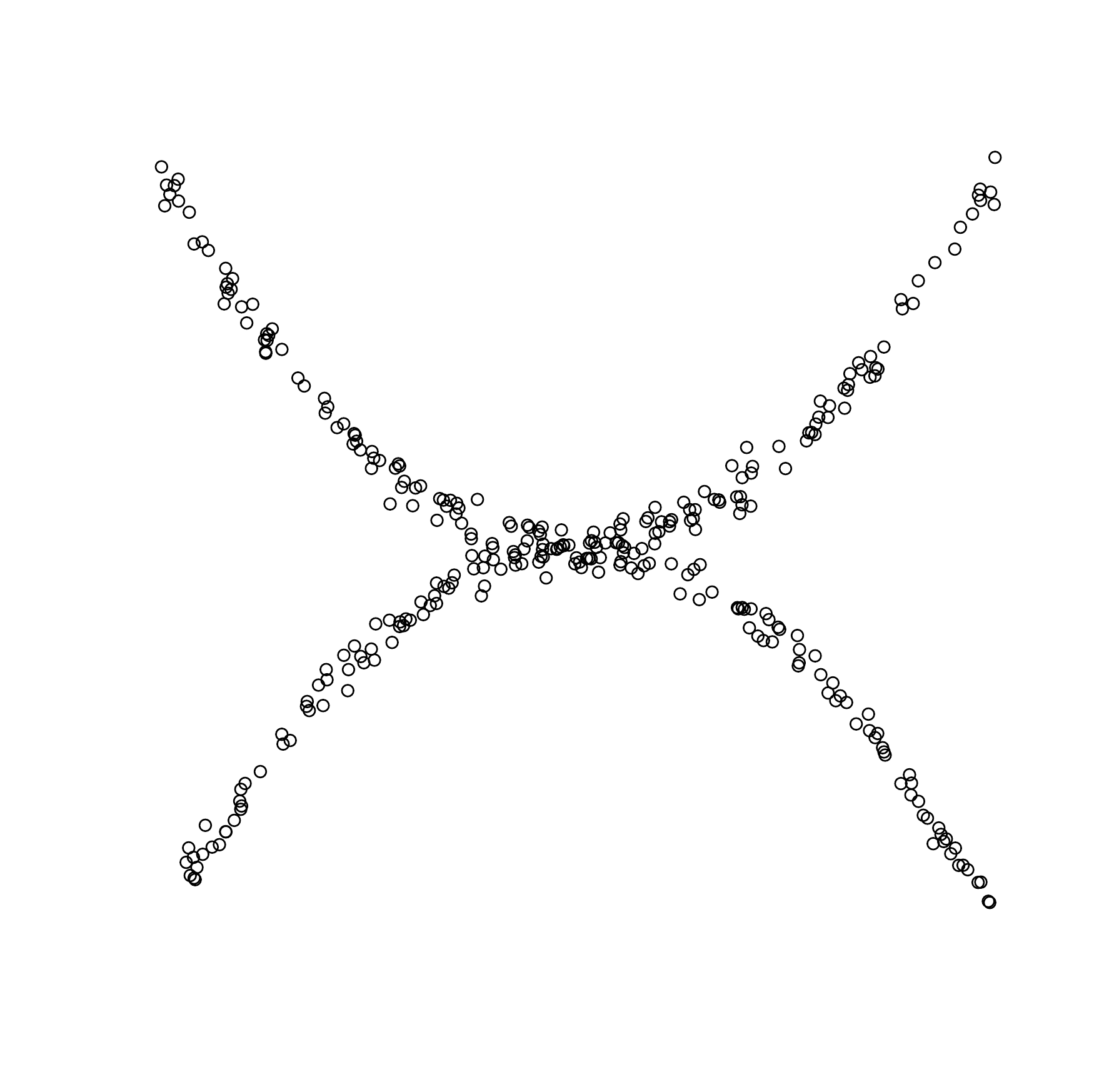} \\
 & &   & four clouds &  &  \includegraphics[scale=0.04]{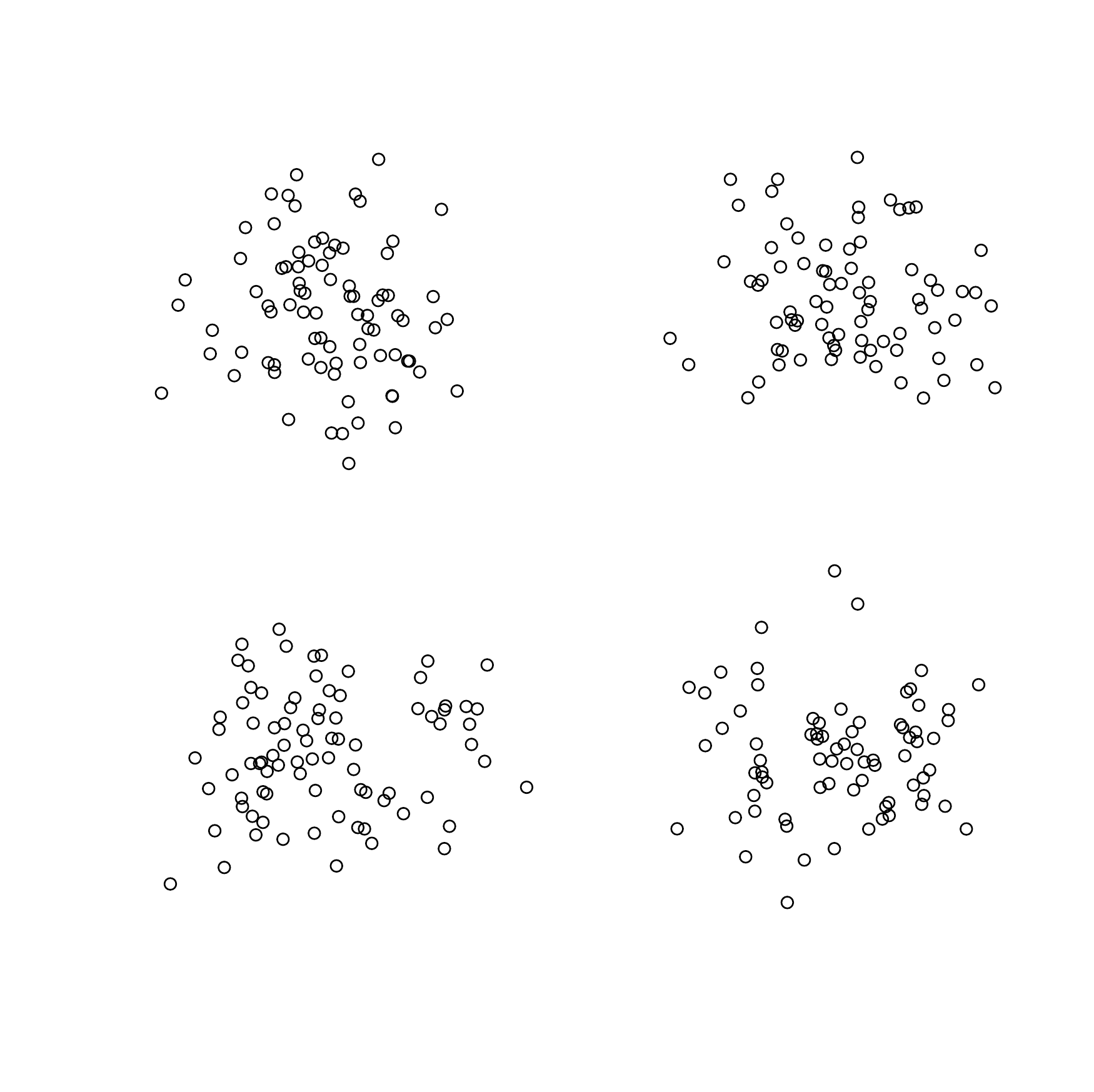} \\
    \hline
\end{tabular}
}
\end{center}
\caption{The functions used in the power comparison simulation. The ``Type" gives the name, $f(x)$ gives the definition, and ``Data" column draws one noisy data set for the type. }
\label{tab:power-data}
\end{table}

\begin{center}
\begin{figure}[htbp]
\includegraphics[scale=0.25]{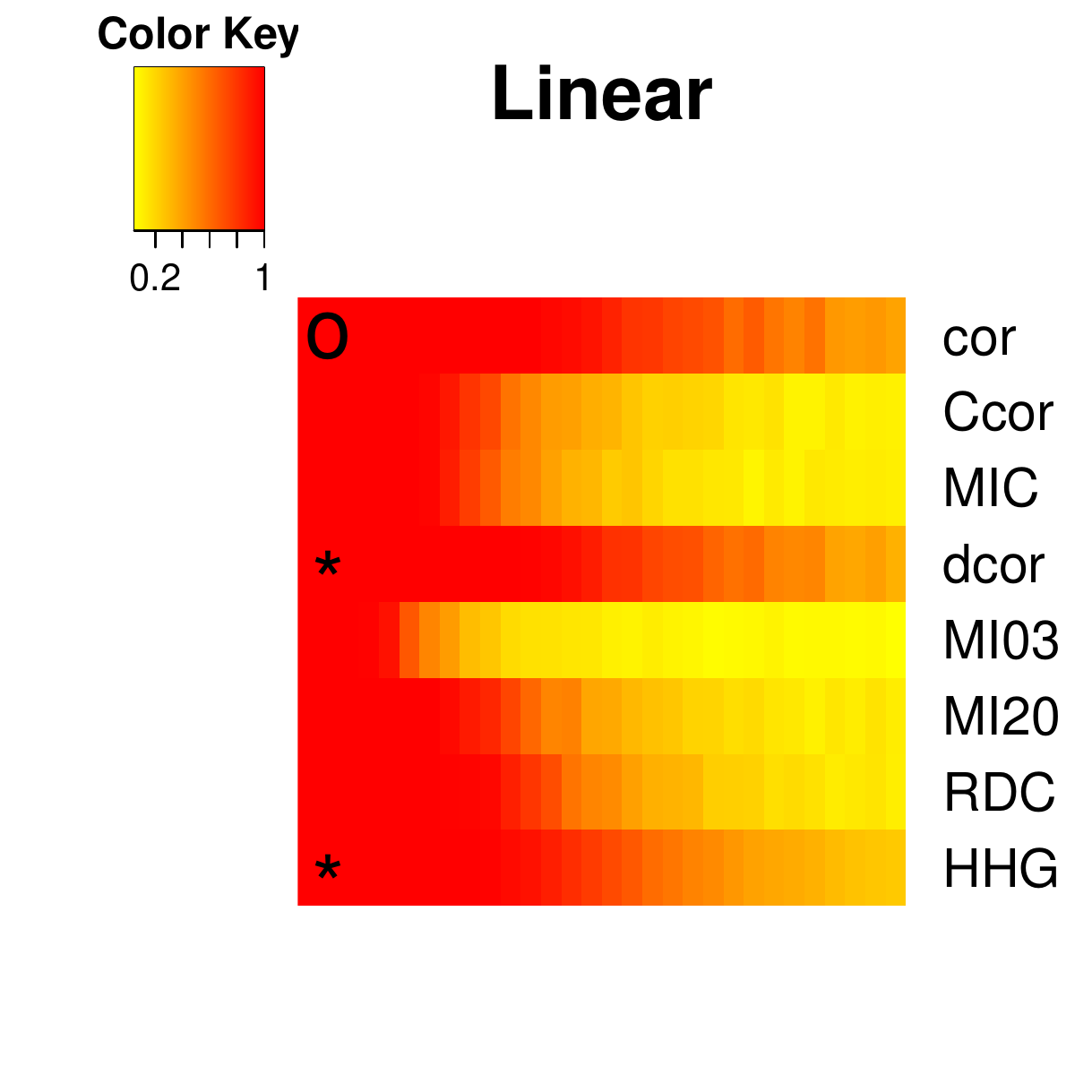}\includegraphics[scale=0.25]{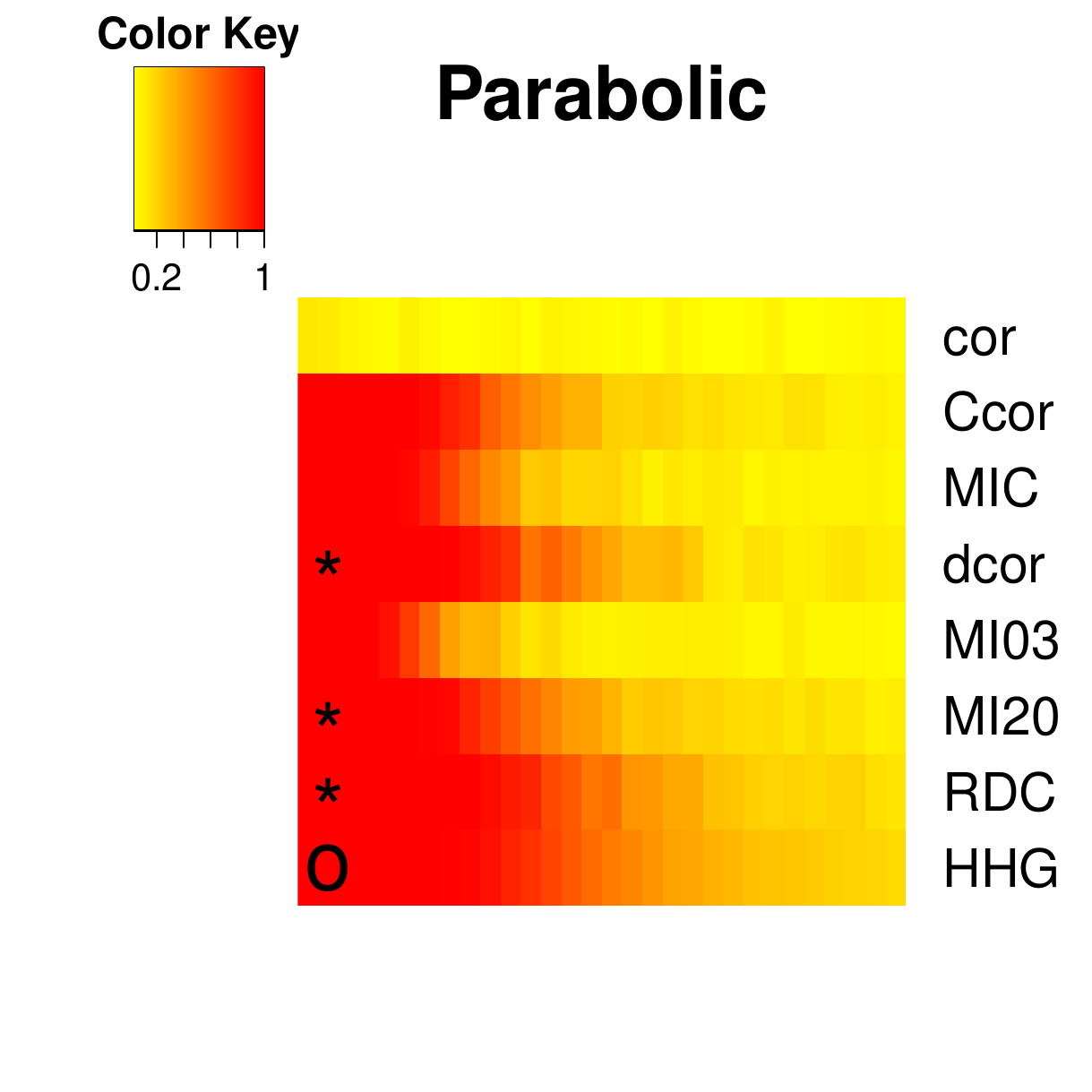}\includegraphics[scale=0.25]{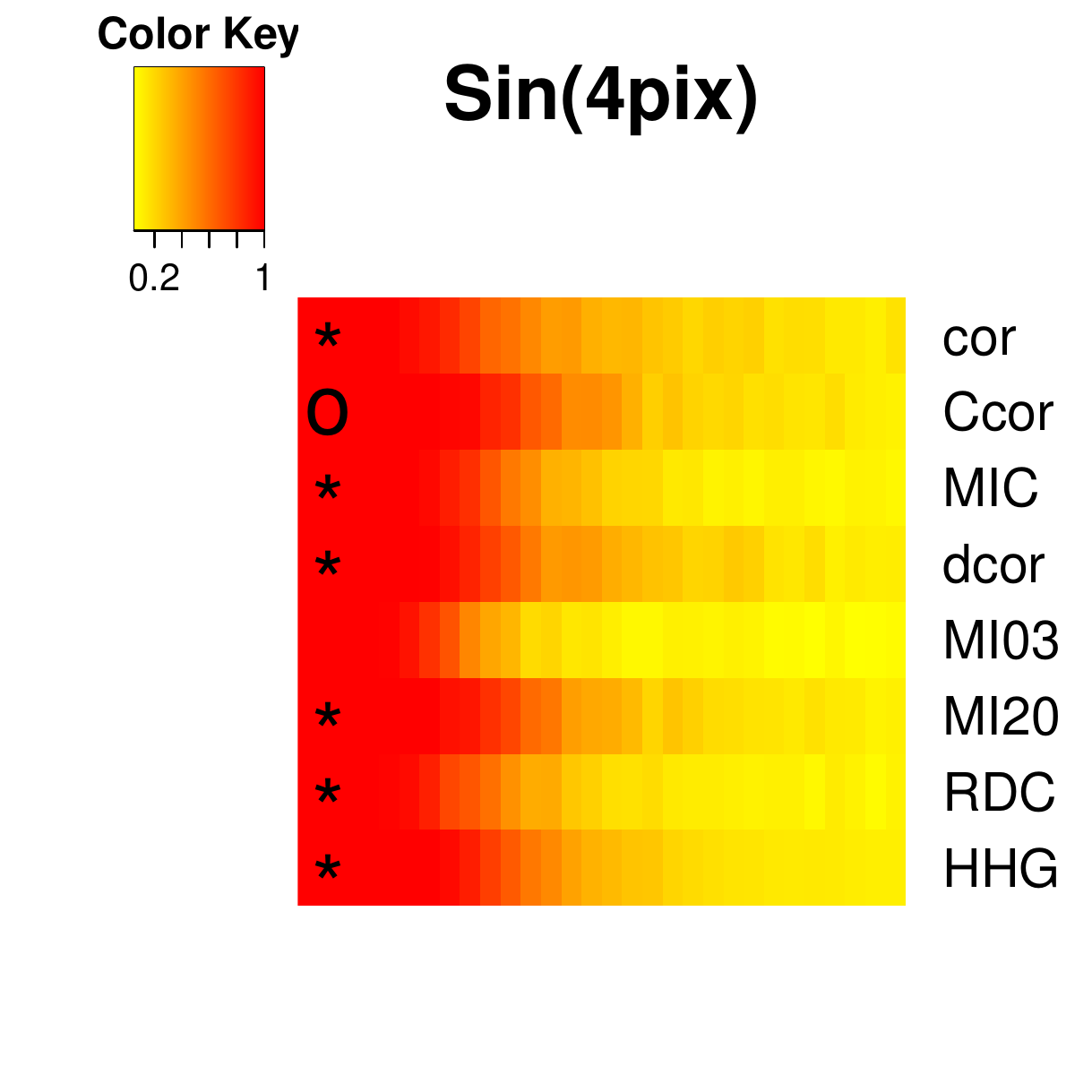}\includegraphics[scale=0.25]{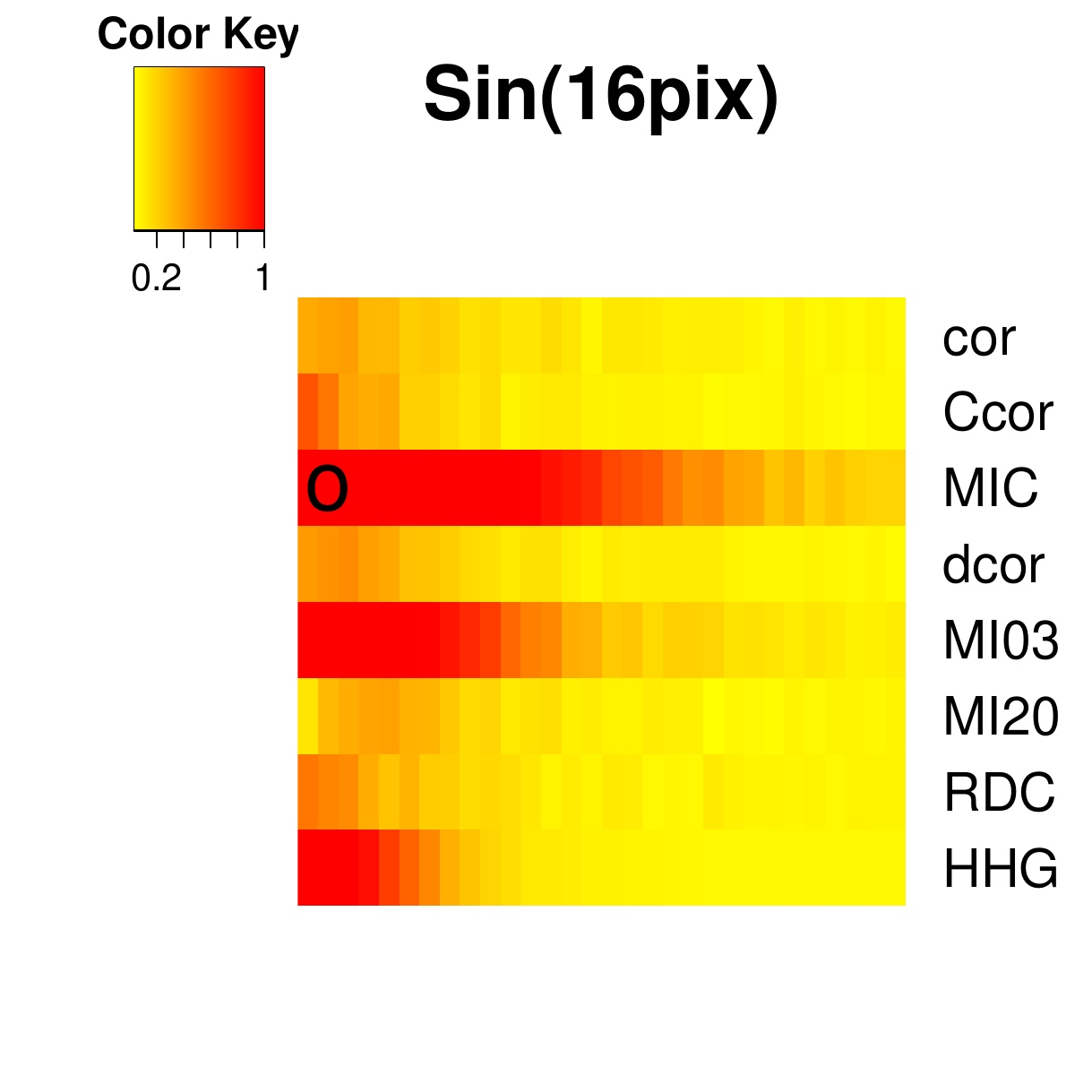}
\includegraphics[scale=0.25]{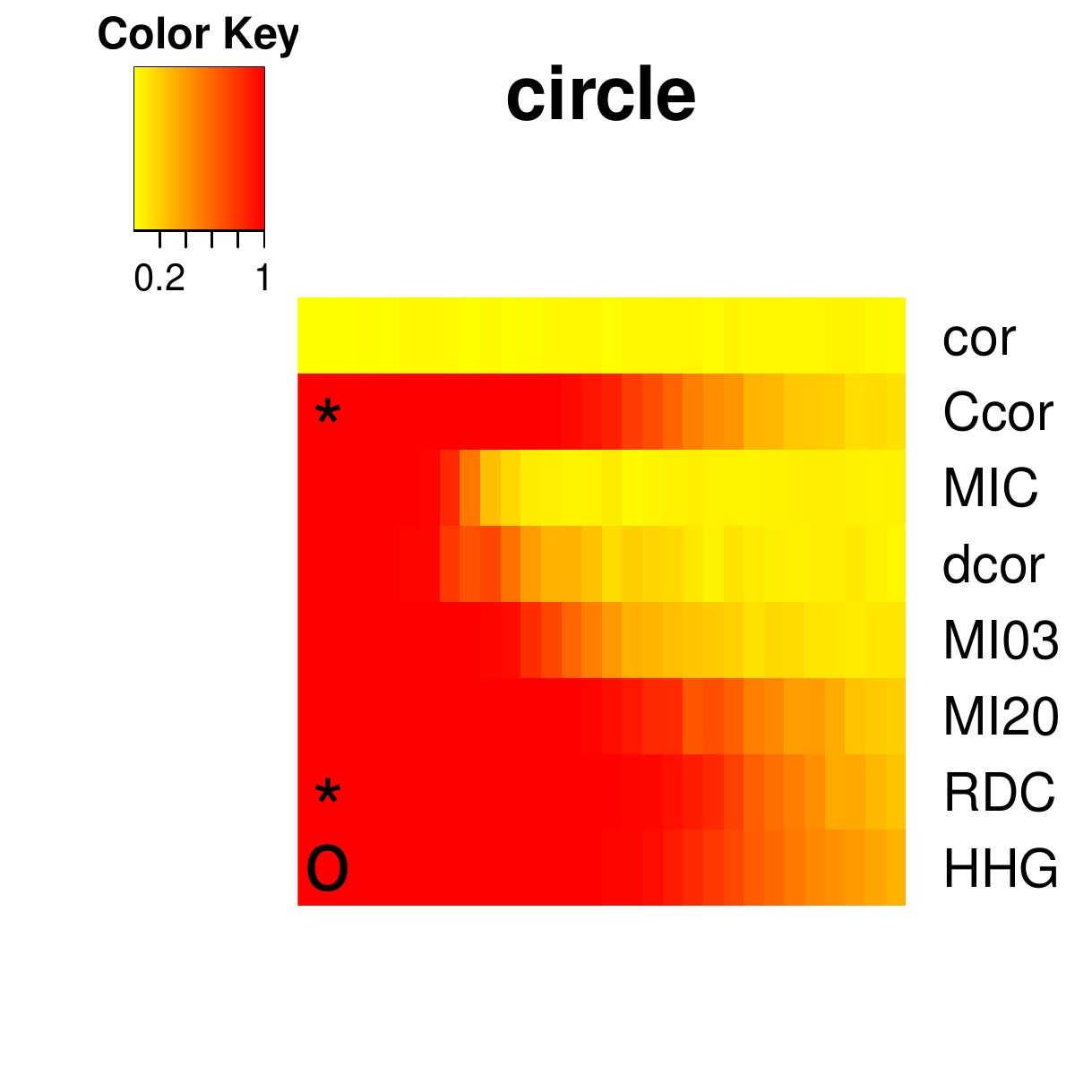}\includegraphics[scale=0.25]{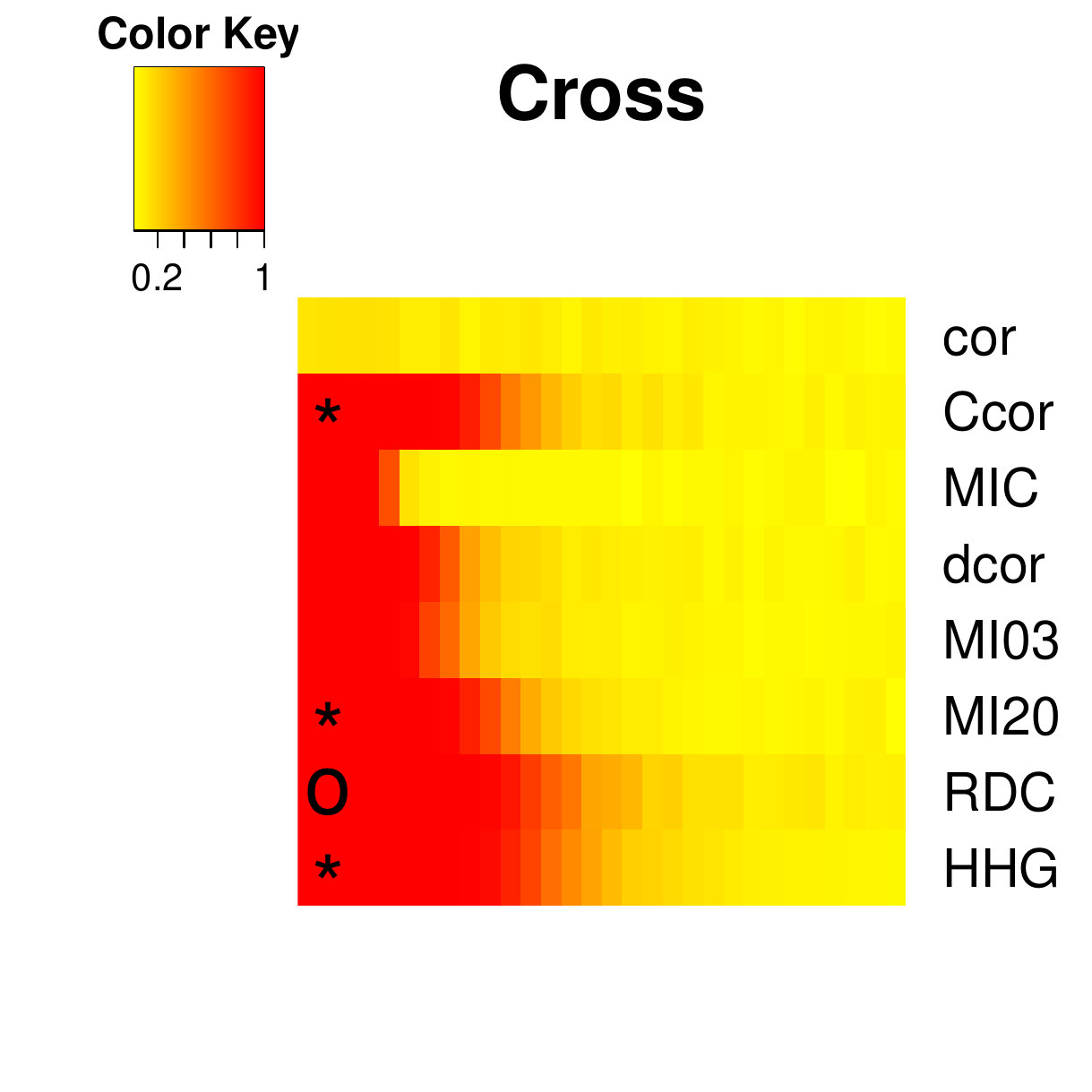}\includegraphics[scale=0.25]{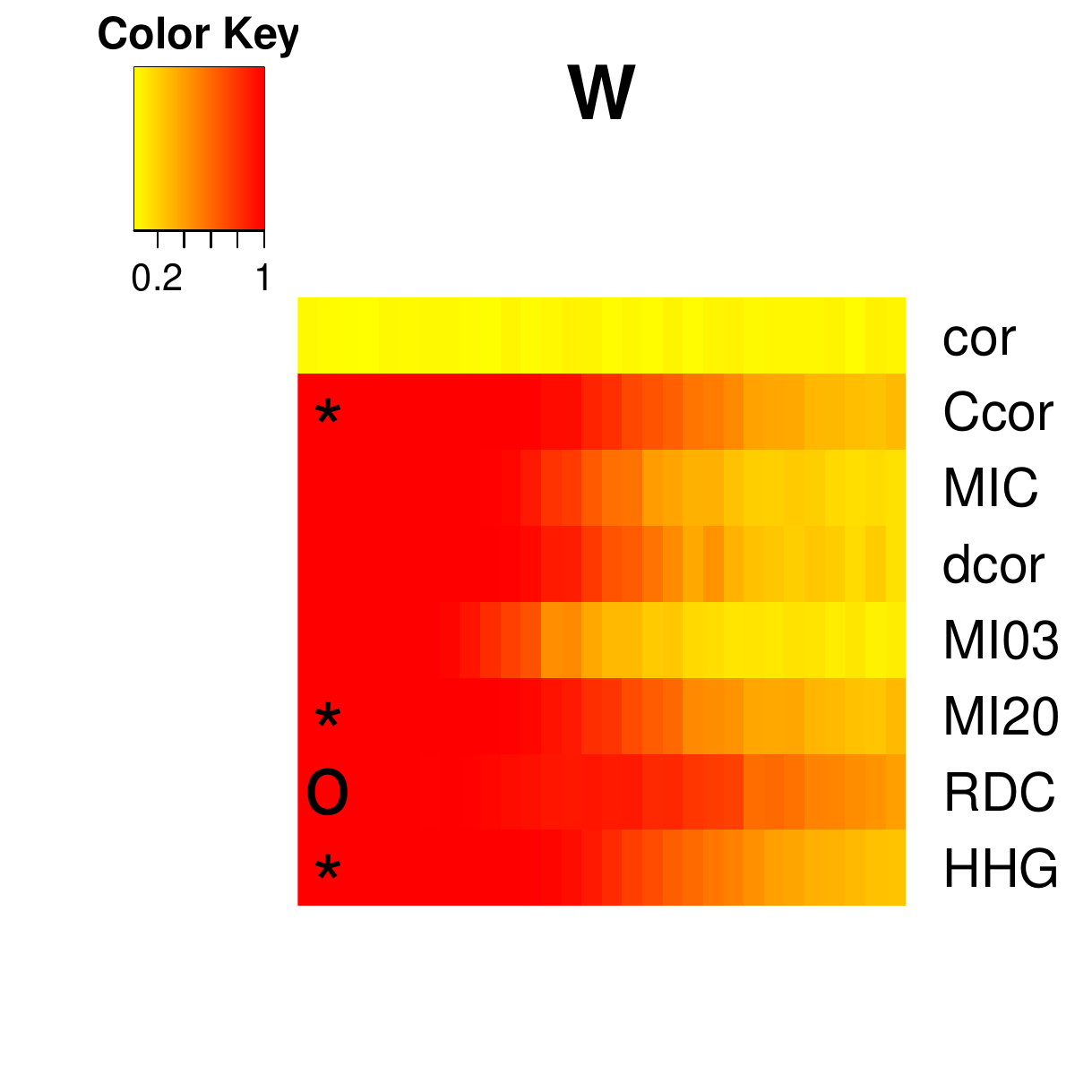}\includegraphics[scale=0.25]{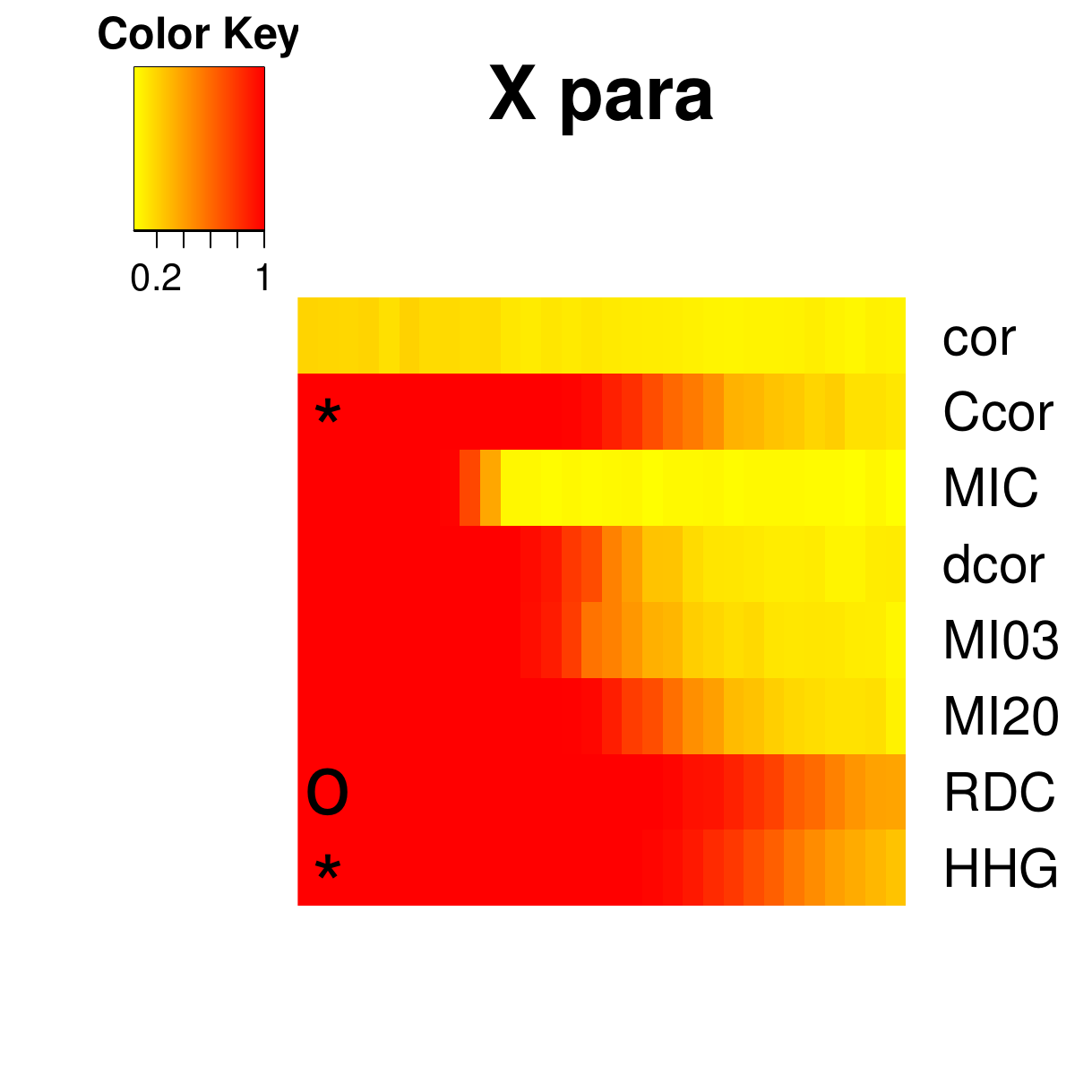}
\caption{Heat maps shows the statistical powers testing independence by various measures. ``O" indicates the test with maximum noise-at-50\% power, ``*" indicates tests with noise-at-50\%-powers within 25\% of the maximum.}
\label{fig:Heatmap}
\end{figure}
\end{center}

Data sets with 30 different increasing noise levels are generated. Similar to \citet{simon2011comment} and \cite{Kinney2014}, we decide the cutoff points as the $95$th percentile from $1000$ ``null" data sets created by randomly permuting the $Y$ values. The test rejects the null hypothesis of independence when the statistic on the simulated data sets exceeds the cutoff point, resulting in a $5\%$ significance level test. The power is calculated from $500$ simulated data sets, and reported in Figure~\ref{fig:Heatmap} across different noise levels and the first eight bivariate relationships. Following \cite{Kinney2014}, we labeled the test with the maximum noise-at-50\%-power and those tests with noise-at-50\%-powers within 25\% of this maximum.

From Figure~\ref{fig:Heatmap}, the linear correlation is best at detecting linear relationships but can have very low power to detect other relationships. There is no single test dominates in power in all cases. The MIC is very good at picking up the high frequency function $sin(16\pi x)$, but low power for other cases. Our Ccor has best power in one case and near best in other four cases. Overall, RDC and HHG have best or near best power in most cases.

We omitted from Figure~\ref{fig:Heatmap} the simulated powers for the last bivariate relationship ``four clouds" in Table~\ref{tab:power-data}. In fact, $X$ and $Y$ are independent in that case. So the simulated powers in that case are actually the Type I error rates, which are indeed close to the nominal $0.05$ level for all tests here. We note that HHG's p-values provided by their package would lead to much higher Type I error rates. Those p-values were not used in our simulation. Instead we decide the cutoff points for HHG test statistics based on simulated ``null" data sets as described above for every other dependence measures.

We note that the power comparison study here is not the best way to assess the dependence measure. The dependence measure should reflect the strength of deterministic relationship in data, which is different from the power of independence test. This can be clarified by the usage of $R^2$ in linear regression settings. The $R^2$ (the square of the linear correlation) measures the strength of linear deterministic relationship in data. Given a fixed sample size $n$, $R^2$ does have a one-to-one mapping to the p-value of its corresponding F-test. However, the F-test p-value measures the ``statistical significance" of the linear relationship, and generally becomes smaller for larger sample size (since we will be able to detect very weak linear relationship given large enough sample size). The $R^2$ reflects the ``practical significance" of the linear relationship. It measures the signal-to-noise ratio in data, and do not keep on increasing with the sample size.

An equitable dependence measure should be an extension of the linear correlation to measure the signal-to-noise ratio in data, regardless if the signal is linear or not. Hence it is more important to assess how well the measure reflect the ``practical significance" of the signal in data. The power is about how well the test captures the ``statistical significance", not the main aim of the dependence measure. It is not surprising that HHG test, tailored for testing independence purpose, has best power in most cases. However, such independence tests do not lead to dependence measures directly. If we try to use the p-values of such tests to rank the strength of deterministic relationships, then they will prefer large sample sizes rather than strong signals in data.

The correct way to judge the equitability of a dependence measure is to check how well it ranks the data according to the strength of signal in data as done in subsection~\ref{sec:equana}. We can see that measures such as MI also prefers large sample size rather than ranking purely based on signal strengths. Ccor is shown to be most equitable there. Therefore, Ccor would be more useful than other dependence measures in selecting variables related to $Y$ among $X_1$, ..., $X_K$, particularly when $X_1$, ..., $X_K$ do not all have the same sample size. The unequal sample sizes occur in practice if some of $X_k$'s are hard or costly to measure. We would not want to choose a weaker related $X_k$ simply because it has more measurements than others.

Another practical issue for applications is the computation time for the dependence measure. We checked the computational times of the independence test statistics on a system with dual Intel E5 2650 CPU's at 2GHz and 128GB  RAM. We simulated data with different sample sizes $n$ and the results are given in Table~\ref{t1}.

\begin{table} [htbp]
\begin{center}
\small{
\begin{tabular}{|c|c|c|c|}
  \hline
    Measures &	n=100&	n=1000&	n=10000\\\hline
	cor&	0.001	&0.001	&0.001 \\
	Ccor&	0.020 &	0.049	&0.437 \\
	MIC	&  0.314	&1.742 & 80.41	\\
	dcor&	0.006	&0.457	& 30.52 \\
	MI03&	0.001 &	0.001 &	0.023\\
	MI20&	0.001 &	0.002 &	0.055\\
	RDC	& 0.005	& 0.012  &	0.262 \\
	HHG	& 0.539	& 27.87 & 3786.9\\
  \hline
\end{tabular}
}
\end{center}
\caption{The computation times of all test statistics (in seconds).}
\label{t1}
\end{table}

As we could see from Table \ref{t1}, MIC, dcor and HHG become very computational intensive for large sample size. That would restrict their usefulness in mining large data sets. Ccor took significantly less time than those three. Ccor does take more time than the other dependence measures, but its computational time is acceptable.

In summary, the computational time and power for Ccor are good but not best among all dependence measures. It is clearly the most equitable measure, providing best ranking of data sets based on the strengths of deterministic relationships. Overall, Ccor performs very well as a dependence measure in these simulation studies.

\subsection{Analysis Of WHO Data}\label{sec:WHO}
We now apply the new measure Ccor to the WHO data set. We repeat the analysis in~\citet{Reshef2011MIC} by calculating the pairwise correlations among the $357$ variables in the data set. The first variable contains the ID numbers of the countries: from $1$ to $202$. These numerical values have no real intrinsic meaning. Hence the correlations between the first variable with other variables are rather senseless. We drop the first variable and only calculate the pairwise correlations among the rest $356$ variables. There are many missing data in the data sets. For some pairs of variables the available sample size is very small. Since our estimator for Ccor uses the copula density estimation, its accuracy under a very small sample size is suspectable. Therefore we calculate the measure Ccor only on those pairs with at least $n=50$ common observations. This results in $49286$ pairwise correlations in total.

\begin{center}
\begin{figure}[htbp]
\includegraphics[scale=0.7]{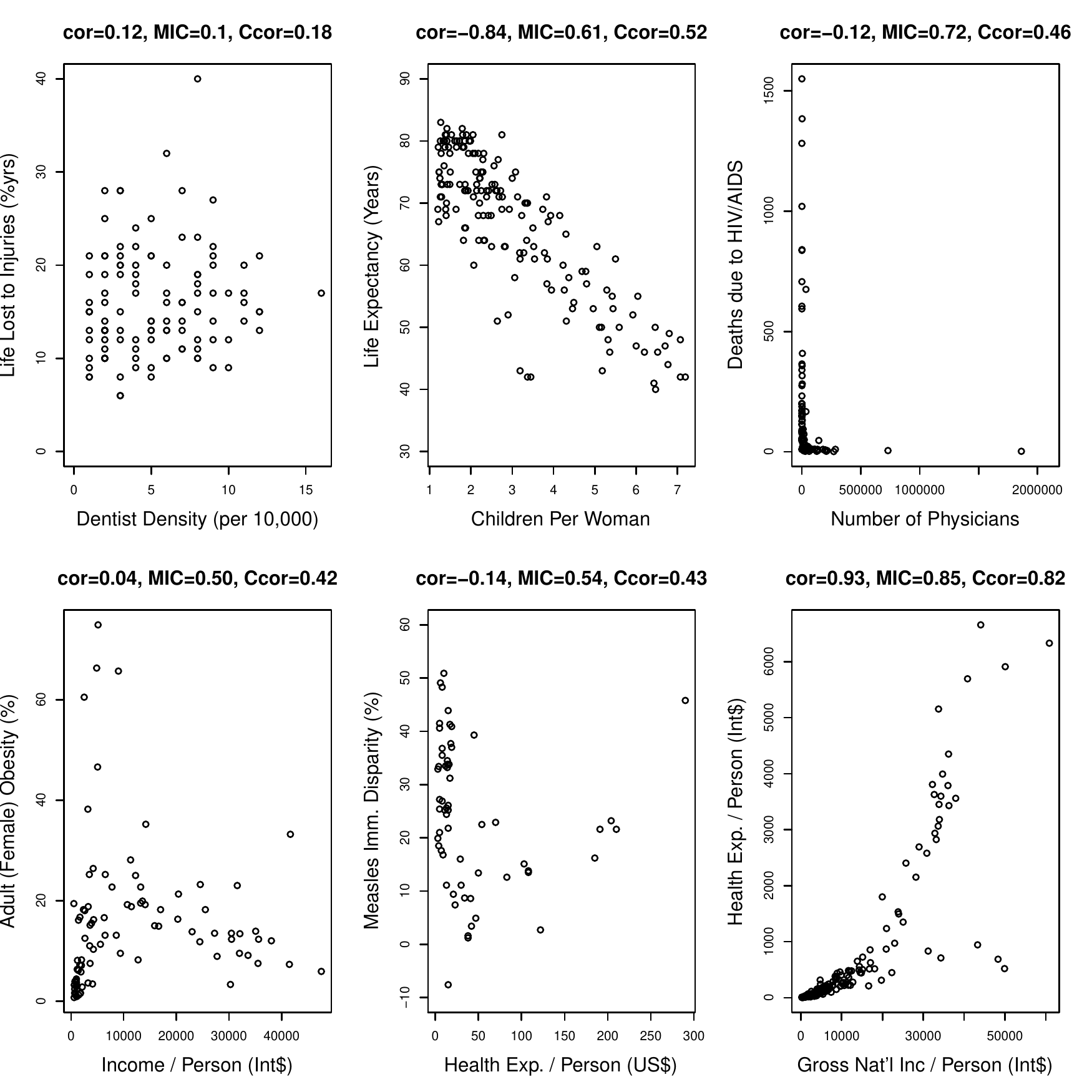}
\caption{The raw data and estimated correlation measures for several example cases in~\citet{Reshef2011MIC}.}
\label{PlotScience}
\end{figure}
\end{center}

We first look at some pairs of variables studied by~\citet{Reshef2011MIC}. Figure~\ref{PlotScience} plots the data along with linear correlation (cor), MIC and Ccor values for the examples $4$C-$4$H in~\citet{Reshef2011MIC}.
We can see that Ccor and MIC qualitatively give the same conclusion in those examples. They both give low correlations to the first case. They both detect some clear nonrandom relationships with weak linear correlations ($cor$). They give lower correlation values than $cor$ in the two cases with high linear correlations, but big enough to detect the relationship. There are some differences in the numerical values between Ccor and MIC. The biggest difference occurs for the third case in the first row, with $MIC=0.72$ and $Ccor=0.46$.

\begin{center}
\begin{figure}[htbp]
\includegraphics[scale=0.7]{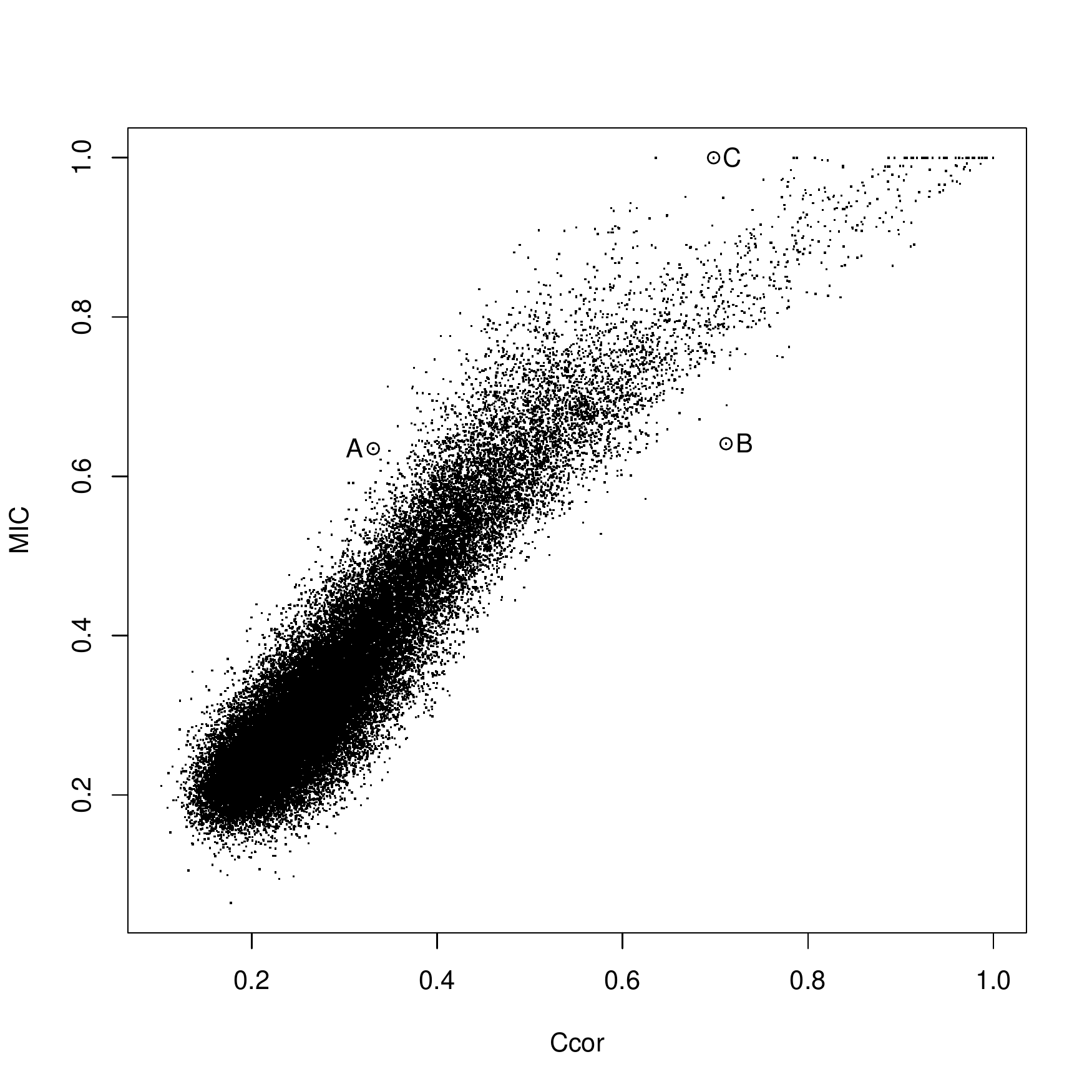}
\caption{The $Ccor$ and $MIC$ values for all pairs in the WHO data. Three cases labeled on the graph is shown in detail in the Figure~\ref{CcorMICcompare}}
\label{CcorMICcurve}
\end{figure}
\end{center}

To compare the estimates for Ccor and MIC, we plotted their values for all $49286$ pairs on the WHO data sets in Figure~\ref{CcorMICcurve}. We can see that the values fall in a band around the diagonal. This means that Ccor and MIC generally rate the pair-wise dependence similarly.

\begin{center}
\begin{figure}[htbp]
\includegraphics[scale=0.7]{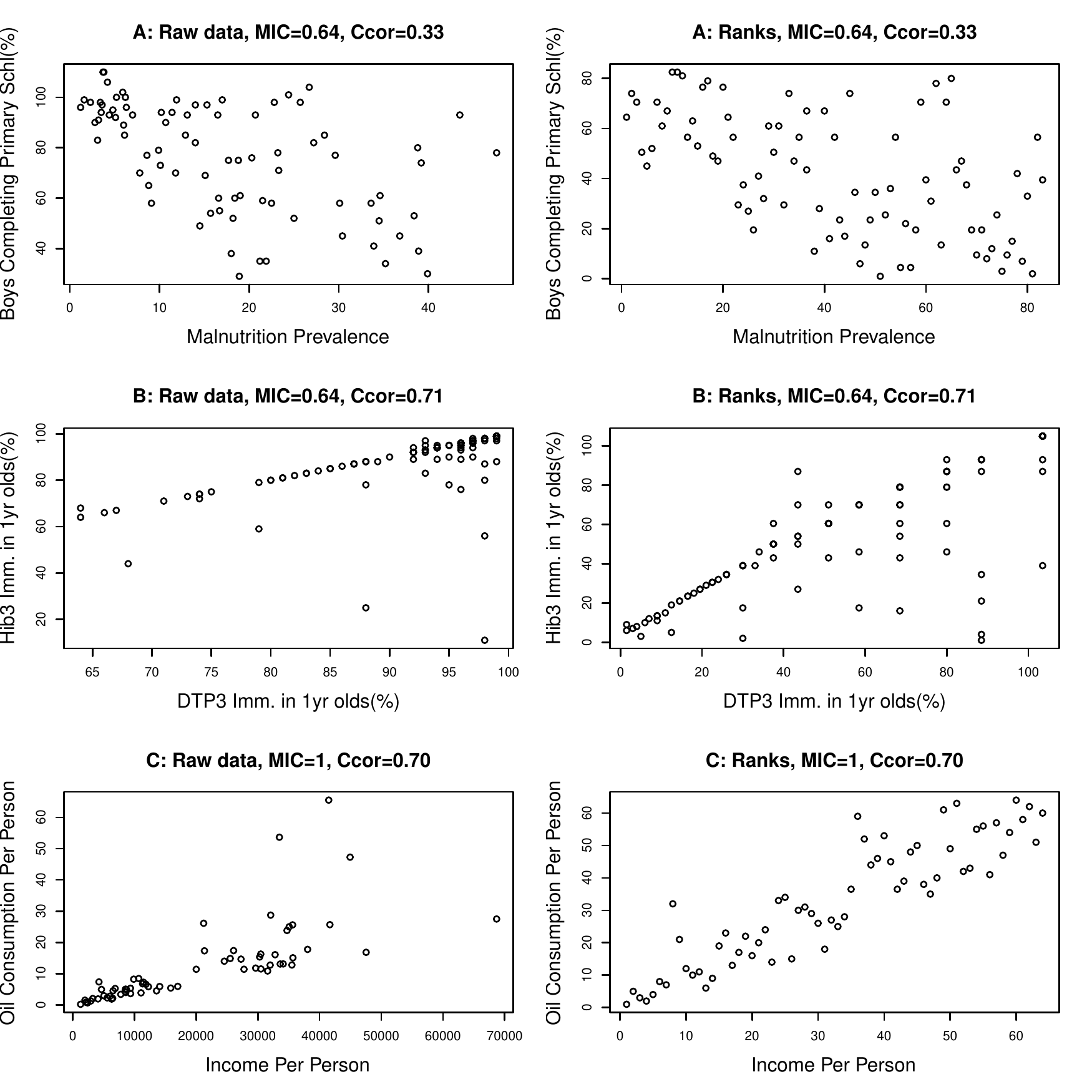}
\caption{The comparison of $Ccor$ and $MIC$ on three example cases.}
\label{CcorMICcompare}
\end{figure}
\end{center}

\begin{center}
\begin{figure}[htbp]
	\centering
\begin{center}
        \begin{subfigure}[b]{0.225\textwidth}
               \includegraphics[width=\textwidth]{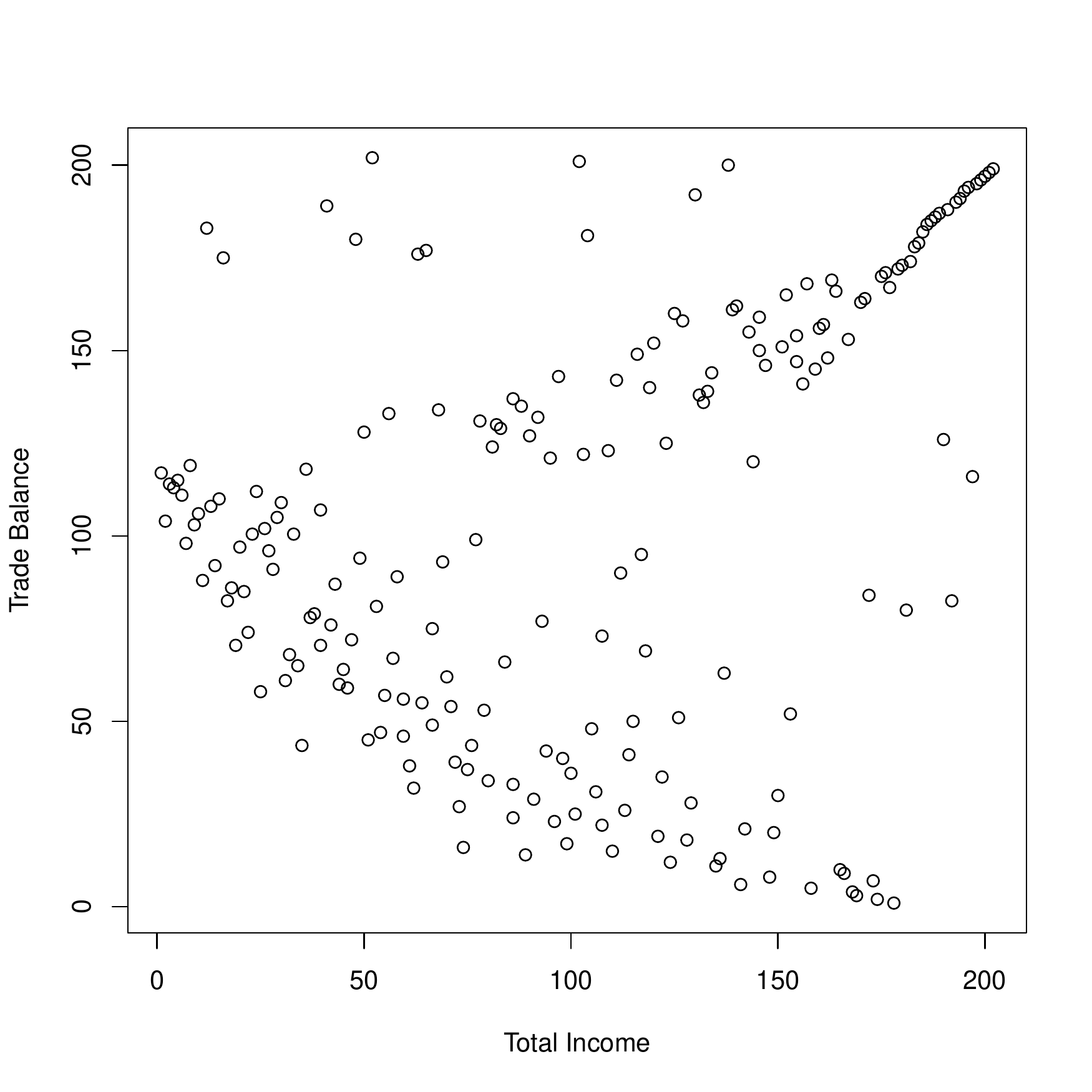}
                \caption{Trade Balance vs. Total Income of Residents}
                \label{fig:top1}
        \end{subfigure}
 	\quad
         \begin{subfigure}[b]{0.225\textwidth}
               \includegraphics[width=\textwidth]{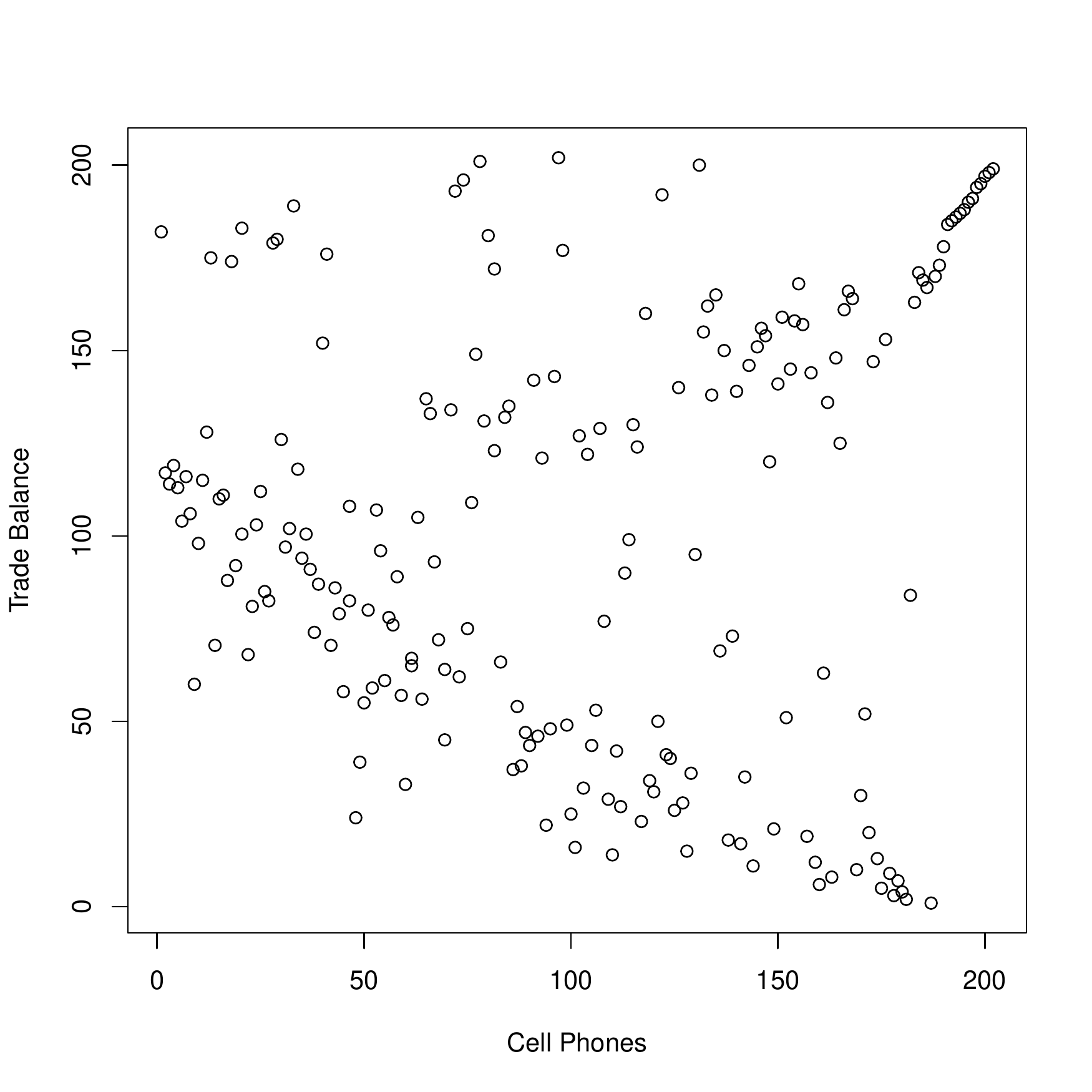}
                \caption{Trade Balance vs. Cell Phones Owned}
                \label{fig:top2}
        \end{subfigure}
	\quad
         \begin{subfigure}[b]{0.225\textwidth}
               \includegraphics[width=\textwidth]{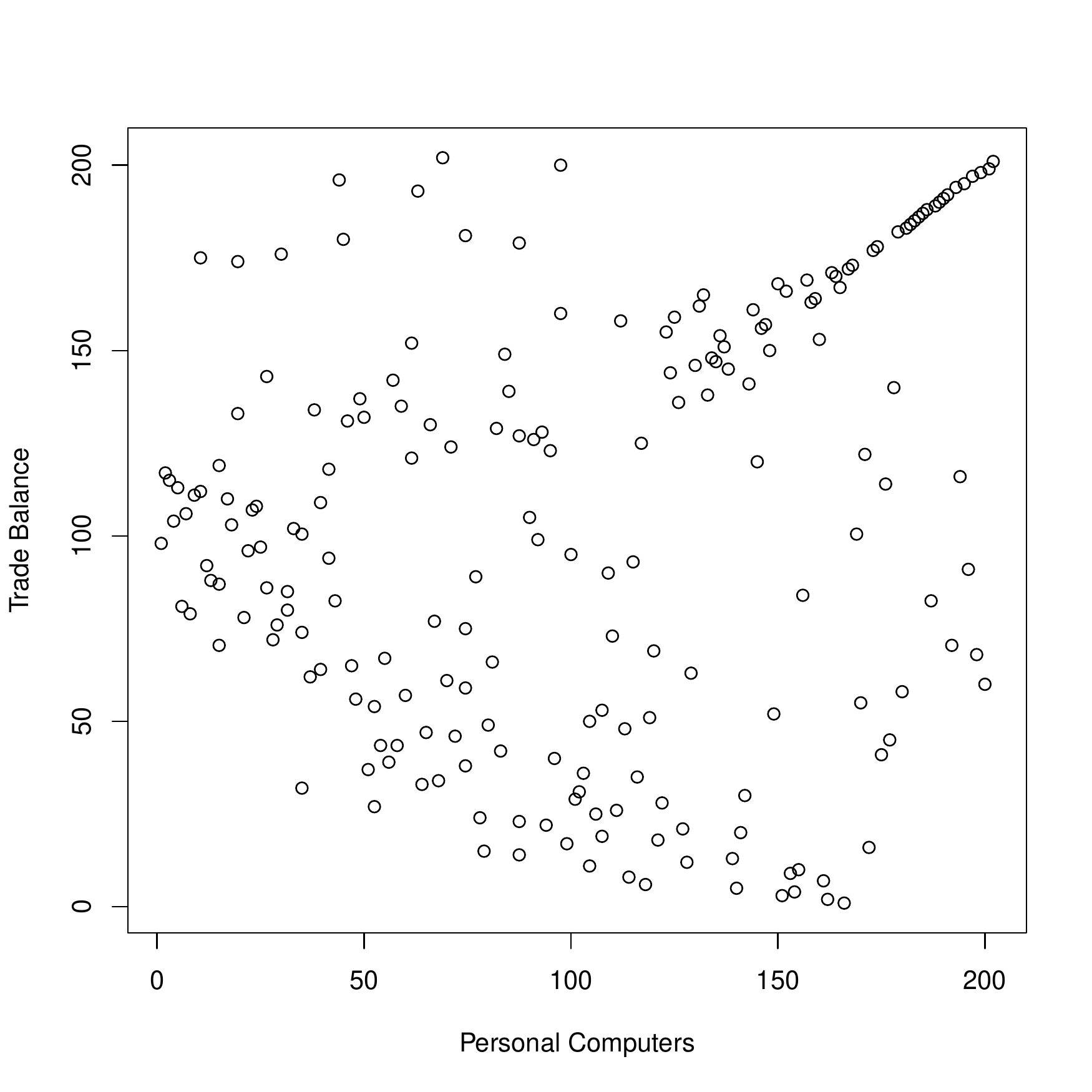}
                \caption{Trade Balance vs. Personal Computers Owned}
                \label{fig:top3}
        \end{subfigure}	
        \quad
         \begin{subfigure}[b]{0.225\textwidth}
               \includegraphics[width=\textwidth]{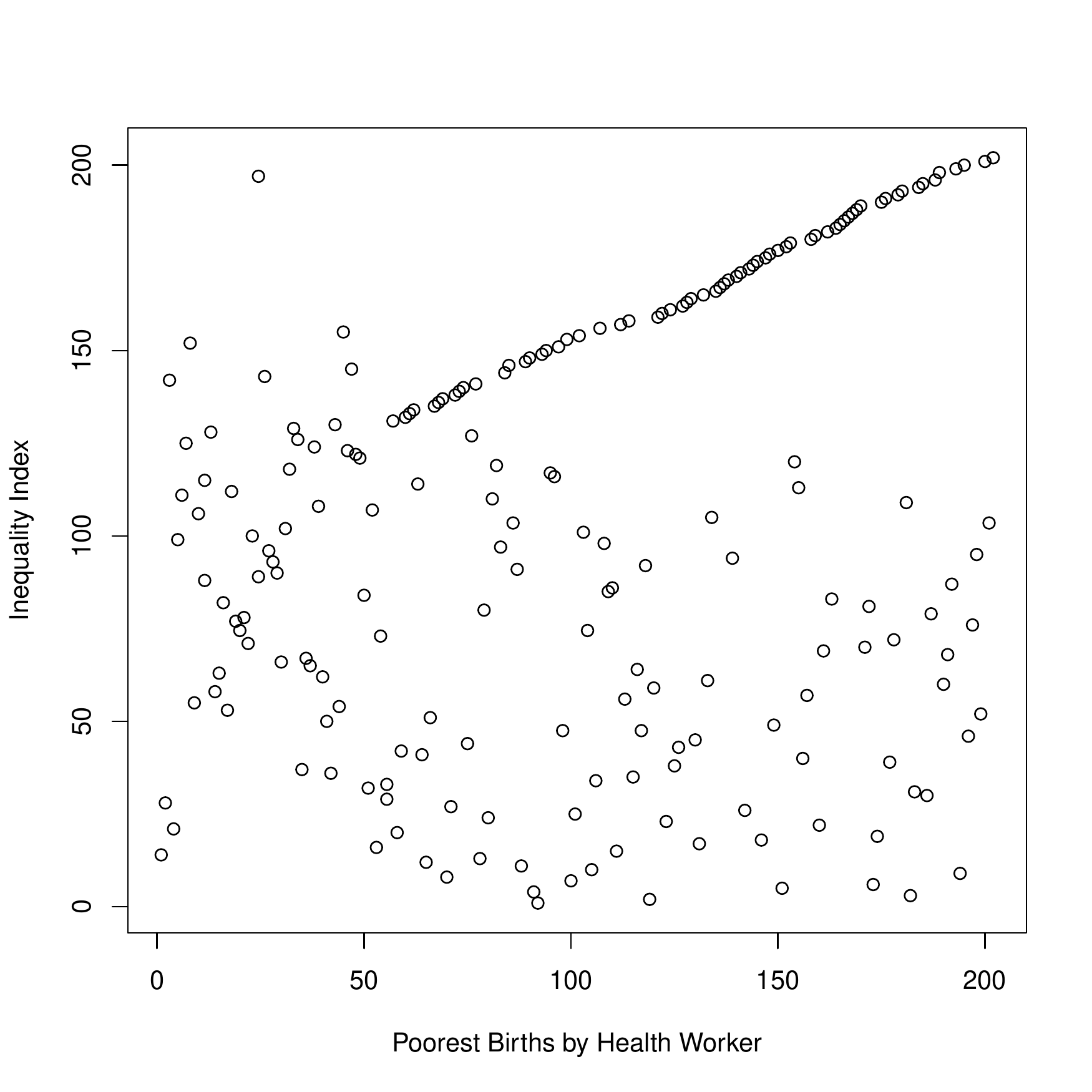}
                \caption{Inequality vs. Infant Med. Service for Poorest}
                \label{fig:top16}
        \end{subfigure}
        \end{center}
 	\caption{The top nonlinear relationships selected by $Ccor$ on WHO data set: (a)-(c) the top 3, (d) the top $16$th.}
 \label{Rank.Ccor.WHO}
\end{figure}
\end{center}

To investigate the different rankings by these two measures, we investigate three pairs of variables that have very similar values in one measure but big difference in the other measure. These three pairs are labeled as A, B and C on the graph of Figure~\ref{CcorMICcurve}. We plot the data for these variables in the Figure~\ref{CcorMICcompare}. Since Ccor and MIC are both rank-based, we also plot these data in the ranks to avoid any specious pattern due to the scales on the variables.
As we can see from Figure~\ref{CcorMICcompare}, the later two cases (B and C) both seem to have strong linear relationships with some noise. While the noise patterns are different in Figure~\ref{CcorMICcompare}B and ~\ref{CcorMICcompare}C, the average noise amount looks about the same. The first case Figure~\ref{CcorMICcompare}A clearly is much noisier than the later two cases. This pattern is correctly reflected by Ccor which assigns similar correlation to the latter two cases while giving the first case a much lower correlation value. However, MIC assigns about the same correlation value to the first two cases and a much higher correlation value to the third case. This certainly does not agree with the observed data patterns. Particularly, MIC assigns a correlation value of $1$ to the case \ref{CcorMICcompare}C which is far from a noiseless deterministic relationship. From these observations, Ccor better reflects the noise level than MIC. Thus Ccor is a better equitable correlation measure.
As suggested by \citet{Reshef2011MIC}, we can search for nonlinear relationships in data by checking the cases where the dependence measure $Ccor$ far exceeds the linear correlation $|\rho|$. Figure~\ref{Rank.Ccor.WHO}(a)-(c) show the top three relationships ranked by $Ccor-|\rho|$. All three (and the next nine top ranked ones not plotted here) are the ``$<$" shaped relationship between variable ``{\it Trade Balance}" against several other variables. These two-branches ``$<$" type relationships are also ranked in the top by $MIC-|\rho|$. For example, the relationship between ``{\it Trade Balance}" and ``{\it Total Income}" is ranked as the top one by  $Ccor-|\rho|$ and as the top second by $MIC-|\rho|$.
$Ccor-|\rho|$ also lead us to discovery of some relationships not found by other measures. The variables ``{\it Inequality Index}" and ``{\it Births Attended by Skilled Health Workers in the Poorest Quintile Residents}" has clearly a proportion of deterministic increasing relationship mixed with more noisy data (Figure~\ref{fig:top16}). This relationship is ranked $16$th by $Ccor-|\rho|$, but ranked very low by other dependence measures. It was ranked in the 268th, 531th and 253th respectively by $MIC-|\rho|$, $dcor-|\rho|$ and $MIcor-|\rho|$.

\section{Proofs}\label{sec:proofs}

\subsection{Proof of Theorem~\ref{thm:MI_rate}.}\label{sec:proof.mi}

To prove the theorem, we use \citet{lecam1973convergence}'s method to find the lower bound on the minimax risk of the estimating mutual information $MI$. To do this, we will use a more convenient form of Le Cam's method developed by \citet{Donoho1991GeoRateII}. Define the module of continuity of a functional $T$ over the class $\mathbf{F}$ with respect to Hellinger distance as in equation (1.1) of \citet{Donoho1991GeoRateII}:
\beq\label{mod-cont}
w(\e) = sup\{|T(F_1) - T(F_2)| : H(F_1,F_2) \le \e, F_i \in \mathbf{F} \}.
\eeq Here $H(F_1,F_2)$ denotes the Hellinger distance between $F_1$ and $F_2$. Then the minimax rate of convergence for estimating $T(F)$ over  the class $\mathbf{F}$ is bounded below by $w(n^{-1/2})$.

We now look for a pair of density functions $c_1(u,v)$ and $c_2(u,v)$ on the unit square for distributions that are close in Hellinger distance but far away in their mutual information. This provides a lower bound on the module of continuity for mutual information MI over the class $\mathfrak{C}$, and hence leads to a lower bound on the minimax risk. We outline the proof here.

We first divide the unit square into three disjoint regions $R_1$, $R_2$ and $R_3$ with $R_1 \cup R_2 \cup R_3 = [0,1] \times [0,1]$. The first density function $c_1(u,v)$ puts probability masses $\delta$, $a$ and $1-a-\delta$ respectively on the regions $R_1$, $R_2$ and $R_3$ each uniformly. The $a$ is an arbitrary small fixed value, for example, $a=0.01$. For now, we take $\delta$ to be another small fixed value.  The area of the region is chosen so that $c_1(u,v)=M$ on region $R_2$ and $c_1(u,v)=M^*$ on region $R_1$ for a very big $M^*$. The second density function $c_2(u,v)$, compared to $c_1(u,v)$, moves a small probability mass $\e$ from $R_1$ to $R_2$. We will see that the Hellinger distance between $c_1$ and $c_2$ is of the same order as $\e$, but the change in MI is unbounded for big $M^*$. Hence module of continuity $w(\e)$ is unbounded for mutual information MI. Therefore the MI can not be consistently estimated over the class $\mathfrak{C}$.

\begin{center}
\begin{figure}
\includegraphics[scale=0.6]{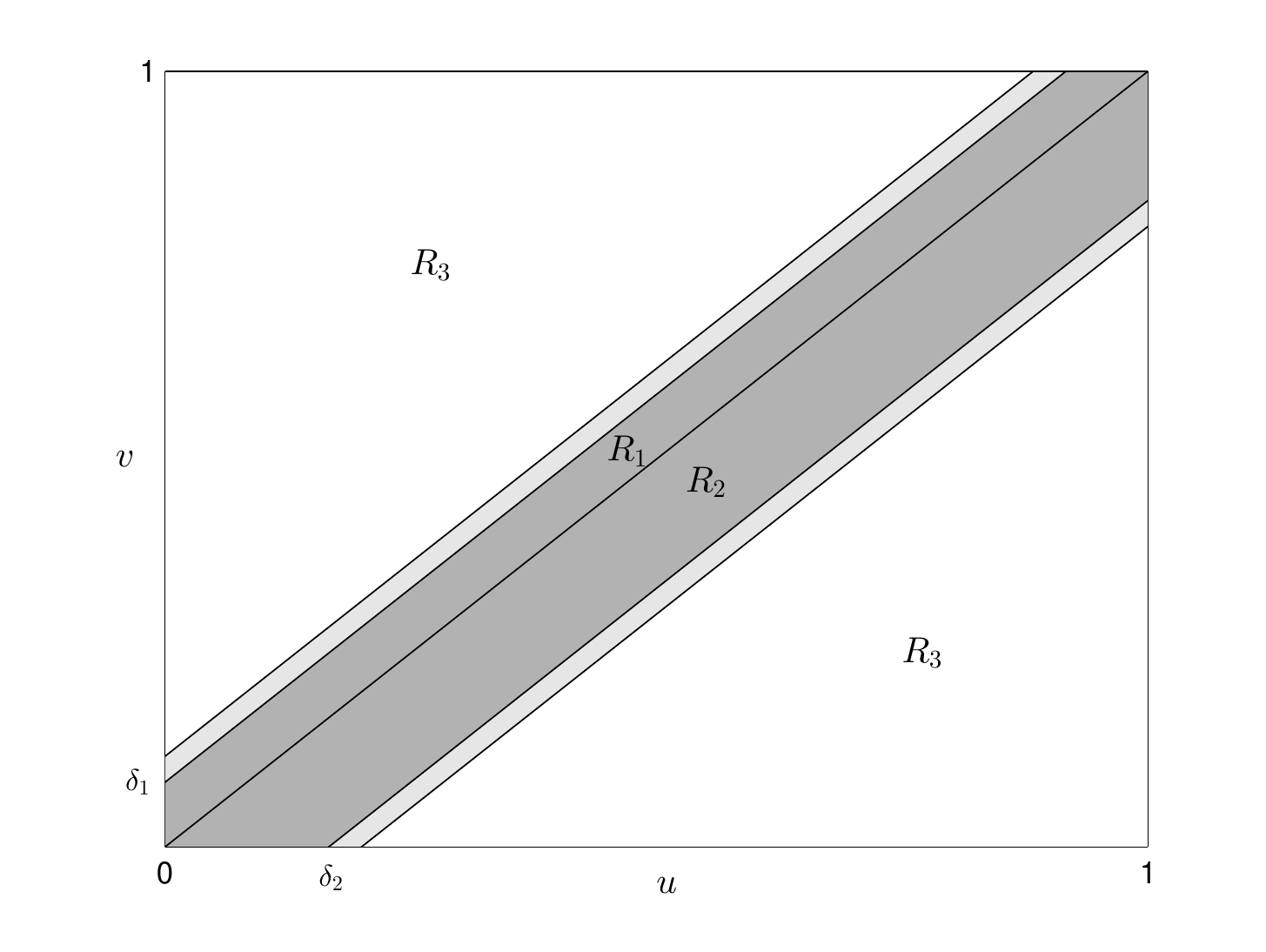}
\caption{The plot shows the regions $R_1$, $R_2$ and $R_3$. The other two narrow strips neighboring $R_1$ and $R_2$ are for the continuity correction mentioned at the end of the proof.}
\label{fig:proof}
\end{figure}
\end{center}

Specifically, the region $R_1$ is chosen to be a narrow strip immediately above the diagonal, $R_1=\{(u,v): - \delta_1 < u-v <0 \}$; and $R_2$ is chosen to be a narrow strip immediately below the diagonal, $R_2=\{(u,v): 0 \le u-v <\delta_2 \}$. The remaining region is $R_3 = [0,1] \times [0,1] \setminus (R_1 \cup R_2)$. The values of $\delta_1$ and $\delta_2$ are chosen so that the areas of regions $R_1$ and $R_2$ are $\delta/M^*$ and $a/M$ respectively. Then clearly $c_1(u,v)=M^*$ on $R_1$; $c_1(u,v)=M$ on $R_2$; $c_1(u,v)=(1-a-\delta)/(1 - a/M -\delta/M^*)$ on $R_3$. And $c_2(u,v)=M^*-\e(M^*/\delta)$ on $R_1$; $c_2(u,v)=M+\e(M/a)$ on $R_2$; $c_2(u,v)=c_1(u,v)$ on $R_3$. See the Figure~\ref{fig:proof}.
Then we have
$$
\ba{cl}
2 H^2(c_1,c_2) & = \iint (\sqrt{c_2(u,v)}-\sqrt{c_1(u,v)})^2 dudv \\
& = (\sqrt{M^*-\e(M^*/\delta)}-\sqrt{M^*})^2 \delta/M^* + (\sqrt{M+\e(M/a)}-\sqrt{M})^2 a/M \\
& =  \delta (\sqrt{1-\e /\delta}-1)^2 + a (\sqrt{1+\e /a}-1)^2 \\
& = \delta ({\e}/{2\delta})^2 + a ({\e}/{2a})^2  + o(\e^2) \\
& =  \e^2 (\frac{1}{4\delta} + \frac{1}{4a}) + o(\e^2).
\ea
$$
Hence the Hellinger distance is of the same order as $\e$:
$$
H(c_1,c_2) = \e \sqrt{\frac{1}{8\delta} + \frac{1}{8a}} + o(\e).
$$
On the other hand, the difference in the mutual information is
\beq \label{MI-diff}
\ba{cl}
& MI(c_1)-MI(c_2) \\
= & \delta \log(M^*) + a \log(M) - (\delta - \e) \log[M^*-\e(M^*/\delta)] -  (a + \e) \log[M + \e(M/a)]  \\
= & \e \log(M^*) - \e \log(M) - (\delta - \e) \log(1 -\e/\delta) - (a + \e) \log(1  + \e/a).
\ea
\eeq Here $M$, $\delta$ and $a$ are fixed constants. Hence when $M^* \to \infty$, this difference in $MI$ also goes to $\infty$. For example, if we let $M^* = e^{1/(\e)^2}$, then the module of continuity $w(\e) \ge O(1/\e)$. That means, the rate of convergence is at least $O( w(n^{-1/2})) = O(n^{1/2}) \to \infty$. In other words, MI can not be consistently estimated.

The small difference in Hellinger distance of $c_1$ and $c_2$ can lead to unbounded difference in $MI(c_1)$ and $MI(c_2)$ since $MI$ is unbounded. After the transformation $MIcor=\sqrt{1-e^{-2MI}}$ in (\ref{MIcor}), the mutual information correlation is bounded. The difference between $MIcor(c_1)$ and $MIcor(c_2)$ in the above example is actually small since the $MI$ are big for both $c_1$ and $c_2$ (leading to corresponding $MIcor$s close to zero). However, $MIcor$ is also very hard to estimate over the class $\mathfrak{C}$. To see this, we follow the same reasoning above but modify the example of $c_1$ and $c_2$. First, we notice that for any pair of densities $c_1$ and $c_2$,
$$
\ba{cl}
|MIcor(c_1)-MIcor(c_2)|
& =  |\sqrt{1 - e^{-2 MI(c_1)}} - \sqrt{1 - e^{-2 MI(c_2)}}|  \\
& = |\frac{[1 - e^{-2 MI(c_1)}] - [1 - e^{-2 MI(c_2)}]}{\sqrt{1 - e^{-2 MI(c_1)}} + \sqrt{1 - e^{-2 MI(c_2)}}}|\\
& \ge \frac{1}{2}|e^{-2 MI(c_1)} - e^{-2 MI(c_2)}| \\
& =  \frac{1}{2} e^{-2 MI(c_1)} |1 - e^{-2[MI(c_1) - MI(c_2)]}|.
\ea
$$
For the difference $MIcor(c_1)-MIcor(c_2)$  to be the same order of the difference $MI(c_1) - MI(c_2)$, we need to set $MI(c_1)$ at constant order when $\e \to 0$.

Therefore, we modify the above $c_1$ to have probability mass $\delta=2 \e$ in region $R_1$, varying with the $\e$ value instead of fixed as before. And we set $M^*=e^{1/\e}$, leading to
$$
\ba{cl}
& MI(c_1) \\
= & \delta \log(M^*) + a \log(M) + (1 -a - \delta) \log[(1- a - \delta)/(1 - a/M - \delta/M^*)] \\
= & 2 + a \log(M) + (1 -a - 2\e) \log[(1- a - 2\e)/(1 - a/M - 2\e e^{-1/\e})],
\ea
$$
which converges to a fixed constant $a_1 = 2 + a \log(M) + (1-a) \log[(1- a )/(1 - a/M)]$ as $\e \to 0$. Using (\ref{MI-diff}), recall that $\delta=2 \e$ and $M^*=e^{1/\e}$, we have
$$
\ba{cl}
& MI(c_1)-MI(c_2) \\
= & \e \log(M^*) - \e \log(M) - (\delta - \e) \log(1 -\e/\delta) - (a + \e) \log(1  + \e/a) \\
= & 1 - \e \log(M) - \e \log(1/2) - (a + \e) \log(1  + \e/a),
\ea
$$
which converges to $1$ as $\e \to 0$. Hence we have
$$
\lim_{\e \to 0} w(\e) \ge \lim_{\e \to 0} \frac{1}{2} e^{-2 MI(c_1)} |1 - e^{-2[MI(c_1) - MI(c_2)]}| = \frac{1}{2} e^{-2 a_1}(1-e^{-2(1)}),
$$
a positive constant $a_2=e^{-2 a_1}(1-e^{-2})/2 $. Therefore, $MIcor$ can not be estimated consistently over the class $\mathfrak{C}$ either.

The above outlines the main idea of the proof, ignoring some mathematical subtleties. One is that the example densities $c_1$ and $c_2$ are only piecewise continuous on the three regions, but not truly continuous as required for the class $\mathfrak{C}$. This can be easily remedied by connecting the three pieces linearly. Specifically we set the densities $c_i(u,v)=M$, $i=1,2$, on the boundary between $R_1$ and $R_3$, $\{(u,v): u- v = - \delta_1\}$, and on the boundary between $R_2$ and $R_3$, $\{(u,v): u- v = \delta_2\}$. Then we use two narrow strips within $R_3$, $\{(u,v): - \delta_3 \le u-v \le - \delta_1 \}$ and $\{(u,v): \delta_2 \le u-v \le \delta_4 \}$ to connect the constant $c_i(u,v)$ values on the rest of region $R_3$ with the boundary value $c_i(u,v)=M$ continuously through linear (in $u-v$) $c_i(u,v)$'s on the two strips that satisfies the H\"{o}lder condition (\ref{eq:holder}). By the H\"{o}lder condition (\ref{eq:holder}), the connection can be made with strips of width at most $(M-1+a+\delta)/M_1$. This continuity modification does not affect the calculation of the difference $MI(c_1)-MI(c_2)$ above as $c_1$ and $c_2$ only differ on regions $R_1$ and $R_2$. Within regions $R_1$ and $R_2$, the densities $c_1$ and $c_2$ can be further similarly connected continuously linearly in $u-v$. As there is no H\"{o}lder condition on $A_M^c$, the connection within $R_1$ and $R_2$ can be as steep as we want. Clearly the order obtained through above calculations will not change if we make these connections very steep so that their effect is negligible.

Another technical subtlety is that the $c_1$ and $c_2$ defined above are only densities on the unit square but not copula densities which require uniform marginal distributions. However, it is clear that the marginal densities for $c_i$s are uniform over the interval $(\delta_3,1-\delta_4)$ and linear in the rest of interval near the two end points $0$ and $1$. The copulas densities $c_i^*$'s corresponding to $c_i$'s can be calculated directly through Sklar's decomposition (\ref{sklar}). It is easy to see that the order for the module of continuity $w(\e)$ remains the same for using the corresponding copula densities $c_i^*$'s.

\subsection{Proof of Theorem~\ref{thm:Ccor}.}\label{sec:proof.ccor}

Let $M_2$ be a constant between $1$ and $M$, say, $M_2=(M+1)/2$. Denote $A_{M_2} = \{(u,v): c(u,v) \le M_2\}$. Then we denote $T_1(c)= \iint\limits_{A_{M_2}} [1-c(u,v)]_+ du dv$ and $T_2(c) = \iint\limits_{A_{M_2}^c} [1-c(u,v)]_+ du dv$ so that $Ccor = T_1(c) + T_2(c)$.

For a density estimator $\hat c_n(u,v)$, we have the corresponding copula correlation estimator by plugging $\hat c_n(u,v)$ into the Ccor expression. Hence $\widehat{Ccor} = T_1(\hat c_n) + T_2(\hat c_n)$. We now bound the errors in estimating $T_1$ and $T_2$ separately.

$T_1$ involves the integral over the $(u,v)$ points in $A_{M_2}$ only. Those points are contained in the set of low density points where the H\"{o}lder condition holds. Hence we can apply the usual bounds for kernel density estimation. Particularly, let $\bar c_n(u,v) = E[\hat c_n(u,v)] = \iint K(s)K(t) c(u+hs,v+ht)ds dt$ denote the expectation of the density estimator $\hat c_n$. Then the bias in density estimation is bounded by
$$
|\bar c_n(u,v) - c(u,v)| \le \int\limits_{-1}^1 \int\limits_{-1}^1 K(s)K(t) |c(u+hs,v+ht) - c(u,v)| ds dt.
$$
For $(u,v) \in A_{M_2}$, $c(u+hs,v+ht) \in A_M$ for $h \le (M-M_2)/(\sqrt{2}M_1)$, $|s| \le 1$ and $|t| \le 1$. Since the support of $K(\cdot)$ is $[-1,1]$, for small enough $h$, the bias is bounded using the H\"{o}lder condition by
\beq\label{bias.bound}
\ba{cl}
\int\limits_{-1}^1 \int\limits_{-1}^1 K(s)K(t) M_1 h (|s| + |t|) ds dt & \le 2 M_1 h \int\limits_{-1}^1 \int\limits_{-1}^1 K(s)K(t) ds dt \\
&= 2 M_1 h.
\ea
\eeq The variance of $\hat c_n$ is given by
$$
\ba{cl}
Var[\hat c_n(u,v)] & = \frac{1}{n} Var[\frac{1}{h^2}K(\frac{u - U_1}{h})K(\frac{v - V_1}{h})] \\
& \le \frac{1}{nh^2} \int\limits_{-1}^1 \int\limits_{-1}^1 K^2(s)K^2(t) c(u+hs,v+ht) ds dt.
\ea
$$
Hence by the same arguments above, for small enough $h$, the variance is bounded by
\beq\label{var.bound}
\ba{cl}
& \frac{1}{nh^2}\int\limits_{-1}^1 \int\limits_{-1}^1  K^2(s)K^2(t) [c(u,v)+ M_1 h (|s| + |t|)] ds dt \\
\le & \frac{1}{nh^2} \mu_2^2 [c(u,v) + 2 M_1 h],
\ea
\eeq where $\mu_2 = \int_{-1}^1 K^2(t) dt$. Combining (\ref{bias.bound}) and (\ref{var.bound}), we get
\beq \label{den.bound1}
E\{[\hat c_n(u,v) - c(u,v)]^2\} \le 4 M_1 h^2 + \frac{1}{nh^2} \mu_2^2 [c(u,v) + 2 M_1 h].
\eeq The integration of the right hand side over the region $A_{M_2}$ is bounded by its integration over the whole unit square: $(u,v) \in [0,1] \times [0,1]$. For $h$ small enough, since $2 M_1 h \le 1$,  we get
\beq \ba{cl}
& E\{\iint\limits_{A_{M_2}}[\hat c_n(u,v) - c(u,v)]^2dudv\} \\
\le & \int\limits_{0}^1 \int\limits_{0}^1 \{ 4 M_1 h^2 + \frac{1}{nh^2} \mu_2^2 [c(u,v) + 1]\}dudv \ \ = 4 M_1 h^2 + \frac{2\mu_2^2}{nh^2}.
\ea
\eeq Hence
$$
\ba{rl}
\{E\iint\limits_{A_{M_2}}|\hat c_n(u,v) - c(u,v)|dudv\}^2 \le & E\{\iint\limits_{A_{M_2}}[\hat c_n(u,v) - c(u,v)]^2dudv\} \\
\le &  4M_1 h^2 + \frac{2\mu_2^2}{nh^2} \le (2\sqrt{M_1} h + \frac{2\mu_2}{\sqrt{n}h})^2.
\ea
$$
That is,
\beq \label{T1error}
|T_1(\hat c_n) - T_1(c)| \le E\iint\limits_{A_{M_2}}|\hat c_n(u,v) - c(u,v)|dudv \le 2\sqrt{M_1} h + \frac{2\mu_2}{\sqrt{n}h}.
\eeq
Now we look at the error bound on $A_{M_2}^c$. Since the H\"{o}lder condition does not hold here, we can not control the error in $\hat c_n$ on $A_{M_2}^c$. Notice that
$$
Var[\hat c_n(u,v)] = \frac{1}{n} Var[\frac{1}{h^2}K(\frac{u - U_1}{h})K(\frac{v - V_1}{h})]
$$
may be unbounded since $c(u,v)$ is unbounded on $A_{M_2}^c$. However,
$$
Var[\hat c_n(u,v)] \le \frac{1}{nh^2} \int\limits_{-1}^1 \int\limits_{-1}^1 K^2(s)K^2(t) c(u+hs,v+ht) ds dt \le \frac{1}{nh^2} M_K^2 E[\hat c_n(u,v)],
$$
where $M_K=\max\limits_{0\le t \le1} K(t)$.

Let $\mathds{1}\{\hat c_n(u,v) <1\}$ be the indicator variable for where $\hat c_n<1$. Then
$$
Pr[\hat c_n(u,v) <1] = E[\mathds{1}\{\hat c_n(u,v) <1\}] \le \frac{Var[\hat c_n(u,v)]}{[\bar c_n(u,v)-1]^2}
$$
by Chebyshev's inequality.

Let $M_3$ be a constant between $1$ and $M_2$, say $M_3 = (1+M_2)/2 >1$. Then for any point $(u,v) \in A_{M_2}^c$, when $h$ is small enough, the $h$-square centered at $(u,v)$ are contained in $A_{M_3}^c$. Hence  $\bar c_n(u,v) =  \iint K(s)K(t) c(u+hs,v+ht)ds dt \ge M_3$. Since the function $x/(x-1)^2$ is strictly decreasing on $[1,\infty)$, let $M_4=M_3/(M_3-1)^2$, then
$$
E[\mathds{1}\{\hat c_n(u,v) <1\}] \le \frac{Var[\hat c_n(u,v)]}{[\bar c_n(u,v)-1]^2} \le \frac{1}{nh^2} M_K^2 \frac{\bar c_n(u,v)}{[\bar c_n(u,v)-1]^2} \le \frac{1}{nh^2} M_K^2 M_4.
$$
Hence,
\beq \label{T2error}
\ba{cl}
& |T_2(\hat c_n) - T_2(c)| = |T_2(\hat c_n)| \\
 = & E|\iint\limits_{A_{M_2}^c} [1-\hat c_n(u,v)]_+ dudv|  \\  \le & \iint\limits_{A_{M_2}^c} E[\mathds{1}\{\hat c_n(u,v) <1\}] du dv \ \ \le \frac{1}{nh^2} M_K^2 M_4.
\ea
\eeq
Combining (\ref{T1error}) and (\ref{T2error}),
$$|\widehat{Ccor} - Ccor| \le 2\sqrt{M_1} h + \frac{2\mu_2}{\sqrt{n}h} + \frac{1}{nh^2} M_K^2 M_4.$$
This is (\ref{Ccor.err}) with $M_5=M_K^2 M_4$.

\section{DISCUSSIONS AND CONCLUSIONS}\label{sec:disc}
We have proposed a new equitability definition for dependence measures that reflect properly the strength of deterministic relationships in data. The copula correlation is proposed as the equitable extension of Pearson's linear correlation. Theoretically we proved that Ccor is robust-equitable and consistently estimable. Its good performance is demonstrated through simulation studies and a real data analysis. Based on these studies, Ccor will be a very useful new tool to explore complex relations in big data sets.

For simplicity of presentation, we focused on bivariate continuous distributions. The multivariate extensions of Ccor are provided in Section~\ref{sec:multivariate}. In higher-dimensions, we need to explore Ccor estimators other than the KDE-based estimator. One possible direction is to develop KNN-based estimator for Ccor, similar to what was done for MI. It may also be worthwhile to explore the connection to dependence measures based on  the reproducing kernel Hilbert space~\citep{Gretton05,Poczos12}. One such measure is in fact $CD_2$ in equation (\ref{CD})~\citep{fukumizu2007kernel}. Because of the good theoretical properties of Ccor proven in this paper, developing better estimators for it deserves more research attention.

\bibliographystyle{imsart-nameyear}
\bibliography{./citations}

\section{Supplemental Materials}\label{sec:supplemental}
\subsection{The Six Functional Relationships Used in the Numerical Equitability Analysis}\label{sec:functions}

The following tables lists the six function relationships used in the equitability analysis in section 5.1 of the main text. We provide the function expressions, the plots of the functions and  their corresponding singular copula $C_s$.

\begin{table} [htbp]
\begin{center}
\small{
\begin{tabular}{c|cccccc}
    \hline
Functions & Linear & Parabolic & Cosine &  2-branches & circle & cross \\
$f(x)$ & $y=x$ & $4(x-\frac{1}{2})^2$ & $cos(4\pi x)$ &  $\pm x$ & $\pm \sqrt{x-x^2}$ & $\pm (x-\frac{1}{2})$ \\
Copula $C_s$   & {\includegraphics[scale=0.08]{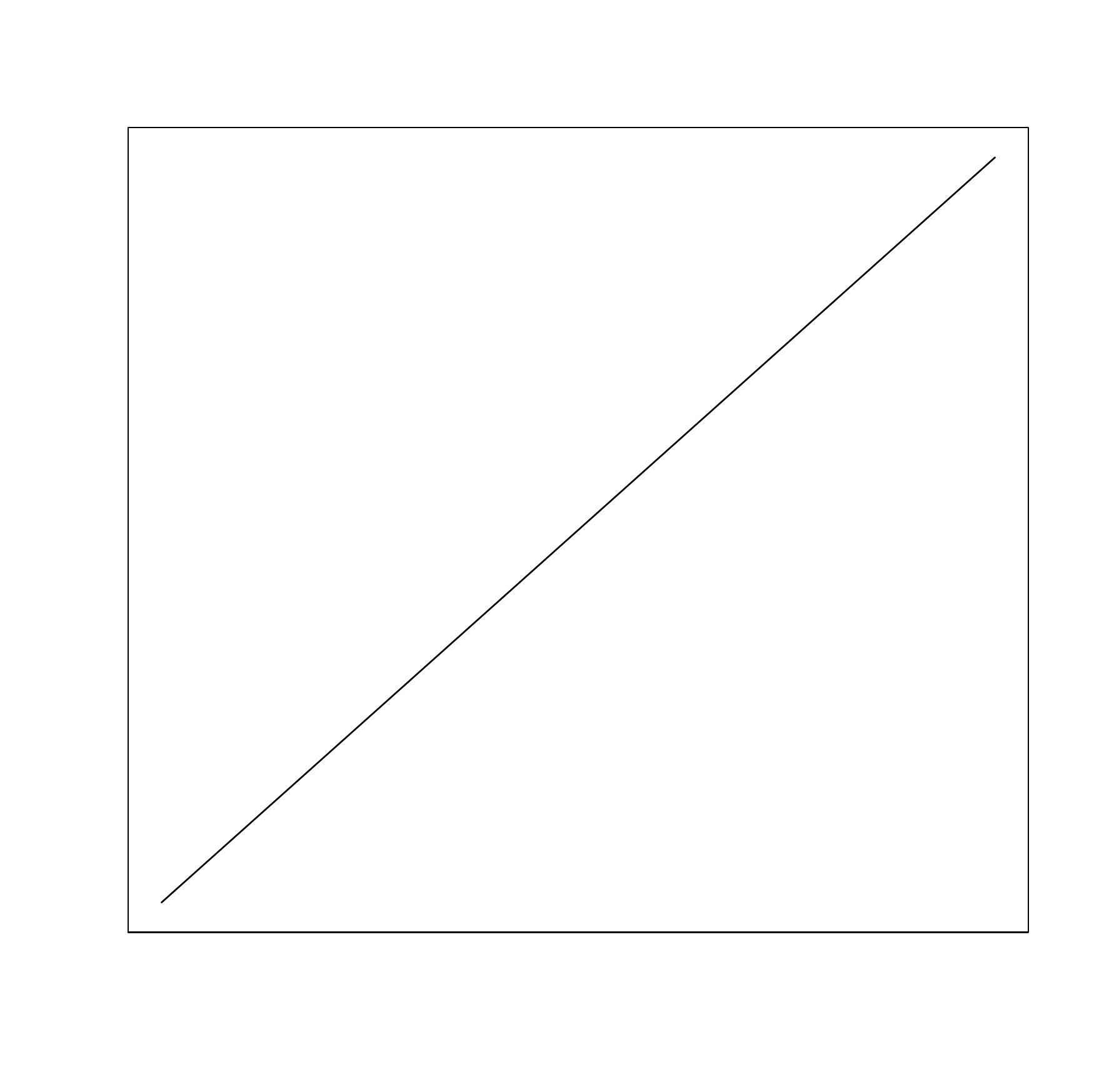}}  & {\includegraphics[scale=0.08]{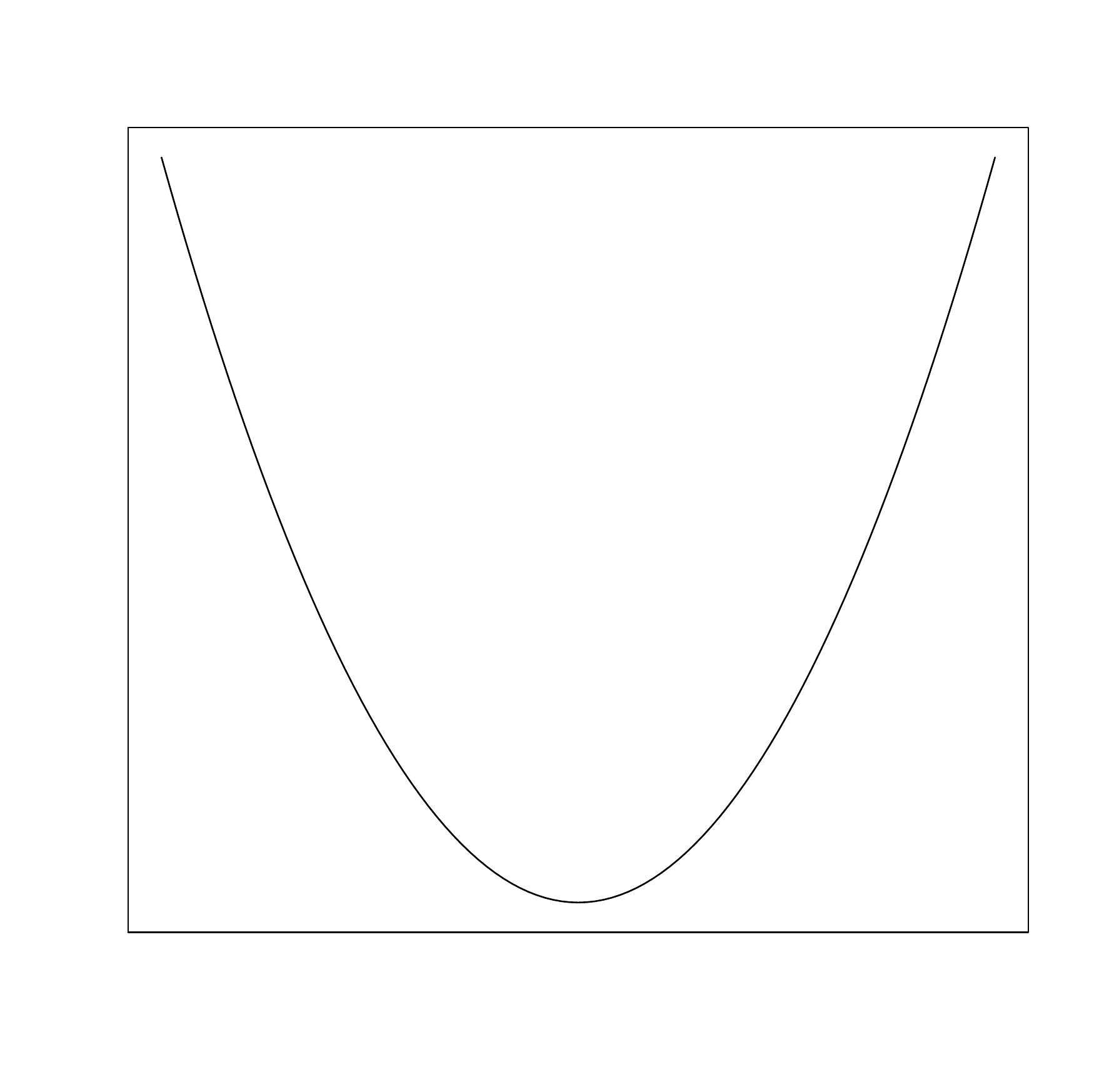}} & {\includegraphics[scale=0.08]{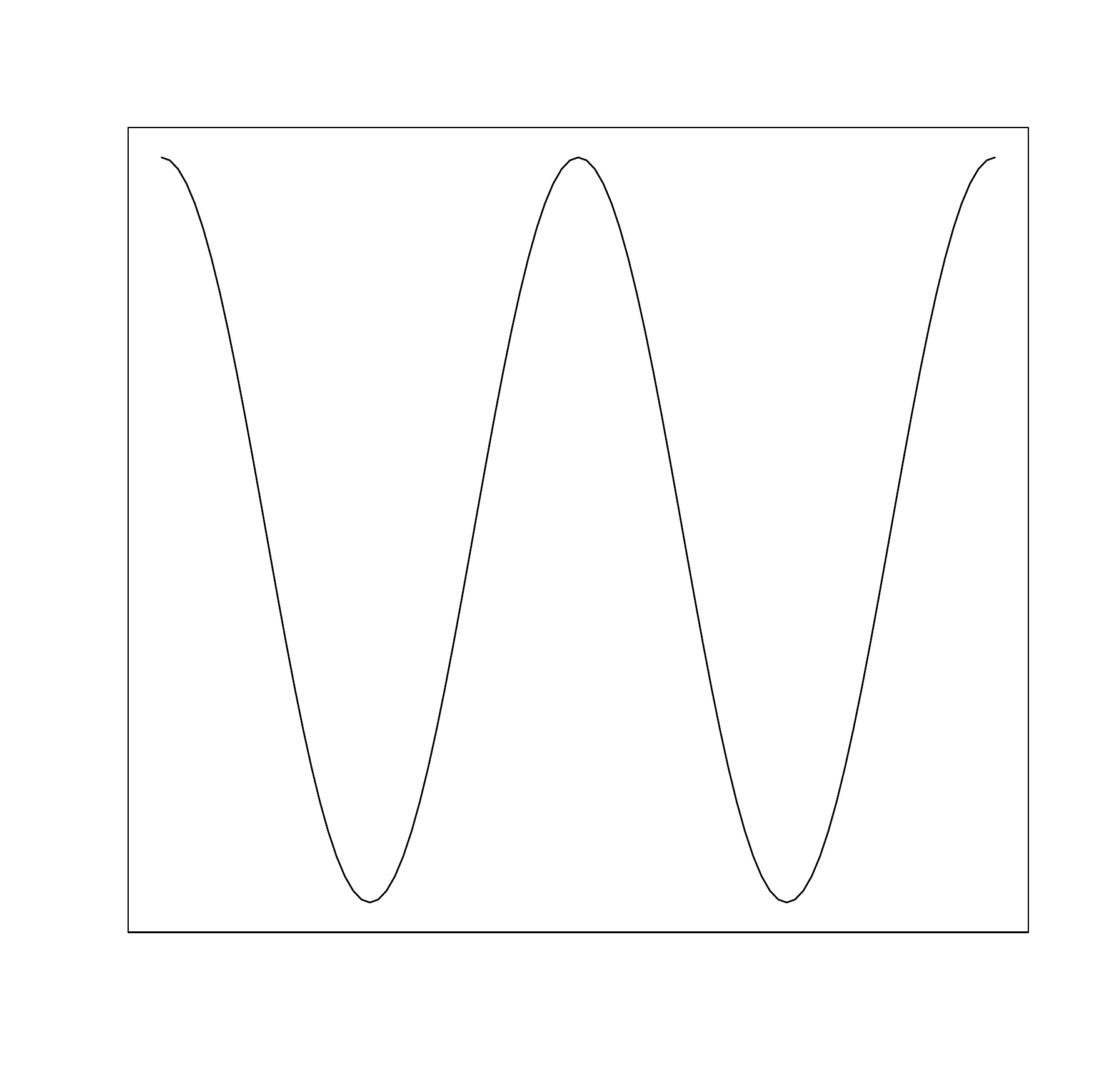}} & {\includegraphics[scale=0.08]{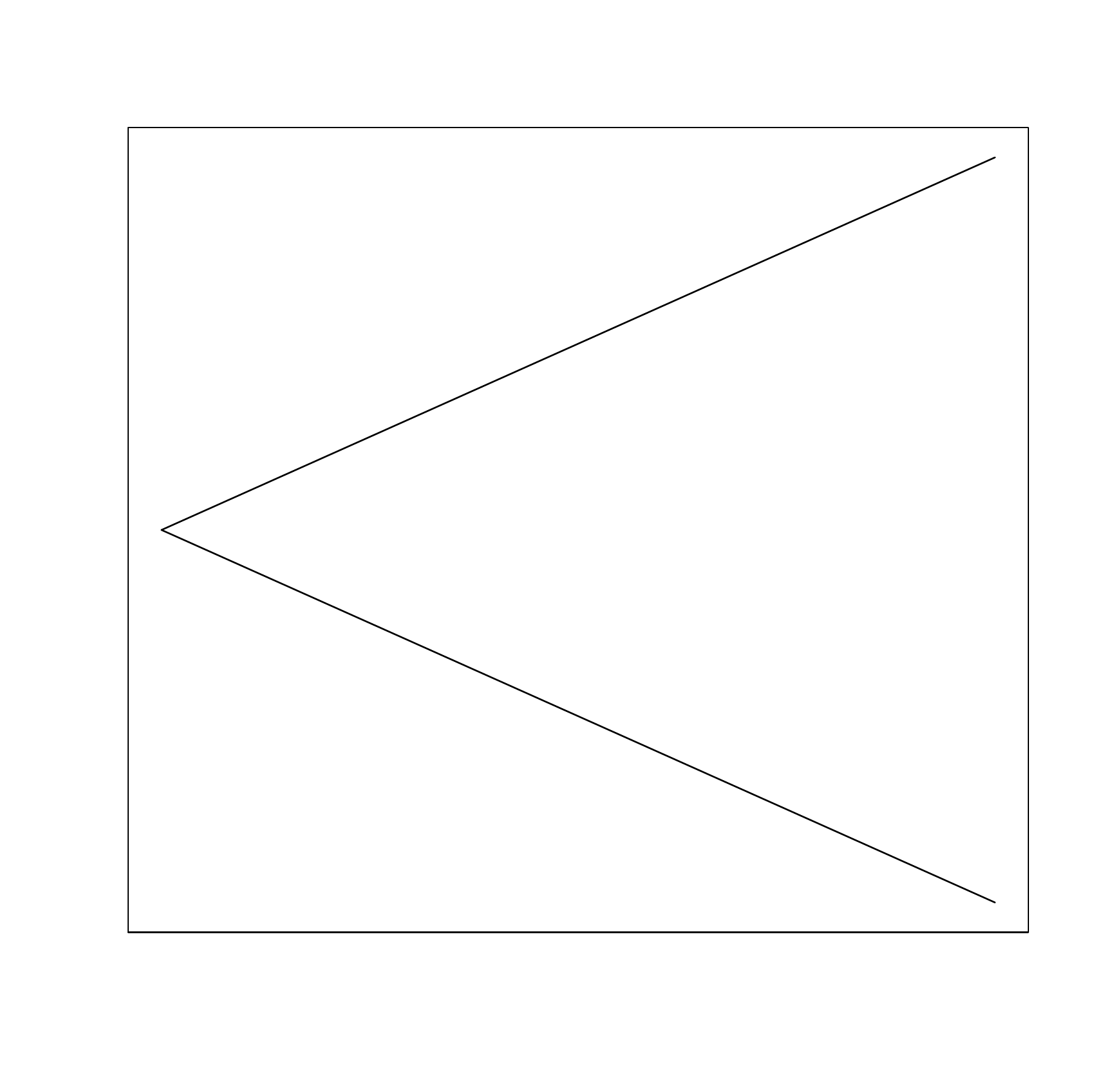}} & {\includegraphics[scale=0.08]{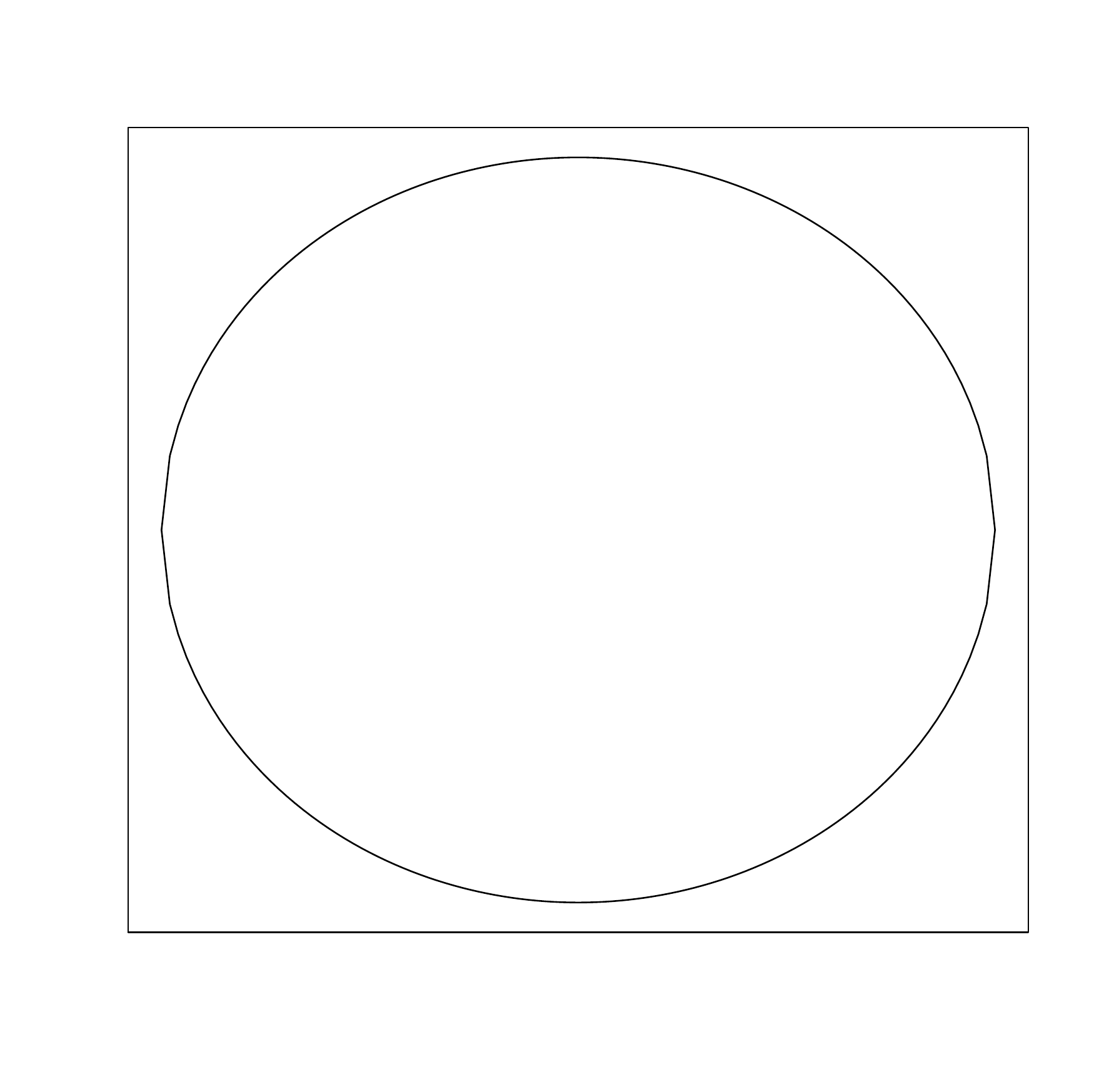}} & {\includegraphics[scale=0.08]{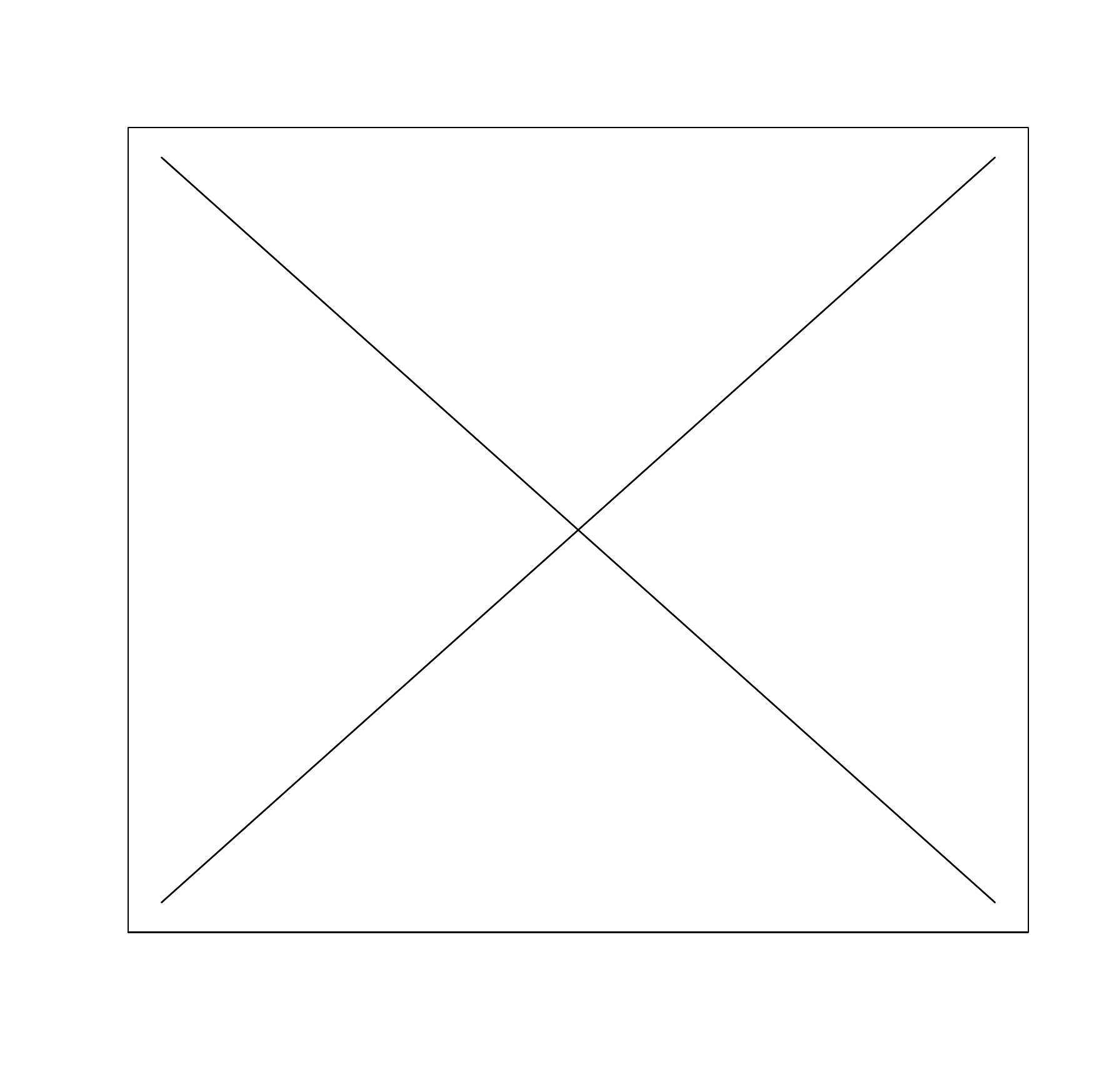}}  \\
 & {\includegraphics[scale=0.08]{lin}}  & {\includegraphics[scale=0.08]{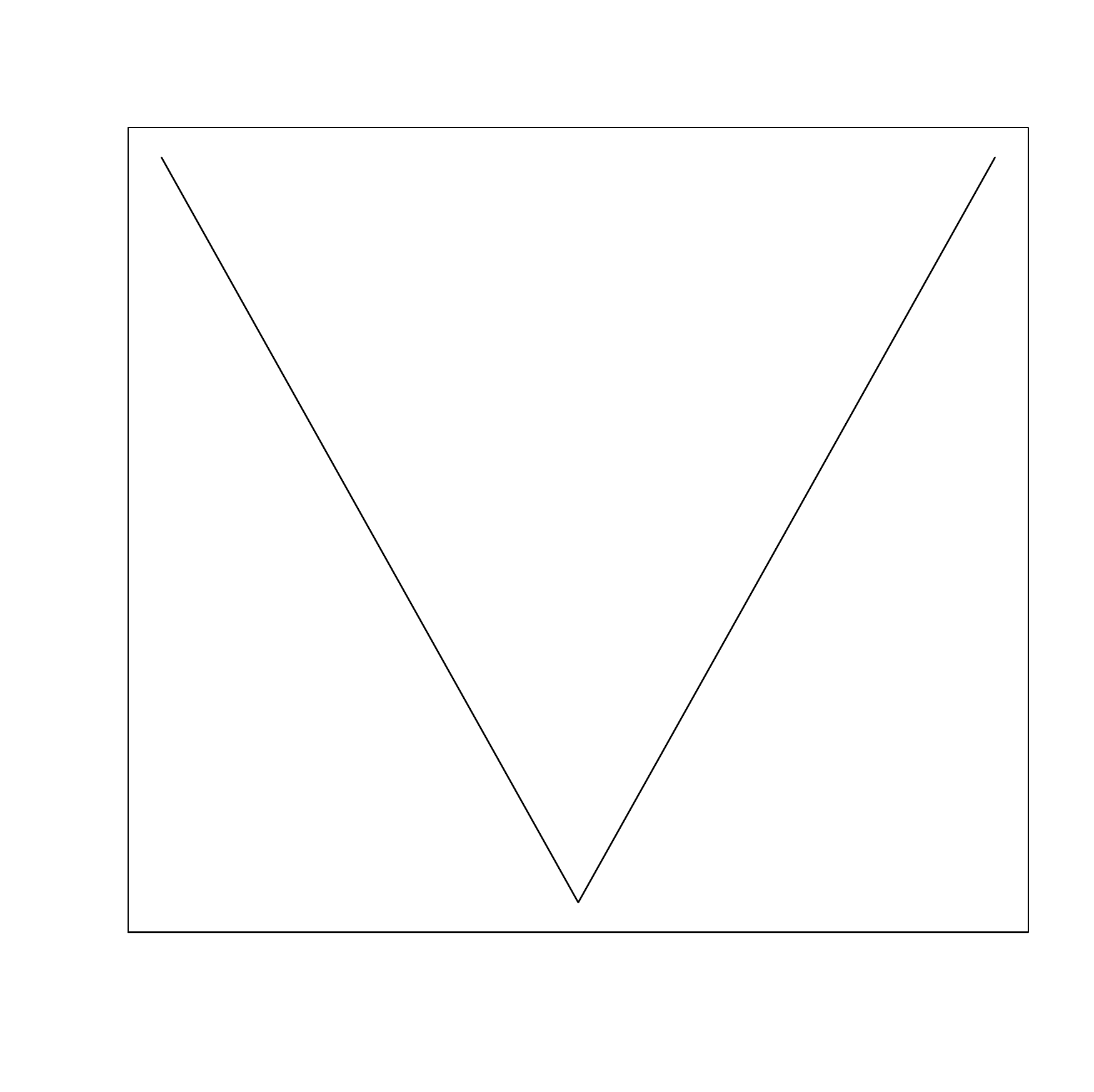}} & {\includegraphics[scale=0.08]{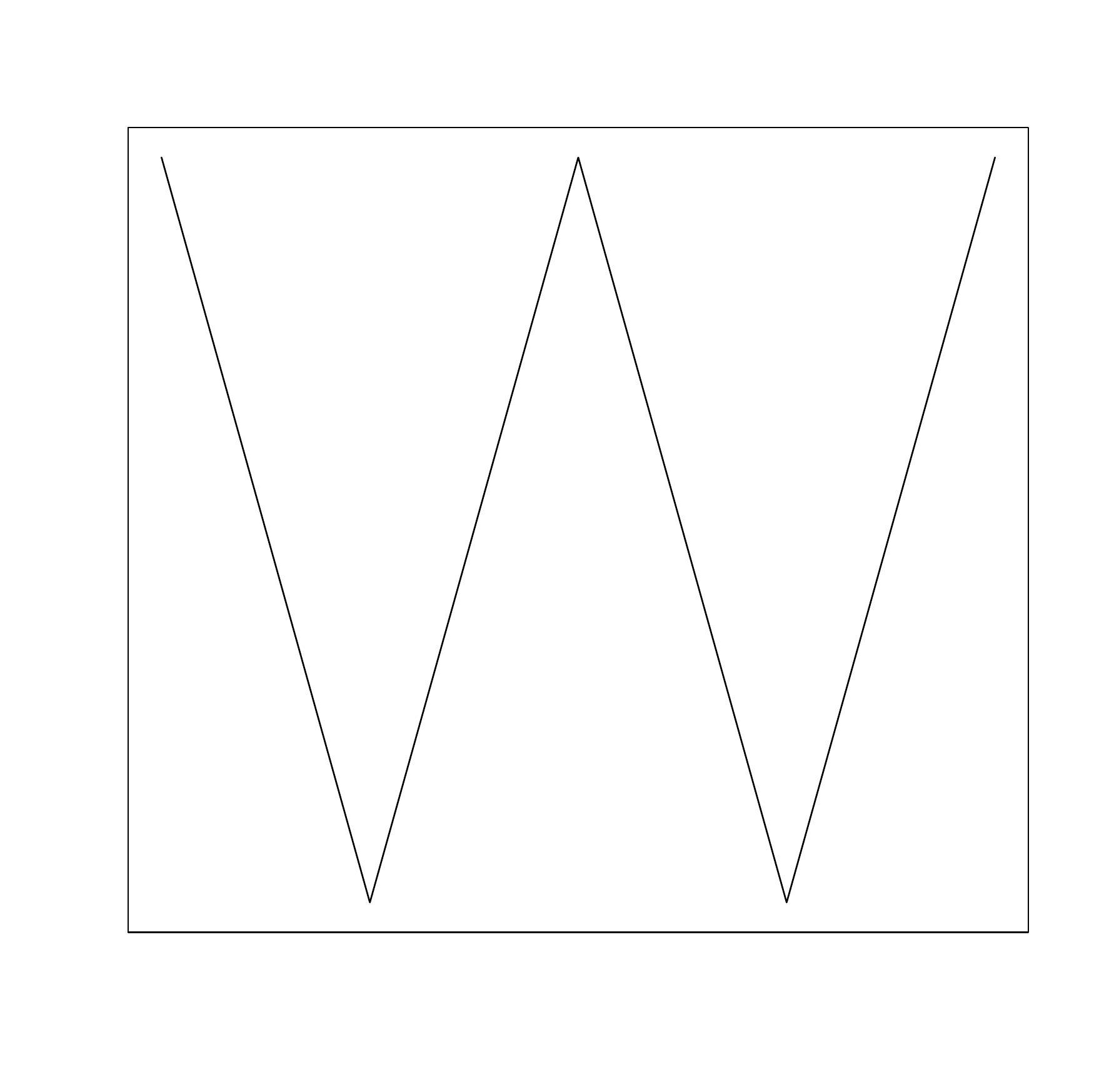}} & {\includegraphics[scale=0.08]{2branches}} & {\includegraphics[scale=0.08]{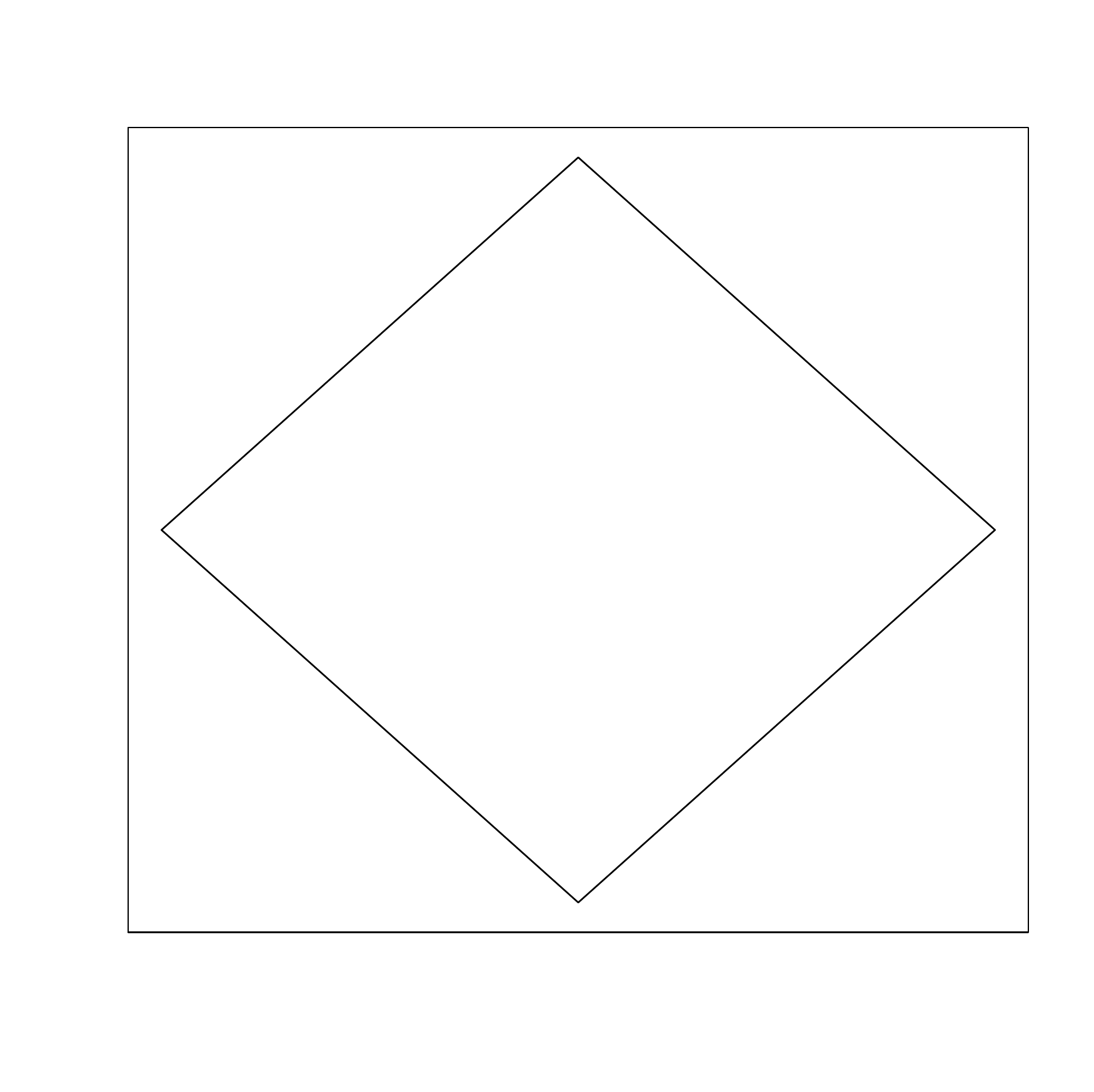}} & {\includegraphics[scale=0.08]{cross}}  \\
    \hline
\end{tabular}
}
\end{center}
\label{tab:Func}
\end{table}

\subsection{Ccor Estimation With Kernel Copula Density Estimator: Bandwidth Selection And Finite Sample Correction}\label{sec:bandwidth}
We estimate Ccor using the plug-in estimator of equation (20) in the main text. For the compact support kernel $K(\cdot)$, we take the constant function on $[-1,1]$. That is, $K(x) = 1/2$ for $-1 \le x \le 1$. Hence the resulting bivariate kernel is simply a square $(u \pm h, v \pm h)$.

We first make a finite-sample correction to $\widehat{Ccor}$. For any fixed sample size $n$ and fixed bandwidth $h$, the estimator $\widehat{Ccor}$ can never reach the value of $1$ and $0$. This problem diminishes for large sample size as $\widehat{Ccor}$ converges to the true value by Theorem~2 in the main text. However, this can be a serious problem for real applications where the sample size is always finite. We make a linear correction of
\beq\label{Ccor.est.corrected} \widetilde{Ccor} =  (\widehat{Ccor} - Cmin)/(Cmax-Cmin). \eeq  Here $Cmax$ and $Cmin$ are the maximum and minimum possible values of $\widehat{Ccor}$ and are functions of $n$ and $h$. $Cmax$ is the $\widehat{Ccor}$ value on perfectly matched $U$ and $V$: $U_i=V_i$, $i=1,...,n$. $Cmin$ is calculated on the most evenly distributed possible case of $(U_i,V_i)$'s. That is, for $U_i$ arranged in increasing order, $V_i$'s are arranged in evenly distributed columns with the neighboring $V_i$s separated by $2h$ distance within each column. The reported values in the numerical studies throughout the paper is for this finite-sample corrected estimator.

We now turn attention to the choice of bandwidth. Theorem~2 suggested the bandwidth $h=b \cdot n^{-1/4}$ for a constant $b$. While asymptotically any $b$ value works, for any finite sample different $b$ values make a big difference. There have been extensive literature on bandwidth selection for density estimations. \citet{wand1993JASAbandwidth} and \citet{wand1994multivariateBandwidth} provided plug-in formulas for choosing bandwidth in multivariate density estimation. However, those formulas can not be directly used here since they are calculated under conditions inappropriate for copula density estimation as argued in the main text. They were calculated for other types of kernels and a Gaussian reference distribution which is not a copula distribution. Also, minimizing estimation error of Ccor is different from minimizing the error in density function $c(u,v)$. In any case, we first still tried to plug into $\widehat{Ccor}$ the bivariate density estimation using the function $KDE2d()$ in R with default bandwidth. This is similar to what is done with $MI$ estimation by \citet{Ganguly2007MI} and \citet{Reshef2011MIC}. The resulting estimator $\widehat{Ccor}$ is ok for big sample size, but can be much improved upon for the mediate sample sizes smaller than thousands.

Therefore, we used an empirical approach to decide on the constant $b$ for bandwidth selection. For the nine functions listed in Table~\ref{tab:FunctionsVariation}, we calculated the true values of Ccor at various noise levels.
\begin{table} [htbp]
\begin{center}
\small{
\begin{tabular}{c|ll}
    \hline
A  & Linear & $y=x$  \\
B  & Quadratic & $y=x^2$  \\
C  & Square Root & $y=\sqrt{x}$  \\
D  & Cubic & $y=x^3$  \\
E  & Centered Cubic & $y=4(x-1/2)^3$  \\
F & Centered Quadratic & $y=4x(1-x)$  \\
G  & Cosine (Period 1) & $y=[cos(2\pi x)+1]/2$  \\
H  & Circle & $(x-1/2)^2+y^2=1/4$  \\
I  & Cross & $y=\pm (x-1/2)$  \\
    \hline
\end{tabular}
}
\end{center}
\caption{ The function relationships used in Figures~\ref{fig:bandwidth}, \ref{fig:bandwidth.small} and \ref{fig:bandwidth.big}.}
\label{tab:FunctionsVariation}
\end{table}
Then we estimated $\widetilde{Ccor}$ on generated noisy data sets using different bandwidth values at sample sizes of $n=10^2$, $10^3$, $10^4$ and $10^5$. The averages of $\widetilde{Ccor}$ from $100$ randomly generated noisy data sets are compared to the true $Ccor$ values to decide on an optimal $b$ value. From this simulation, we decided on the bandwidth $h=0.25n^{-1/4}$. Figure~\ref{fig:bandwidth} plots the simulation results using $h=0.25n^{-1/4}$. We can see that the performance of $\widetilde{Ccor}$ improves as sample size increases, and gives very accurate estimates for $Ccor$ under big sample sizes. For illustration, we showed the plots with bandwidth $h=0.1n^{-1/4}$ and $h=0.5n^{-1/4}$ in Figure~\ref{fig:bandwidth.small} and Figure~\ref{fig:bandwidth.big} respectively. Those bandwidth choices are clearly either too small or too big.

All the reported numerical results in the main text use the plug-in estimator $\widetilde{Ccor}$ in equation (\ref{Ccor.est.corrected}) with a square kernel and bandwidth $h=0.25n^{-1/4}$. This choice works well in the numerical studies. Further investigation of other kernel and bandwidth choices is a future research topic.  Data-based adaptive bandwidth selection~\citep{jones1996surveyBandSelect} could also be investigated.

Another possible future research direction is to consider the Ccor estimator over a range of varying bandwidths. This idea is motivated by the MIC measure. Although theoretically not equitable, \citet{Reshef2011MIC} demonstrated some good attributes of MIC under finite sample. More mathematical investigation of MIC is warranted to understand its behaviour. Studies by \citet{reshef2013equitability} indicate that taking the maximum value of the MI statistics over varying sizes of grids is essential to its stability across different functional relationships in finite samples. It can be proven that taking maximum of the plug-in Ccor estimator over a range of varying bandwidths still results in a consistent estimator. It could be interesting to investigate if such estimators can also take on some good attributes of MIC in finite sample.

\begin{center}
\begin{figure}[htbp]
\includegraphics[scale=0.85]{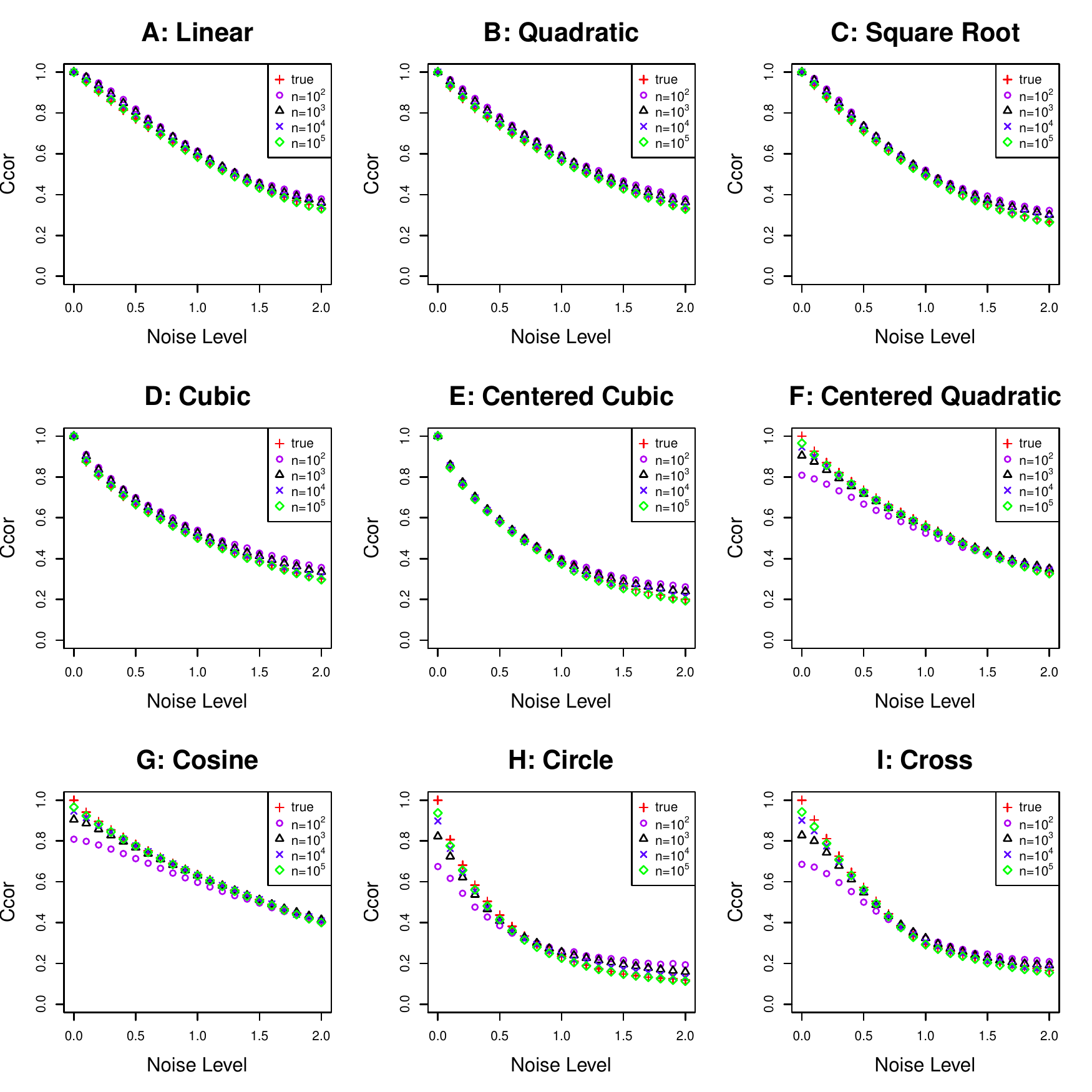}
\caption{The comparison of $Ccor$ with its estimated values under different sample sizes. This estimator uses the square kernel density estimator with bandwidth $h=0.25n^{-1/4}$. }
\label{fig:bandwidth}
\end{figure}

\begin{figure}[htbp]
\includegraphics[scale=0.85]{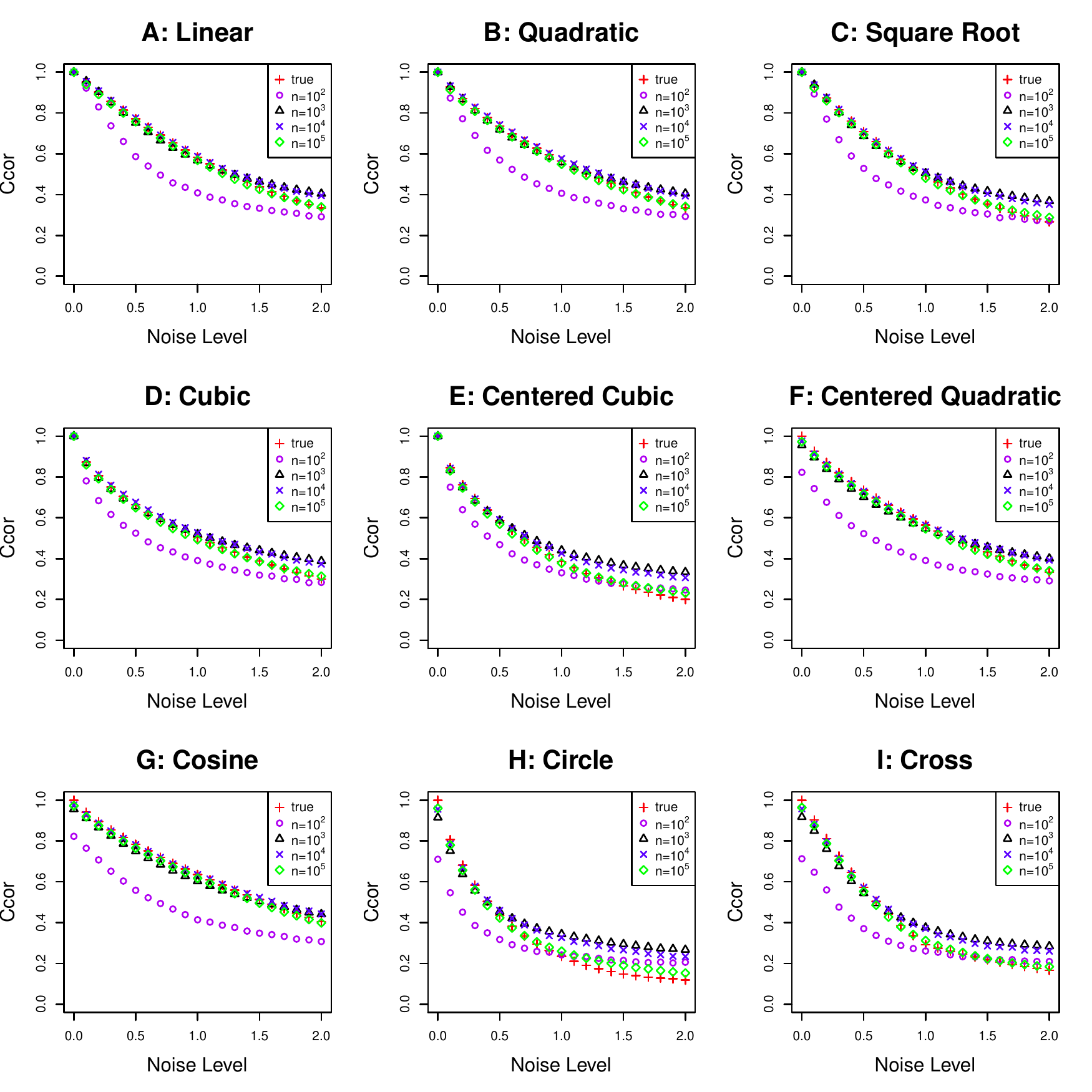}
\caption{The comparison of $Ccor$ with its estimated values under different sample sizes. This estimator uses the square kernel density estimator with bandwidth $h=0.1n^{-1/4}$. }
\label{fig:bandwidth.small}
\end{figure}

\begin{figure}[htbp]
\includegraphics[scale=0.85]{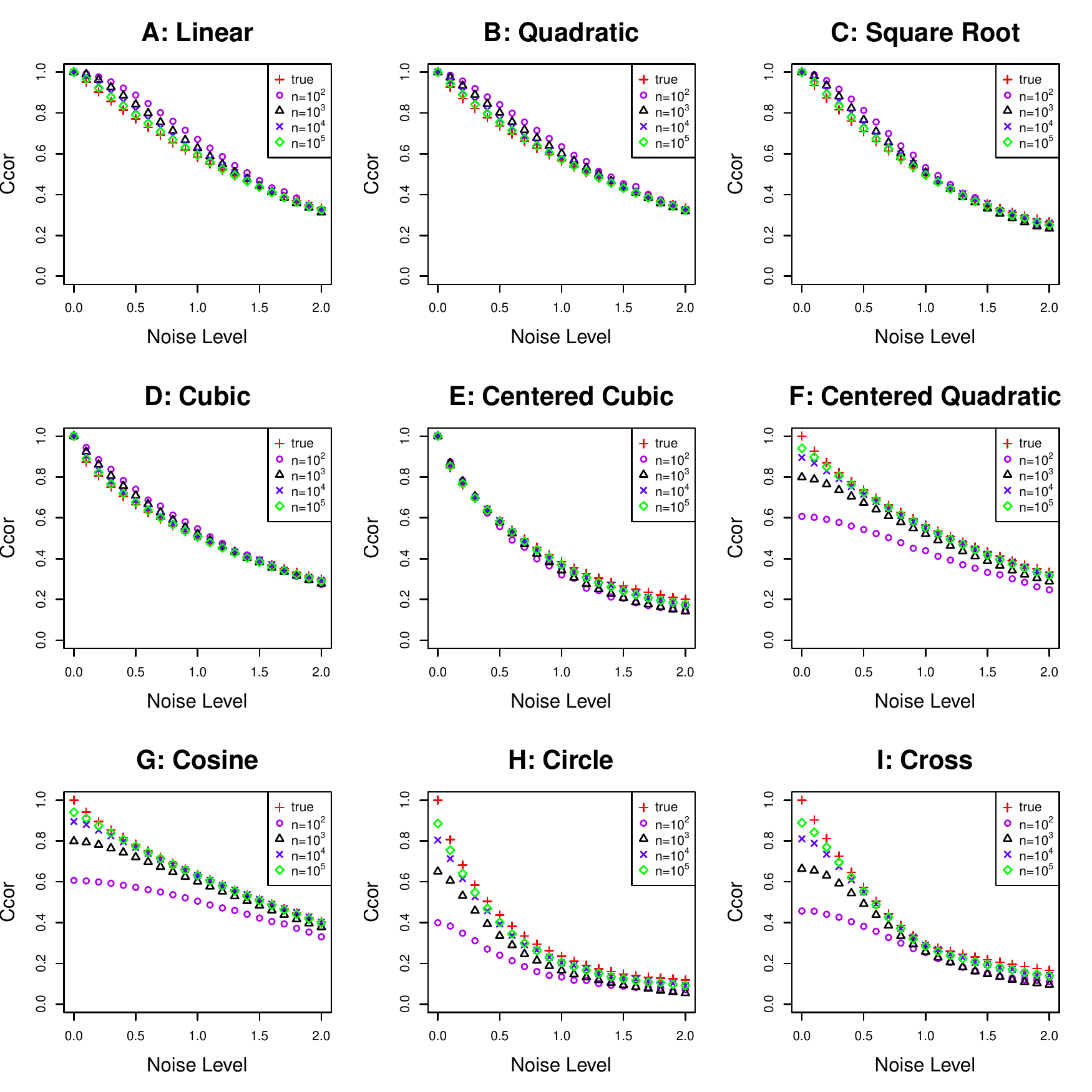}
\caption{The comparison of $Ccor$ with its estimated values under different sample sizes. This estimator uses the square kernel density estimator with bandwidth $h=0.5n^{-1/4}$. }
\label{fig:bandwidth.big}
\end{figure}
\end{center}

\end{document}